\documentclass[a4paper,fleqn,usenatbib,letter]{mnras}

\usepackage{newtxtext,newtxmath}

\usepackage[T1]{fontenc}
\usepackage{ae,aecompl}

\usepackage{graphicx}	
\usepackage{amsmath}	
\usepackage{amssymb}	

\newcommand{\msun}{{M$_\odot$}}

\renewcommand{\deg}{^\circ}

\newcommand{\rva}[1]{{ #1}}
\newcommand{\rvb}[1]{{#1}}

\title[Close binary black holes in powerful jet sources]{How frequent are close supermassive binary black holes in powerful jet sources?}

\author[M. G. H. Krause et al.]{
Martin G. H. Krause$^{1,2}$\thanks{E-mail: M.G.H.Krause@herts.ac.uk},
Stanislav S. Shabala$^{1}$,
Martin J. Hardcastle$^{2}$,\newauthor
\:\,Geoffrey V. Bicknell$^{3}$, 
Hans B\"ohringer$^{4}$, 
Gayoung Chon$^{4}$, \newauthor
\:\,Mohammad A. Nawaz$^{5}$, 
Marc Sarzi$^{2,6}$ and 
Alexander Y. Wagner$^{7}$\\
$^{1}$School of Natural Sciences, Private Bag 37, University of Tasmania, Hobart, TAS, 7001, Australia\\
$^{2}$Centre for Astrophysics Research, School of Physics, Astronomy and Mathematics, University of Hertfordshire, 
College Lane, \\ \:\,Hatfield, Hertfordshire AL10 9AB, UK\\
$^{3}$Research School of Astronomy and Astrophysics, The Australian National University, ACT 2611, Australia\\
$^{4}$Max-Planck-Institut f\"ur extraterrestrische Physik, Giessenbachstrasse, Garching D-85748 Germany\\
$^{5}$Escola de Artes, Ciencias e Humanidades, Universidade de Sao Paulo, Rua Arlindo Bettio, 1000, Sao Paulo, 
SP 03828-000, Brazil\\
$^{6}$Armagh Observatory and Planetarium, College Hill, Armagh, BT61 9DG, UK \\
$^{7}$Center for Computational Sciences, University of Tsukuba, 1-1-1 Tennodai, Tsukuba, Ibaraki, 305-8577 Japan
}

\date{Accepted XXX. Received YYY; in original form ZZZ}

\pubyear{2017}

\begin{document}
\label{firstpage}
\pagerange{\pageref{firstpage}--\pageref{lastpage}}
\maketitle

\begin{abstract}
Supermassive black hole binaries may be detectable by an upcoming
suite of gravitational wave experiments. Their binary nature can
also be revealed by radio jets via a short-period precession driven
by the orbital motion as well as the geodetic precession at
typically longer periods. We have investigated Karl G. Jansky Very
Large Array (VLA) and MERLIN radio maps of powerful jet sources for
morphological evidence of geodetic precession. For perhaps the best
studied source, Cygnus A, we find strong evidence for geodetic
precession. Projection effects can enhance precession features,
\rva{for which we find indications in} strongly projected sources.
For a complete sample of 33 3CR radio sources we find strong
evidence for jet precession in 24 cases (73 per cent). The
morphology of the radio maps suggests that the precession periods
are of the order of $10^6-10^7$~yr. We consider different explanations
for the morphological features and conclude that geodetic precession
is the best explanation. 
The frequently observed gradual jet angle changes in samples of 
powerful blazars can be explained by orbital motion. 
Both observations can be explained simultaneously by postulating that 
a high fraction of powerful radio sources have sub-parsec
supermassive black hole binaries.
We consider complementary evidence and discuss if any jetted
supermassive black hole with some indication of precession could 
be detected as individual gravitational wave source in the near future.
This appears unlikely,
with the possible exception of M87.
\end{abstract}

\begin{keywords}
black hole physics –- galaxies:jets -- galaxies: quasars: supermassive black holes -- 
radio continuum: galaxies -- gravitational waves
\end{keywords}


\section{Introduction}\label{sec:intro}

Black holes with masses of millions to billions of solar masses reside in the centers of most galaxies 
\citep[e.g.,][]{Magea98,HR04,CaDoToMe17}. 
The hierarchical growth of galaxies by merging therefore inevitably leads to binary systems of supermassive 
black holes \citep*{BBR80}. 
Plausible cases of wide separations are known \citep[e.g.,][]{Kmsea03}, evidence for supermassive black hole 
binaries at separations of a parsec and less is accumulating \citep{Grahea15,Liea16,Bansea17} and possible hints for actual 
black hole mergers have been suggested \citep{MeEk02,Chiabea17}. 
\rva{Recently, a double radio nucleus with sub-parsec separation has been resolved with  
Very Long BaseLine Interferometry \citep[VLBI,][]{KLM17}. }

Thus, we may be able to to extend the mass range of direct detection of black-hole merging via gravitational waves
\citep[e.g.,][]{Abbottea16a} to that of supermassive black holes. The lower mass end of the supermassive
regime is expected to become accessible with the Laser Interferometer Space Antenna \citep[LISA, up to about
$10^7$~\msun,][]{Amarea12}, whereas the Square Kilometre Array pulsar timing array will be able to constrain 
the highest known black hole masses in the cosmic neighbourhood \citep[e.g.,][]{Zhuea15}.

Detailed simulations of the galaxy merging process with zoom simulations
of the central parsec including post-Newtonian corrections predict that any contained black holes
are driven to sub-parsec separations within $10^{7-8}$~yr, where they stay for another $10^{7-9}$~yr,
these timescales being
shorter at higher redshift \citep{Kahnea16,Mayer17b}. The latter case can be probed with radio observations
if at least one of the black holes is spinning and produces a jet. 

If a jet-producing supermassive black hole is a member of a close binary system, one expects two observable 
effects \citep[for a review]{BBR80,BBR84}. 
First, the orbital motion imposes a widening helical pattern on the jet. For sub-parsec supermassive black 
hole binaries, orbital periods from less than a year to many thousands of years are expected, corresponding 
to structure on scales of parsecs up to about a kiloparsec. Secondly, both spin axes, likely coincident with the 
jet axes \citep{BBR80,MeEk02} precess around the orbital angular momentum vector due to the geodetic precession. The effect 
is caused by the motion of the masses, similar to the magnetic force being caused by the motion of electric 
charges. Massive black holes with sub-parsec separations can produce periods of the order of $10^6$ years, 
significantly less than the typical ages of observed radio sources, $10^7$-$10^8$ years
\citep*[e.g.,][]{Krause2005a,EHK2016,Harwea17,Turnea18b}, such that jet-lobe
asymmetries and jet curvature would be visible on radio maps of 100~kpc-scale jets.

Orbital motion and jet precession have been discussed and detections have been claimed for many 
individual jetted supermassive black hole systems.
\citep[e.g.][]{Ekea78,GowHut82,Gowea82,Bary83,HutPriGow88,AbrahamCar98,Morgea99,SteenBlun08,Kharbea14,Kunea14,Kunea15,Ekers16,Britzea17}.
Derived periods are indeed typically of the order of 10-1000 years for VLBI and $10^5$-$10^7$ years for 
shorter baseline observations (VLA), as expected for parsec-scale supermassive
black hole binaries and reflecting the sensitivity range given by the scale of the respective radio maps.


%
Here, we address the question of how common the evidence for binary black holes is among powerful extragalactic radio sources. 
We first investigate the object with probably the best available data on all spatial scales, Cygnus~A, and show
that both the 100-kpc-scale and parsec-scale morphology shows evidence for perturbations of the jet consistent
with a \rva{sub-parsec} supermassive binary (Sect.~\ref{sec:cyga}).
Because one prediction from relativistic aberration theory
is that precession effects become more pronounced in more strongly projected sources, we investigate the two
broad line radio galaxies with high quality radio maps 
in the 2Jy sample in Sect.~\ref{sec:2Jy-blrgs}. They show clear evidence of precession,
pointing to close binary supermassive black holes. In a complete sample of radio galaxies 
\rvb{(derived from the 3CRR sample)} we find morphological
evidence for close binary supermassive black holes in 73~per cent of the sources (Sect.~\ref{sec:sample}).
In Sect.~\ref{sec:disc}, we discuss our findings in the context of other work, in particular the
possibility of direct detections of jetted supermassive binary black holes via gravitational waves
(Sect.~\ref{sec:gw}). We conclude in Sect.~\ref{sec:conc} that subparsec supermassive binary black holes appear
to be a common feature for powerful jet sources, but will in general be challenging to
detect with gravitational wave instruments. 

\section{Data analysis}

\subsection{Radio maps and X-ray data}\label{sec:data}
We use published radio maps throughout, except where otherwise stated. 
The 5~GHz VLA map of Cygnus~A was previously 
published in \citet{CarBar96}\footnote{Kindly provided in electronic
format by Chris Carilli and Rick Perley.}. We also used the VLBI maps from 
\citet{Boccardiea14,Boccardiea16a,Boccardiea16b}\footnote{We adopted the radio map from 
\citet{Boccardiea14} in Fig.~\ref{fig:cygA5GHz} under the Creative Commons Attribution -- Non Commercial Share Alike Licence.}. 
\rva{We used archival Chandra ACIS-I observations for Cyg A taken in VFAINT mode and performed 
standard data analysis with CIAO~4.8 with calibration database (CALDB)~4.7.1. 
This includes flare cleaning by filtering light curves of source-free regions, VFAINT mode, 
gain, and charge transfer inefficiency corrections \citep{Chonea12}. 
The total cleaned exposure amounts to 3.2~Ms for the image presented in this paper. 
An ACIS blank sky observation was used for our background estimate after the 
normalisation was adjusted. We used the 9-12.5 keV count rate to get the appropriate 
normalisation for the background exposure.}

We made use of the 2Jy~sample \citep{WaPe85} and associated data 
compilation\footnote{C. Tadhunter, R. Morganti, D. Dicken (2014), The 2Jy Sample: 
\url{https://2jy.extragalactic.info/The\_2Jy\_Sample.html}}.
The 5GHz data for 3C~17 and 3C~327.1 were downloaded from the VLA archive and 
reduced with standard data 
analysis methods (AIPS). The 3C~17 image uses archival data at A, B and C configurations, 
and the 3C~327.1 one uses A and B configurations. 
\rva{The project IDs are AS179, AD419, AM548 and AM270.}

We also analysed a complete sample \citep{Blackea92,Hardea97,Hardea98,Lea97,Gilbea04,Mulea06,Mulea08} 
of 98 radio galaxies based on the 3CRR catalogue \citep*{Laingea83}. 
The selection criteria were \citet[FR]{FR74} class II (edge-brightened, powerful radio galaxies), redshift 
less than unity. Radio flux at 178 MHz $>10.9$~Jy, declination $>10\deg$ and Galactic latitude 
$|b| > 10\deg$ are inherited from the 3CRR sample.   The data are available from: \url{https://zl1.extragalactic.info}. The radio maps have a 
resolution of typically one arcsecond or better and have been obtained with the VLA and MERLIN radio telescopes. 
We considered \rvb{the} sources where the database shows at least one definite jet detection. 
Exceptions are 4C12.03, 3C219 and 3C341 where we decided after careful re-assessment of 
the original data that the previously identified jets are more likely thin lobes of a restarted radio 
source. Therefore, we did not consider these three sources. The final sample we used for the 
present study comprises 33 radio sources with high quality radio maps 
\rva{for which jets had previously been identified.}
\rvb{We show the radio maps in Appendix~\ref{ap:data}.}

\subsection{Assessment of jet precession}\label{sec:assessment}

To determine whether a jet is precessing or not, one needs to consider the intrinsic features of precessing jets
as well as possible complications such as the jet-environment interaction.

\subsubsection{Jet stability and deflections}\label{sec:jstab}
In the simplest case, an FR~II radio source inflates a lobe that is { of comparable density with respect to}
the jet \citep[e.g.,][]{mypap03a,EHK2016}. A relativistic, super-magnetosonic
jet is then safe from disruption by instabilities \citep[e.g.][]{Appl96,Alea99,RHC99,
Peruea04b,Peruea10}, at least as long as the jet is not too strongly magnetised
\citep[e.g.,][]{Appl96,OBB12}. Jets attain their asymptotic velocity already close to the core
\citep[e.g.][]{Komea07,PF10}. In any case, magnetic acceleration, which requires a finite opening angle, 
will not work on kpc scales, where
the jets are expected to be collimated \citep[e.g.][]{Krausea12b}. Each piece of plasma in the jet
will therefore to first order follow a ballistic trajectory at constant velocity. The projection onto
the sky plane due to relativistic aberration has been studied \rvb{for example} by \citet{Gowea82}.

\rva{Possible complications} include deflection at dense clouds 
in the centre of the host galaxy. 
Jet-cloud interactions are strong when such sources are young and the host galaxy is gas-rich, 
as shown in 3D
simulations \citep{Gaiblea12,WBU12,Mukhea16}. The lobe length asymmetry in some sources,
e.g., Cygnus A, probably still witnesses this early phase \citep{GKK11}. 

The simulations show, however, that the jet path is cleared when the jet has traversed
the scale height of the host galaxy ($\lesssim$~kpc). 
Therefore, in comparatively old FR~II sources, like Cygnus A and the sample considered here, 
jet-cloud interactions are unlikely to be much of an issue: 
A realistic giant molecular cloud, 
comparable in size to the jet diameter (say, 100 pc) can reasonably be expected to 
be of the order of \rva{$\eta^{-1}>10^5$} times denser than the jet \citep{CWR07}. For a jet 
with bulk velocity $v_\mathrm{j}$, momentum flux conservation then
determines its head advance speed to be approximately $v_\mathrm{j}\sqrt{\eta}$
\citep[e.g.,][]{mypap01a}.  
The jet would have overcome such a cloud in at most $10^5$ years, a small fraction 
of the source age of 100~kpc scale radio sources. 
Interstellar clouds move so slowly that they are being ablated as 
they move into the jet path. 
Further, such deflections would 
lead to bright radio hot spots close to the core, which generally are not seen in the maps we show here,
in good agreement with the expectation that any clouds in the jet path have been cleared out 
in the early evolution of the sources. 

\subsubsection{Dynamical or asymmetric environments}\label{sec:env}
Another effect that is important in this context is the motion of the host galaxy relative to the 
surrounding hot ambient gas, either because the host galaxy is not the central member of the
group or cluster, or because the latter \rva{has} had a recent merger with another galaxy cluster or
substructure, which happens comparatively frequently \citep{ChonBo17}.  
Simulations of jets with crosswinds show that the jets may be bent and are found towards
the windward side of the lobes \citep{BalNorm92}. Such cases may be recognised, however,
because one expects a plane-symmetric effect for both jets \rva{(compare, e.g., 3C~98, Fig.~\ref{fig:3c98}),} 
as opposed to S-symmetry for precession \rva{(e.g., 3C~334, Fig.~\ref{fig:3c334}).} 
The 3D simulations of precessing jets with crosswinds of \citet{Rodrea06}
show that the S-symmetry should still be observable in such cases.

If a hydrostatic galaxy halo has elliptical or triaxial symmetry and 
a non-precessing jet propagates
in a direction that is not aligned with one of the symmetry axes of the halo we 
also expect an S-symmetric
morphology, similar to the morphology of precessing jets.
In a 3D simulation with such a setup, \citet{Rossiea17} find that the lobes adjust to the
ambient pressure profile whereas the jets remain straight. This indeed produces an S-symmetric 
appearance (their Fig.~9), comparable to what is expected from precession. 
This may confuse the interpretation of individual sources, particularly,
when the precession is slow enough, so that the jets appear straight. 
However, 
for non-precessing jets in triaxial halos we expect no 
strong dependence 
of the misalignment angle between jet direction and lobe axis on the viewing angle,
because the jet direction should not be correlated with the axes of the dark matter halo. 
For a precessing jet, this misalignment angle will be much smaller for edge-on sources than 
for smaller inclination. In the extreme case, when the line of sight is within the precession cone,
{ the angle between jet and lobe axis is unconstrained.}

This issue can therefore be addressed by 
examining strongly projected sources like broad line radio galaxies (BLRGs).
We { investigate} the broad line radio galaxies  in the 2~Jy flux-limited sample 
in Sect.~\ref{sec:2Jy-blrgs} finding outstanding misalignments between jets and lobes.
\rva{In a statistical analysis, we confirm this trend, but with low significance.}
We also note that the X-ray isophotes in Cygnus~A outside of the jet-influenced region suggest that the halo
is approximately spherically symmetric \citep{Sea01}. Massive elliptical galaxies,
the typical hosts of powerful jets, tend to be round, slow rotators \citep[e.g.,][]{Weijmea14},
which also suggests that the underlying gravitational potential may often have close 
to spherical symmetry.
\rva{Overall, this suggests that precession is more important than triaxiality 
\rvb{of the hydrostatic galaxy halo} for powerful jets.}

We conclude that for large-scale ($\gtrsim10$~kpc) FR~II radio sources, 
\rva{stability and jet-environment interaction} are 
reasonably well understood, so that it appears possible to pick up a precession signal
in the radio maps.

\subsubsection{Signatures of precessing jets}\label{sec:pcrits}
Three-dimensional hydrodynamics simulations of precessing jets 
have demonstrated the  
morphological features that are specific to precessing jets 
\citep*[e.g.,][]{CGS91,Moncea14,DonSmi16}, especially, when 
compared to simulations of the interaction of non-precessing jets with their environment 
\citep*[e.g.,][]{WB11,EHK2016}. 
Distinctive features of precessing jets include gradually curving jets with S-symmetry if both jets are detected. 
In general, the jet direction differs from the symmetry axis of the lobes. This is, because 
the jet fluid moves at much higher velocity than the lobe advances into the environment.
Thus, even if the jet ejection direction changes very little during the time it takes a piece of jet
to travel from the jet formation region to the tip of the lobe, this current direction will still be
misaligned with the lobe axis, which records a much longer term average.
The impact of the jet on the lobe boundary causes a high-pressure hotspot 
\citep[e.g.,][]{EHK2016}, which will advance the lobe boundary, most strongly
in the current jet direction. For slowly precessing jets, this may lead to lobe extensions.
Faster precession might carve out (partial) rings near the tip of the lobe.

Hydra~A is to our knowledge the only precessing extragalactic radio source, 
for which a dedicated 3D hydrodynamics simulation 
study has been carried out. \citet{Nawea16a} were able to reproduce the radio morphology of the source with
high accuracy. With a precession period of 1~Myr, they achieved a close match between the simulated
and the observed radio maps, whereas for periods of 5~Myr or longer, the jet became too straight
 (more details in Appendix~\ref{ap:hydra}). 

\begin{figure*}\centering
	\includegraphics[width=0.7\textwidth]{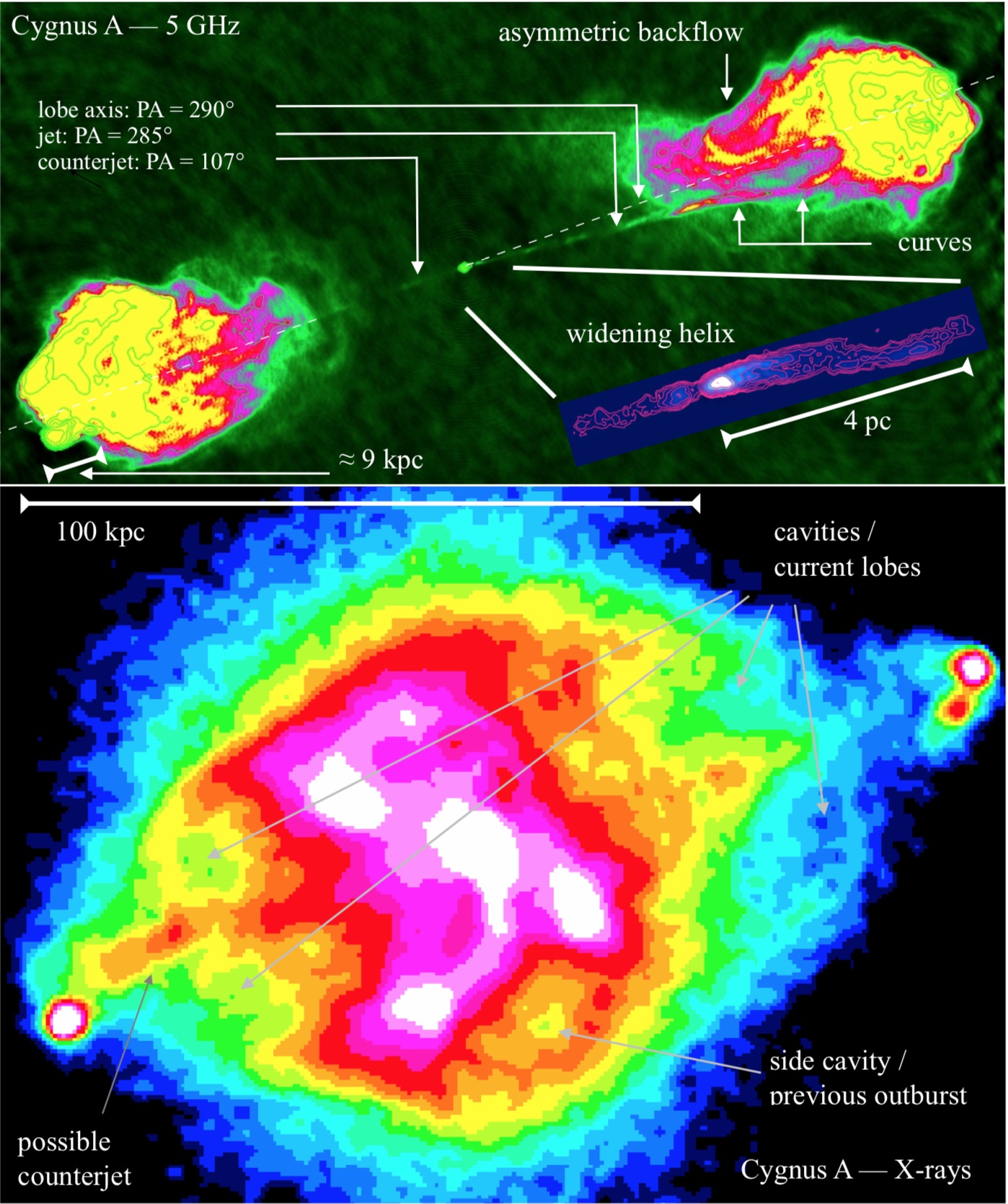}
	\caption{Cygnus A radio galaxy. 
	Top: 5~GHz VLA image. Jet (right) and counterjet (left) are seen to emanate from the 
	bright core in the middle of the image. The jet is first directed towards position angle 
	$PA=285\deg$ and then curves first westwards and then northwest towards the hotspots. 
	A fainter counterjet extends initially towards 
	$PA=107\deg$. The position angle of the jet in particular differs from the lobe axis. 
	\rva{Credits: NRAO/AUI, Chris Carilli and Rick Perley, \citet{CarBar96}. 
	Inset: 43 GHz VLBI observation
	of the core region, showing helical motion in the parsec-scale jet \citep{Boccardiea14}.	 
	Bottom: Chandra X-ray colour map in the 0.5-2~keV band. Outer contours are close to spherically symmetric.
	Labelled: cavities coincides with the current radio lobes, a side cavity, possibly related to a previous outburst,
	a possible counterjet.}
    	\label{fig:cygA5GHz}}
\end{figure*}

Relativistic aberration effects \citep{Gowea82} may affect the intrinsic symmetry between jet and counterjet,
{ i.e. precessing sources are not necessarily exactly S-symmetric.}

The criteria we used to assess the evidence for binary black holes for the presence of precession are:
\begin{enumerate}
\item[(C) --] Jet curvature. A precessing jet curves and the shape of the jet depends on its 
precession parameters (precession angle, precession period). However, it is not straightforward 
to understand precession of jets from their observed curvature since the apparent jet structure 
strongly depends on the viewing direction. A precessing jet can appear with strong or mild curvature 
or even as a straight jet. {The counterjet is usually more strongly curved than
the approaching jet  \citep[relativistic aberration, compare][]{Gowea82}. 
While jet curvature may also be produced by temporal pressure
imbalances or 3D instabilities, especially near the high pressure hot spot \citep{HN90}, 
curvature is clearly a prime characteristic of a precessing jet \citep{Scheuer82,CGS91}.
We show in Appendix~\ref{ap:hydra} that even for the comparatively weak jet source Hydra~A,
a ballistic model for a precessing jet can reproduce the curvature in the jet well.}
\rva{Curved jets are particularly well visible in 3C~17, 3C327.1 (Fig~\ref{fig:2JyBLRGs}), 
3C~388 (Fig.~\ref{fig:3c388}), and 3C~401 (Fig.~\ref{fig:3c401}). }
\item[(E) --] Jet at edge of lobe. In a precessing source the lobes conserve the long-term average jet direction
(i.e. the orbital angular momentum direction for precessing jets from binary black holes), whereas 
the jet indicates the current direction of the jet-producing black hole's spin axis. 
The result can be visible as a jet at the lobe-edge, that is, a misalignment 
between the lobe axis and the jet direction will be introduced. 
\rva{Misalignment is well visible in Cygnus~A (Fig~\ref{fig:cygA5GHz}). Clear cases for jets at the edge of the lobe 
are, e.g., 3C~327.1 (Fig~\ref{fig:2JyBLRGs}) or 3C~334 (Fig.~\ref{fig:3c334}).}
Additionally, the back flow caused by a precessing jet is not isotropic. Therefore, a precessing jet will exhibit an asymmetry in lobe brightness, i.e., the lobe will be brighter in the side where the back flow is strong and fainter on the other side.
\rva{Cygnus~A is an excellent example for this.}
\item[(R) --] Wide / Multiple terminal hotspots possibly with ring-like features. A fast-precessing jet 
interacts with a wide area of environment at the jet-head location. 
As a result, multiple hotspots, wide hotspots or a trail of hot back flow in a different direction 
than the jet may be visible.
Sometimes any of the above features can appear as a shape of a ring or a partial ring.
{ Such morphologies have been reproduced in detail in simulations \citep{WG85,CGS91}.}
\rva{Good examples for multiple hotspots include Cygnus~A (Fig~\ref{fig:cygA5GHz}) 
and 3C~20  (Fig.~\ref{fig:3c20}). A clear ring-like hotspot feature is seen in 3C~47 (Fig.~\ref{fig:3c47}).}
\item[(S) --] S-symmetry of jets and hotspots. Jets / hotspots are found on opposite sides of a symmetry axis through the lobes.
\rva{Good cases are 3C~200  (Fig.~\ref{fig:3c200}) and 3C~334  (Fig.~\ref{fig:3c334}).}
\end{enumerate}

For solid precession cases, we require that at least two precession characteristics are present.
\rva{This accommodates the fact that there is a chance that factors such as interaction with a triaxial environment
might also occasionally produce signatures comparable to the above. In particular, (C) and (E) can be produced by a crosswind,
but unlikely in connection with S-symmetry, (S). Also, crosswinds are usually identifiable from the radio maps, and we indicate
this in the analysis below. We report curvature (C) in all sources, but make sure that a single occurence of curvature
does not bring a source with crosswind in the precessing jets category (the jet \rvb{in} 3C~382 curves twice in opposite directions).
(R) appears unlikely to be caused by circumstances other than precession. (S) could in principal occur if a jet
interacts with a triaxial halo, but we argued above that we expect this to be rare.}


\section{Evidence for jet precession from kpc-scale jets}\label{sec:kpc-jets}
\rva{An assessment of jet precession requires high-resolution radio maps of good quality. This is 
best available for powerful jet sources. We therefore first investigate what is perhaps the
best studied powerful radio
source, Cygnus~A. The 2-Jy sample and the complete sample of \citet*{Mulea08},
\rvb{which is derived from the 3CRR sample,} contain powerful 
radio sources with excellent imaging data. The 2-Jy sample, however has many Southern sources
without VLA coverage. We therefore use the 2-Jy sample for exemplifying the effects of projection and
the complete sample of \citet{Mulea08} for statistical results. }

\subsection{Cygnus~A}\label{sec:cyga}
The 5~GHz VLA radio map of Cygnus~A, one of the best studied extragalactic radio sources, is shown in 
Fig~\ref{fig:cygA5GHz}. 
To first order, the source shows point reflection symmetry regarding the bright hotspots and the jets. 
The position angle of the jet differs by about $5\deg$ from the lobe axis (drawn here by eye). The backflow on the
jet side at 5~GHz (highlighting relatively recently accelerated electrons)
{ is clearly asymmetric with almost the entire backflow to the north of the jet.} This is, 
however, a short term feature, because the cavities in the Chandra X-ray map (Fig~\ref{fig:cygA5GHz}, bottom)
show that lobe plasma displaces X-ray gas also southwards of the jet. This is also confirmed in radio maps
at lower frequency \citep{Lazioea06,McKeanea16}. Both sides feature multiple hotspots { in the 5~GHz map
\citep[see also][]{Cea99}. }
On the jet side,
these are also observed in X-rays { \citep[for details]{WYS00}}. The western jet shows curvature.
{ \citet{CGS91} argue that in this case, the curvature likely results from interaction with an asymmetric
backflow. We follow their analysis and conservatively do not assign the C-criterion to this source.}
The criteria E, R and S we defined in Sect.~\ref{sec:pcrits}
are satisfied.

The observed jet morphology cannot be explained by interaction with gas clumps, because the timescale 
for ablation is too short, or hydrodynamic instabilities, as the jets are otherwise stable
(compare Sect.~\ref{sec:jstab}). 
While on the scale of the radio lobes, the X-ray gas is strongly morphologically disturbed, the outer
isophotes in the X-ray image (Fig~\ref{fig:cygA5GHz}, bottom) are much more spherically symmetric.
This has been shown rigorously by \citet{Smithea02}. It is therefore very unlikely that
interaction with an asymmetric environment is responsible for morphological features in this source
(compare Sect.~\ref{sec:env}).
\rva{X-ray and radio observations can} therefore most naturally \rva{be} explained by precession. 
 \begin{figure}
	\includegraphics[width=0.47\textwidth]{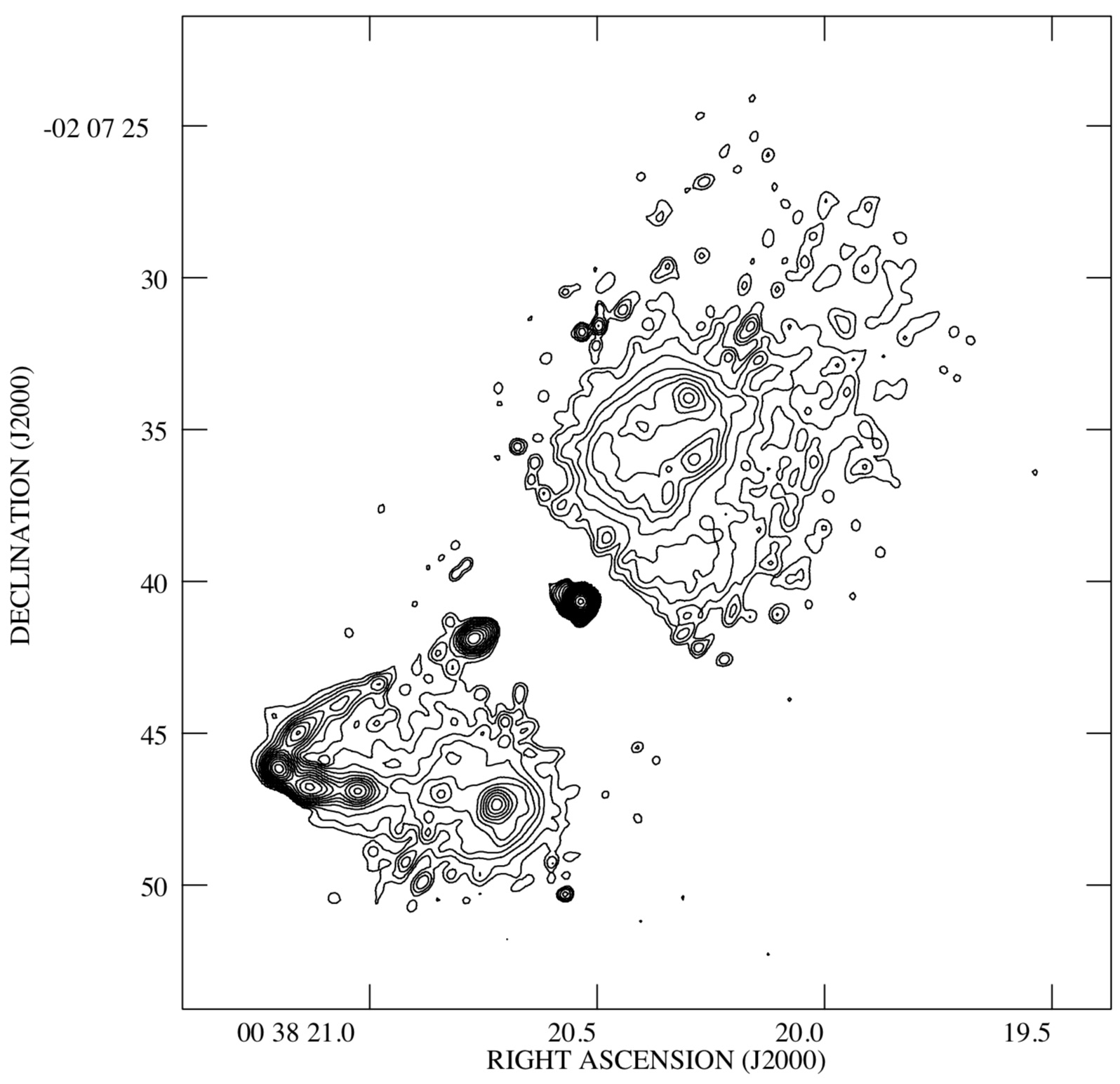}
	\includegraphics[width=0.47\textwidth]{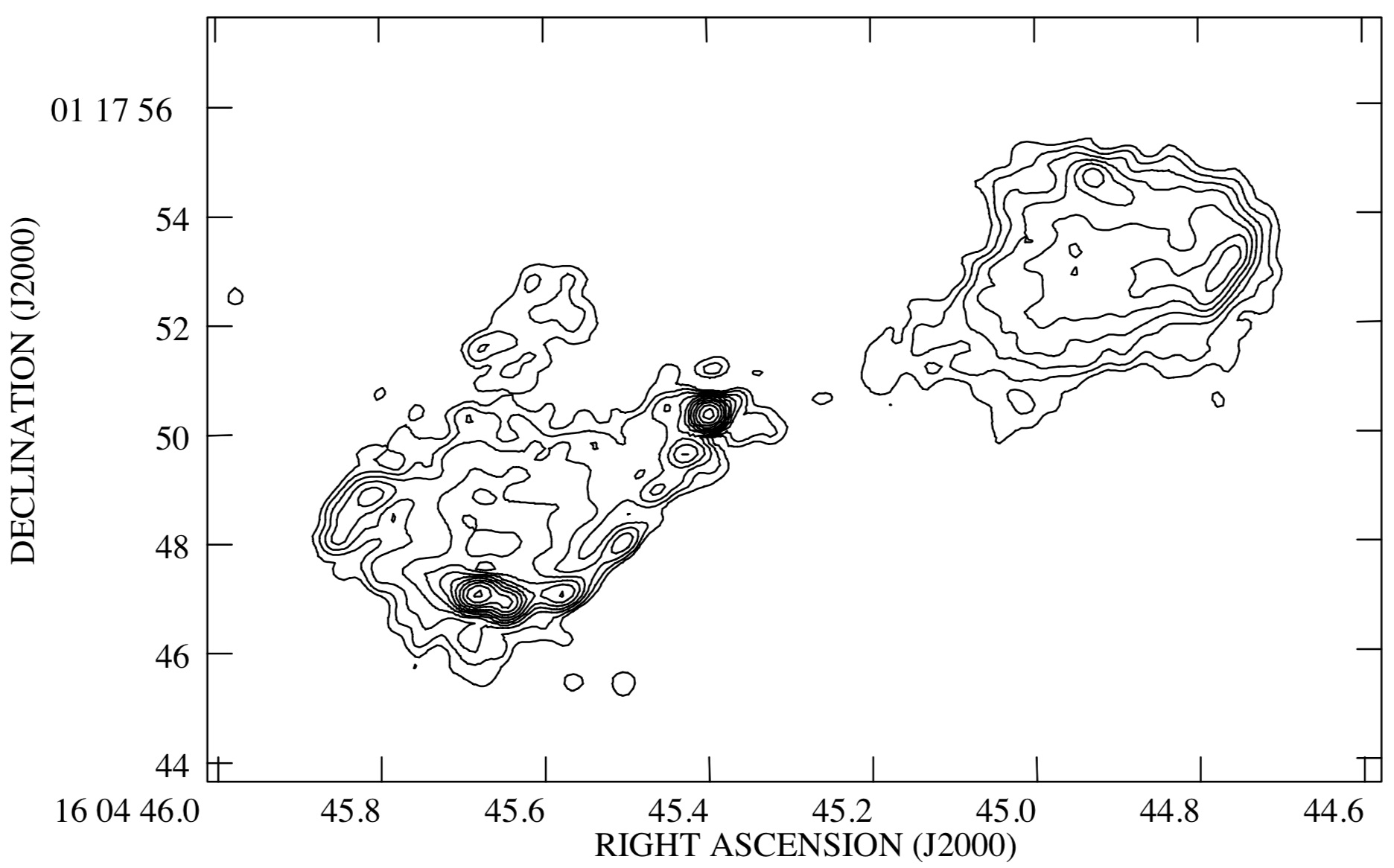}
	\caption{Broad line radio galaxies 3C~17 (top) and 3C~327.1 (bottom) observed with
	 the Very Large Array \citep{Morgea99b}. Shown are 6~cm radio contours at $\sqrt{2}$ 
	 spacings. }
    	\label{fig:2JyBLRGs}
\end{figure}

 The precession period needs to be much shorter than the source age
 of about 24~Myr \citep{Krause2005a}, because otherwise we would not expect
 mirror symmetry in X-ray cavities and low-frequency radio maps, which is, however, observed (see above).
 \rva{This can also be seen when comparing to simulations of precessing jet sources. \citet{DonSmi16} present
 simulations of jets that reach a simulated age comparable to one precession cycle. For example, their Fig.~14
 shows a simulation after 1.5 precession cycles. Here, the lobe shows strong asymmetries, which are not seen in 
 Cygnus~A \citep[also compare the low frequency images][]{Lazioea06,McKeanea16}. They find approximately symmetric lobes
 after about four precession cycles.}
 \rva{For Cygnus~A , we apply the lobe symmetry argument in the following way: In Fig~\ref{fig:cygA5GHz} the lobe has advanced 
 further on the south side of the western jet, consistent with the current orientation of the jet. The eastern lobe has advanced 
 further on the northern side of the jet. The difference is about 9~kpc. From the size and estimated age of the radio source
 we estimate an average lobe advance speed of 3~kpc Myr$^{-1}$ ($\approx3000$~km~s$^{-1}$). 
\rvb{From a comparison of ram pressure and hotspot pressure,
\citet{AP96} argue that} the part where the jet currently impacts advance\rvb{s} at roughly twice 
 this speed.
It would therefore take about 1.5~Myr to produce such an asymmetry via a precessing jet. Hence, if the precession period 
was longer than 3~Myr, the jet would remain on one side of the axis \rvb{for too long} and produce a much larger asymmetry.
 The precession period should therefore be less than 3~Myr. }
 The jet is essentially straight through half of the lobe, i.e., for about 50~kpc.
 For a precessing jet to avoid showing observable curvature, the precession period needs to be much longer than
 the jet travel time for the straight part of the jet. Because the jet is likely 
 relativistic\footnote{\rva{Estimating the Lorentz factor in the kpc-scale jet of Cygnus~A
 is difficult \citep[compare][]{CarBar96}.
 One argument for a Lorentz factor of at least a few comes from a combination of  
 hydrodynamics and radio luminosity. The ratio $\eta$ between jet density and ambient density 
 $\rho_\mathrm{a} = 0.05 \rho_\mathrm{a*} m_\mathrm{p}$~cm$^{-2}$, 
 with  $\rho_\mathrm{a*} \approx1$ \citep{Smithea02},
 is $\eta= 10^{-4} \eta_*$, with $\eta_*\lesssim1$, to explain hotspot advance speed and lobe width \citep{AP96,Rosea99,Krause2005a}.
 The energy flux required to power the observed radio emission and to expand the radio source is  
 $Q_0 = 2\times 10^{46} Q_{0*}$~erg~s$^{-1}$, with $Q_{0*}=1$ \citep{KA99}. 
 The jet radius is $r_\mathrm{j}=0.55 r_\mathrm{j*}$~kpc with $r_\mathrm{j*}\approx 1$\citep{CarBar96}.
 Using the non-relativistic expression
 for the jet power then yields a jet velocity in units of the velocity of light of 
 $\beta_\mathrm{j}=8 \; Q_{0*}^{1/3} r_\mathrm{j*}^{-2/3} \eta_*^{-1/3} \rho_\mathrm{a*}^{-1/3}$, which implies
 that the jet cannot be non-relativistic.} }  \citep{Krause2005a},
 this corresponds to the light travel time, i.e. 0.15~Myr.
 \rva{We investigated precessing jet models using the model of \citet{Gowea82}, varying
 precession period, jet phase, inclination, and precession cone opening angle. To reproduce a 
 straight jet of comparable length to the one in Cygnus~A, we needed a precession period of at least 0.5~Myr.}
 
 The morphological details shown in Fig~\ref{fig:cygA5GHz} therefore suggest a 
 precession period of about \rva{0.5-3}~Myr.

\subsection{Strongly projected radio sources}\label{sec:2Jy-blrgs}
Strongly projected radio sources are of particular interest, because at low inclination,
precessing jets have a high probability to be much more strongly misaligned with the lobe
axis. In the extreme case that the line of sight is inside the precession cone, the jet can have
any angle with respect to the lobe axis. 

%
\rva{We first searched the  2-Jy sample of the brightest radio sources at 2.7~GHz for strongly projected
sources. The data accessible from the sample web page shows only poorly resolved quasars.
Among the broad line radio galaxies, only}
 3C~17 and 3C~327.1, 
\rva{have} high-quality radio maps \rva{with well resolved} jet as well as lobe detections (Fig. ~\ref{fig:2JyBLRGs}). 
Though the lobe axis is now much more difficult to determine, because of the strong
contrast in brightness between the likely Doppler boosted jets and the fainter parts of the
lobes, the misalignment is evident in both cases.
We estimate misalignment angles of
$21\deg$ for 3C~17 and $35\deg$ for 3C~327.1, much more than the $5\deg$
for Cygnus~A.

Both sources show curved jets with a sharp bend at the brightest jet knot (hotspot), suggesting
that the jets may be deflected there before hitting the lobe boundary a second time at the final hotspot.
It appears possible that the double hotspots in Cygnus~A have a very similar origin. 

The sources also have S-symmetry and the backflow on the jet side of 3C~327.1 is strongly asymmetric.
These features are clear indicators for precession according to the criteria defined 
in Sect.~\ref{sec:pcrits}. Because the length scales
are very similar to Cygnus~A, the precession period must also be of the order of $10^6$~yrs.

\rva{There are ten other BLRGs in the 2-Jy sample for which the data quality was not sufficient to 
assess the relative orientation of jet and lobe axis. 3C~17 and 3C~327.1 therefore only demonstrate 
that strong misalignment between jets and lobe axis may occur. We show statistics for the complete 
sample below.}
\begin{table*}
	\centering
	\caption{Evidence for binary black holes in a complete sample of powerful jet sources
	\rvb{derived from the 3CRR catalogue}.\label{t:data}
	Columns from left to right:
	Optical type (see Sect.~\ref{sec:data} for the database webpage): NLRG: narrow line radio galaxy, BLRG: broad line radio galaxy,
	 LERG: Low excitation radio galaxy, Q: quasar. Structure: x-wind for crosswind.
	Precession: letters refer to precession criteria defined in Sect.~\ref{sec:pcrits}. 
	\rva{Misalignment: angle between jet and lobe axis, if lobes are detected and a lobe axis can be assigned. 
	Source size corresponds to database entry "sum of linear lobe lengths". Jet length corresponds to database entry
	"straight jet length". Age is the model age from \citet{Turnea18b}, where available.}}
	\begin{tabular}{llllcccc} 
		\hline
		Source & Opt. t. & Struct. & Prec. & Misalign. & Source size & Straight jet & Age\\
			   &             &            &         &               &   (kpc)         &   (kpc)      & (log, yr) \\ 
		\hline 
3C 20	& NLRG &simple / x-wind &CERS & -            &    139         & 37            &   - \\
3C 22	& BLRG &simple	&ES 		    & -             &    206        & 83             & $7.1\pm0.2$\\
3C 33.1	& BLRG     &simple	&CES	& $0\deg$     &    691         & 156          & $8.4\pm0.2$\\
3C 41	& NLRG &simple	&CERS	& $5\deg$     &    180         & 61            & - \\
3C 47	& Q  	    &simple	&ERS	        &  $6\deg$    &    325        &  166          & $7.8\pm0.2$\\
3C 98	&NLRG &simple / x-wind &C 	&  $0\deg$    &    169        &  62            & $8.8\pm0.4$\\
3C 171	&NLRG &complex	&CER       &     -             &     37         &  12            & - \\
3C 173.1	&LERG & simple	&R           & -                 &   248         &  49            & $7.7\pm0.2$\\
3C 175	& Q     & simple	&R           &    -              &   374         & 150           & $7.4\pm0.2$\\
3C 200	&NLRG &simple	&CERS     & $12\deg$    &   144         &  41            & $7.3\pm0.3$\\
3C 207	&Q      &complex	&ER         &  -                &     88         & 35             &  - \\
3C 220.1	&NLRG &simple	&CER       & $0\deg$     &   216         & 79             & -  \\
3C 228	&NLRG &simple	&-           & $0\deg$     &   300         & 45             &  $7.3\pm0.2$ \\
3C 234	&NLRG &complex	&-           & -                &   295         & 183           &  $8.3\pm0.2$ \\
4C 74.16	&NLRG &simple	&ER         & $6\deg$     &  312          & 76             & - \\
3C 249.1	&Q     & complex	&-           & -                 & 119          & 21             & $7.7\pm0.2$\\
3C 263	&Q     &complex	&E           &-                 &  341          & 16             & - \\
3C 275.1	&Q	  &simple	        &CE        & -                 & 118           & 53             & - \\
3C 285	&LERG &complex / x-wind&E & -                & 260           & 72             &  $8.9\pm0.3$\\
3C 300	&NLRG &simple	&ER         &-                 & 394           & 20             & - \\
3C 303	&BLRG & complex	&ER         & $41\deg$   & 107          & 31             & $8.4\pm0.3$ \\
3C 334	&Q      & simple	&CERS     & $22\deg$   & 326          & 83             & - \\
3C 336	&Q      & simple	&ERS       & -                 & 201          & 38             & $7.2\pm0.2$\\
3C 352	&NLRG & simple	&ER         & $10\deg$   & 96            & 29             & - \\
3C 382	&BLRG  & simple / x-wind&CR& $0\deg$     & 190         & 54             & - \\
3C 388	&LERG &simple	&CR                & $0\deg$     & 72            & 14             & - \\
3C 390.3	&BLRG &simple / x-wind&ER  & $4\deg$     & 243         &   98            & -     \\
3C 401	&LERG &simple	&CER              & $0\deg$     & 79            & 19             & - \\
3C 433	&NLRG & complex	&CER      &-                  & 105         & 8                & - \\
3C 436	&NLRG & simple	&ERS      & $2\deg$      & 368         & 70              & $7.9\pm0.2$\\
3C 438	&LERG &simple	        &CERS    & $4\deg$,$12\deg$&94 & 38              &  - \\
3C 441	&NLRG & simple	&ERS      & -                 & 256         & 62              & - \\
3C 452	&NLRG &simple / x-wind&-  &$0\deg$      & 417          & 71              & $8.5\pm0.2$\\
		\hline
	\end{tabular}
\end{table*}
\begin{table}
	\centering
	\caption{Statistics of table~\ref{t:data}.\label{t:stat1}}
	\begin{tabular}{ll} 
		\hline
		Criterion & Occurence\\
		\hline 
		Curvature (C) & 14 \\
		Jet towards Edge of Lobe (E) & 24 \\
		Ring / multiple / extended hotspot (R) & 23 \\
		S-symmetry (S) & 11 \\
		\hline 
	\end{tabular}
\end{table}
%
\subsection{A complete sample of powerful jet sources}\label{sec:sample}
To see if these are special cases, or if powerful radio sources commonly show similar evidence 
for jet precession, we have analysed the 33 high-resolution radio maps of a complete sample of 
powerful radio sources with redshifts less than one based on the 3CRR catalogue 
\citep{Laingea83}, only requiring a definite jet detection \rva{(for details see Sect.~\ref{sec:data}). 
The jets in this sample have been identified previously, independently of the present analysis.}
We use the four criteria detailed in Sect.~\ref{sec:pcrits} to assess jet precession.
Additionally, we assess the complexity of the source structure, e.g., if the source
is affected by cross winds. If the source only shows features as expected for precessing jets
(compare Sect.~\ref{sec:assessment}), we label it 'simple'.
All radio maps are presented in Appendix~\ref{ap:data}.
The findings are summarised in Table\rvb{s}~\ref{t:data} \rva{and~\ref{t:stat1}.}

\rva{The most common features we find are jets at the edge of the lobe 
(24 occurrences) and ring-like / multiple / extended hotspots (23 occurrences).
Jet curvature and S-symmetry are comparatively less important (14 and 11 occurrences, respectively,
compare Table~\ref{t:stat1}).}

From our sample of 33 radio sources, we found no evidence for precession in four sources. 
Interestingly, three of these sources have a complex source morphology or crosswind, 
which complicated the assessment. In a further five sources, we found one of the above 
features per source. Since any one given feature might have different explanations, 
we regarded only sources with at least two independent indicators as solid precession cases. 
Of such sources, we have 24 in the sample, corresponding to a fraction of 73~per~cent. 
{ 14} sources or { 42}~per~cent show three or four features expected from precession, simultaneously. 
{ The high fraction of sources with multiple indicators of precession in our sample suggests }
that precession is very common among powerful radio sources.

\rva{We estimate the order of magnitude for the precession period for the sample in the following way:
For jets to remain straight for tens of and sometimes over 100~kpc (compare Table~\ref{t:data}),
the precession period needs to be significantly longer than the light travel time through the straight 
parts of the jets. This limits precession periods from below
to be greater than or equal to about 0.5 to 5 Myr. }

\rva{Where the lobes are well observed, the lobe near where
the jet impacts (hotspot) is not further advanced than other parts,
see for example 3C~33.1, 3C~41, 3C~47, 3C~200 (north), 4C~74.16 (south).
Multiple hotspots, where the jet interacts with the lobe boundary,
or ring-like hotspots imply that the jet interacts with the lobe
boundary for a time that is short compared to the source age. Otherwise
the lobe boundary (a sharp rise in density) would be deformed.
These features have been confirmed in simulations of precessing jets by
\citet{CGS91}. Clear symmetric ring features are seen in 3C~47, 3C~388 and
3C~441 (south). Double \rvb{hotspots} which could arise from interaction
with a roughly symmetric lobe boundary are seen in, e.g., 3C~20. A hotspot
on one side of the lobe with a \rvb{bright extension} outlining a symmetric lobe
boundary near the tip of the lobe is seen in 3C~200, 4C~74.16 and
3C~334.
The ages for some of our sources have been modelled by \citet{Turnea18b} and are included 
in Table~\ref{t:data}. They are in the range $10^7$ to $10^9$ yr. The sizes of the sources that have no age estimates 
are similar (Table~\ref{t:data}). Therefore, their ages are most likely also similar. \rvb{These} morphological
features suggest precession periods of about a tenth of the source
ages or less, i.e., a few to a few tens of Myr.}

\rva{In summary, morphological analysis of jets and lobes suggests precession periods of the order of 1-10 Myr, in any case
significantly less than the source age.}

\rva{ We note that mild curvature is seen in about half of the precessing sources.
While this can also be the result of hydrodynamic processes in the lobe
(one reason why we require at least two precession criteria above), it would
be expected if the precession period was of the order of 1-10~Myr, as found above. 
Literature estimates for the precession period for the precessing 
sources in our sample include 0.5~Myr for 3C~388
\citep[from fitting a relativistic, precessing jet model]{Gowea82} and
$>3$~Myr for 3C~390.3 \citep[flow model from spectral analysis]{Alex85}. 
}

\rva{We also repeated the test for orientation dependence of the jet-lobe axis misalignment in this sample
\rvb{of powerful radio sources based on the 3CRR catalogue.} 
As expected, it is more difficult to assess the shape of the radio lobes in the more strongly projected sources
due to a combination of beaming effects and limited dynamic range of the images. We were able to estimate 
the misalignment angle in 2 of the 8 quasars, 4 of the 5 BLRGs, 9 of the 15
narrow line radio galaxies (NLRGs), and 3 of the 5 low excitation radio galaxies (LERGs).
The largest misalignment angle for NLRGs is $12\deg$ whereas among BLRGs and quasars 
values of up to $41\deg$ are found. Low values are observed in both groups.
Mean and standard deviation of the two distributions are $4\deg\pm4\deg$ for the NLRGs
and $12\deg\pm15\deg$ for BLRGs and quasars. The p-value for the Kolmogorov-Smirnov test,
i.e. the probability that both distributions are from the same parent sample, is 72 per cent.
At the current sample size, there is therefore no significant difference between the two distributions.
However, there is an indication that the misalignment in BLRGs and quasars is stronger, as expected,
if precession was a significant factor.}

\section{Discussion: the case for close binary black holes}\label{sec:disc}

\rva{Sub-parsec supermassive black hole binaries with jets may produce observable effects on the 
100~kpc scale via geodetic precession, and at the same time show morphological features
on parsec scale due to \rvb{their} orbital motion. Due to observational limitations,
both effects will rarely be observable in the same source: lobe structure is best studied in 
radio galaxies where the jet has a large angle to the line of sight, whereas well-observed parsec
scale jets are usually strongly Doppler boosted, with small inclinations and thus have poor lobe images
due to limited dynamic range. To make the case, 
we present a rare example with very good observations on parsec as well as 100-kpc scales, 
Cygnus A, and search for evidence of, respectively, orbital and geodetic precession separately in 
samples of 100-kpc scale radio galaxies (Sect.~\ref{sec:kpc-jets}) 
and parsec-scale radio sources from the literature
(Sect.~\ref{sec:pc-scale-jets} below).
We first discuss if the precession observed
in 100~kpc-scale radio sources could be a consequence of geodetic precession.
Based on this likely being the case, we then make the case for a supermassive binary black hole in 
Cygnus~A, where we argue that geodetic and orbital precession are seen. We then discuss 
evidence for orbital precession in parsec-scale, powerful radio sources. Finally, we investigate if any supermassive 
binary black hole candidate inferred from precession in extragalactic radio sources could be 
confirmed via an individual gravitational wave detection in the near future. }

\subsection{\rva{Causes of precession in 100-kpc scale radio sources}}

Jet formation models and explanations of differing radio-loudness of quasars are 
often based on the assumption that jets, in particular powerful ones as studied here
are produced by fast-spinning black holes 
\citep[e.g.,][]{MSL98,Koidea00,dVHK03,KomMcK07,BHK08,Ghiea14,GarDon18}, even though 
direct evidence for this is so far elusive \citep{FenGR10}.
In such models, the instantaneous jet direction is given by the spin axis.
Hence, it is likely that changes in jet direction are caused by corresponding
changes of the spin axis of the jet-producing black hole
\citep[compare also][]{BBR80,BBR84,MeEk02}.
Known mechanisms that may affect the jet direction
include orbital motion and geodetic precession in a black hole binary, 
and interaction with the angular momentum of a disc
around the black hole, which may be the accretion disc.

\subsubsection{Geodetic precession}

A natural explanation for precession of jets is via the geodetic precession of the black hole spins 
in a binary system where the jet is ejected along the spin axis of one of the black holes.
The geodetic precession period in Myr, $P_\mathrm{gp,Myr}$,
for binary black holes with total mass $M_9 10^9 M_\odot$,
mass ratio $r$ and separation of $d_\mathrm{pc}$~pc 
in a circular orbit is given by \citep{BOC75,Stairs03,TP17}:
\begin{equation}\label{eq:geoprec}
P_\mathrm{gp,Myr} = 124\, \frac{(1+r)^2}{r(3r+4)} d_\mathrm{pc}^{5/2} M_9^{-3/2}  \, .
\end{equation}
{ For powerful jets, it is reasonable to assume that the observed jet is produced by the more massive
black hole. Therefore, }
the mass ratio cannot exceed unity. { Equation}~(\ref{eq:geoprec}) can { then
}be written as an upper limit for the binary separation:
\begin{equation}\label{eq:geoprecsep}
d_\mathrm{pc} < 0.18\, P_\mathrm{gp,Myr}^{2/5}   M_9^{3/5}  \, .
\end{equation}
We have argued above that VLA radio maps of powerful jet sources suggest that precession periods 
of the order of Myr are very common. Many of these sources 
\citep[e.g., Cygnus~A,][]{Tadea03} are known from independent
analyses to have central dark masses of the order of $10^9 M_\odot$.
Masses of super-massive black holes in general extend up to about
$10^{10} M_\odot$ \rva{\citep[e.g.,][]{HR04,Peterea04a,Gebhea11,Bogdea18}.} 
It is therefore clear that a measurement of precession periods of the 
order of Myr implies sub-parsec supermassive black hole binaries, if the 
precession is caused by this mechanism.

\subsubsection{Other precession mechanisms}

It is difficult to explain a sustained precession with a period of the order of Myr
with other mechanisms.

If the precession period of the order of Myr we argued for above corresponded to
an orbital period, the separation of the binary would be 
$48 M_9^{1/3}$~pc and the Keplerian velocity $298 M_9^{1/3}$~km~s$^{-1}$.
For a jet at a velocity comparable to the speed of light, this would lead to  
a precession cone opening angle of less than a degree. This would be far too small
to be observable and inconsistent with the jet-lobe axis misalignments in our observations.

Torques caused by a misalignment between the spin axis of a black hole and the 
angular momentum of the accretion disc can also induce jet precession. 
In an overall misaligned system the inner part of the accretion disc is aligned with 
the blackhole spin axis by the combined effect of Lense-Thirring precession 
and internal viscosity of the disc. This process is known as the Bardeen-Petterson effect 
\citep{BarPet75}. The outer disc will maintain its angular momentum direction. 
Through the accretion of material of the outer disc on to the black hole, 
the black hole spin axis will precess and align with the outer disc. 
The precession time-scale is of the same order as the alignment timescale 
\citep{ScheuerF96}.
Two aspects of the disc-induced precession, namely the relatively slow precession rate 
 and the single possible precession cycle in the alignment lifetime 
\citep{ScheuerF96,LoPri06}, make this process less likely to be a cause of 
jet precession for the given radio sources. For example, following \citet{LoPri06},
adopting their numerical values for accretion efficiency $\epsilon$ 
and disc viscosity parameters $\alpha_{1,2}$ and assuming black hole spin 
$a\approx 1$, we obtain a precession period of
\begin{equation}\label{eq:align}
\begin{split}
t_\mathrm{prec} &\approx  t_\mathrm{align} \approx 7\,\,\mathrm{Myr} \\
  &\times \, a^{11/16} \left(\frac{\epsilon}{0.1}\right)^{7/8}
\left(\frac{L}{0.1 L_\mathrm{Edd}}\right)^{-7/8}
M_9^{-1/16} \\
 &\times
\left(\frac{\alpha_1}{0.03}\right)^{15/16}
\left(\frac{\alpha_2}{0.3}\right)^{-11/16} \, ,\\
\end{split}
\end{equation}
This is larger than the precession period we estimated \rva{for Cygnus~A}. 
Simulations show that if the precession timescale becomes
comparable to the source age, very asymmetric lobe morphologies result \citep{DonSmi16}. 
\rva{As discussed in Sect.~\ref{sec:sample}, the sources in our sample do not show such asymmetries,
but often show smooth, ring-like hotspot features that suggest a local impact timescale short compared 
to the source ages and thus} many precession cycles. \rva{This is not consistent with precession induced
by a misaligned accretion disc. A process that produces one precession cycle only, where the period
of the cycle can be much shorter than the source age (eq.~\ref{eq:align}), would have only
a limited probability to be detected at the time a given source is observed. However, we also find
precession frequently in comparatively old sources, e.g., 3C~33.1, 3C~303 and 3C 436. These
are three of the seven sources with modelled ages of log(age/yr)$\ge$7.9. This makes disc-induced precession
less likely for this sub-population.}
%

\rva{A pancake like, dense nuclear star cluster could also cause a precession to a misaligned 
black hole spin \citep{Applea96}. The induced precession period is, however, of the order
of $10^9$~years or longer for plausible parameters of nuclear star clusters,
and hence not of interest in the present context.}

\subsection{The case for a supermassive black hole binary in Cygnus~A}

\begin{figure}
	\includegraphics[width=0.5\textwidth]{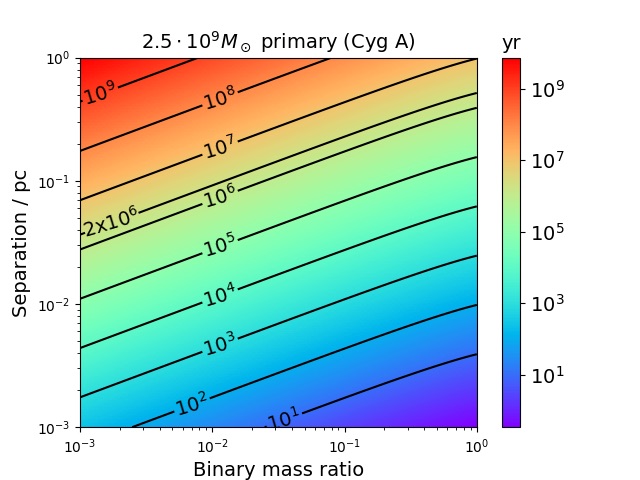}
	\caption{Graphical representation of the geodetic precession period for Cygnus~A as a function of
	mass ratio and binary separation. A precession period of 1-2~Myr, as estimated from the VLA image
	requires a binary separation less than 0.4~pc. 
    	\label{fig:cygA_mratio}}
\end{figure}

For Cygnus A, where $M_9 = 2.5$ \citep{Tadea03}, a precession period of \rva{$0.5-3$~Myr} 
would imply a separation \rva{<~0.5~pc} (compare Fig.~\ref{fig:cygA_mratio}).
Figure~\ref{fig:cygA5GHz} also shows the 43~GHz VLBI image of the center of Cygnus A
\citep{Boccardiea14,Boccardiea16a}. The structure of the jet shows a helical pattern with 
a wavelength of roughly $\lambda = 4$~pc. If this is caused by a jet from a binary black hole, 
the implied Keplerian orbital period is $P_\mathrm{orb} = \lambda/v_\mathrm{j} \approx 18$~yr, 
where we have used a jet velocity of $v_\mathrm{j}=0.7 c$ consistent with multi-epoch 
observations \citep{Boccardiea16a}. This implies a binary separation of { 0.05~pc}. 

In order to investigate this scenario further, we can look at the implied constraint on the mass ratio:
the combination of
total mass, precession period and a binary separation of $0.05$~pc determine the mass ratio 
uniquely as  { $r\lesssim10^{-2}$ (compare Fig.~\ref{fig:cygA_mratio})}. 
Such significant differences in mass are expected from 
cosmological evolution: Black hole mass scales with galaxy bulge mass \citep{HR04}, 
and since minor galaxy mergers occur much more frequently than major ones, 
very different masses should be the rule \citep{Rodrea15}.

If the mass ratio were indeed { $r\lesssim10^{-2}$}, \rvb{orbital motion alon
could only produce} a helical jet \rvb{in} the secondary, because the centre of mass would be much
closer to the primary, such that its orbital velocity would be small, resulting in a
too narrow a precession cone.
Because the high power of the main radio source requires a jet from the bigger
black hole \citep{Krause2005a}, this would support the conclusion that both black holes are currently producing 
a jet: the bigger one would be responsible for the main radio structure,
the smaller one 
\rva{would significantly contribute to the emission on the parsec scale, so that the 
helical morphology could be produced. This would, however, not be expected, as 
one would expect the mass ratio (1:100) to be reflected in the ratio of jet luminosities.}

In an alternative model with two black holes of similar masses and only one jet the separation would 
have to be about 0.5~pc. Hence, the orbital period would be of the order of 1000~yr, corresponding 
to a predicted wavelength for the helical pattern of about 300~pc. This
could therefore not explain the parsec-scale helical structure.

\rvb{The model suggested by \citet{LR05} for 3C~345 could also apply to Cygnus~A.
In this case, the presence of the secondary induces a precession of the accretion disc of the primary 
which causes the jet, ejected prom the accretion disc around the primary, to precess.}

Finally, an alternative method to estimate the binary separation is via the opening angle of the 
helical pattern. The half opening angle $\alpha$ is given by 
$\tan \alpha = v_\mathrm{o} / v_\mathrm{j}$. Since the orbital velocity $v_\mathrm{o}$ is 
related to the Keplerian orbital period $P_\mathrm{K}$ via $v_\mathrm{o} = \pi d/P_\mathrm{K}$, 
and the jet velocity $v_\mathrm{j} = \lambda/P_\mathrm{K}$, we can give the binary separation as 
function of the observables: $d = \lambda \tan \alpha / \pi$. 
Using $\alpha=5\deg$ \citep{Boccardiea16a}, we find d = 0.1 pc,
which supports the $r\approx 10^{-2}$ solution. 
Because of the observed complexity of the flow structure and absorption 
\rvb{likely playing a role} we regard this as a less accurate estimate.

\rva{The VLBI assessment of the mass ratio is therefore inconclusive, but more evidence seems to support a
separation of  0.05-0.1~pc combined with a mass ratio of about 1:100.}

The radio structures produced by the jets in Cygnus A have displaced the X-ray gas in 
certain regions (Fig.~\ref{fig:cygA5GHz}). The X-ray image reveals not only cavities 
produced by the current jet, but also at least one side cavity which is difficult to relate 
to the current jet activity or previous jets with the same orientation. 
Comparing to dedicated 3D simulations, we found that the side cavity is well explained 
by a preceding activity period where a jet was oriented at an angle of roughly $90\deg$ 
\citep{Chonea12}. 
\rva{This could be interpreted in the context of black hole mergers \citep{MeEk02},
which would imply an additional supermassive black hole that has now merged.
However, as shown above, a pause in jet activity for 10~Myr, but with strong accretion
could have re-oriented the jet to the current position without the need of an 
additional black hole.}

Precession models for Cygnus A have been presented before: 
{ \citet{Bary83} based the analysis on a radio map that was not sensitive enough to show the jets.
\citet{SteenBlun08} have aligned the axis of the precession cone with the current jet direction,
and fitted a ballistic jet path to the curvature highlighted in Fig.~\ref{fig:cygA5GHz}.
3D hydrodynamic modelling favours, however, the explanation that these jet curves
are the result of interaction of the jet with the asymmetric backflow \citep{CGS91}. 
Also, their precession model does not explain the misalignment of the jet with the \rva{lobe} axis.}
In contrast, we interpret the overall jet-lobe axis misalignment { as due to precession}, 
and compare to broad line radio galaxies and a complete sample. 
In view of these new results, it appears possible that some of the jet structure 
interpreted as due to precession in 
those previous models are actually due to unrelated local brightness variations. 
However, the precession period
inferred from the previous modelling is even shorter than ours, 
which would re-enforce the case for a sub-parsec binary black hole in Cygnus~A.

{ The recent discovery of a further supermassive black hole candidate in Cygnus~A
\citep[Cyg~A-2][]{Perlea17,Dabbea18} cannot be related to the precession signatures
discussed before. Cyg~A-2 is located 460~pc away from the primary AGN (Cyg~A-1) in projection,
too far away to cause a detectable geodetic precession. The timescale for a supermassive 
black hole with a mass of $m_8 10^8 M_\odot$ at the location of Cyg~A-2 
to approach Cyg~A-1 to within a parsec
is determined by dynamical friction, and is of the order of $10^7 m_8^{-1}$~yr
\citep[further depending on galaxy properties, e.g.,][]{BBR80}.
The timescale for gravitational inspiral and coalescence
for a supermassive black hole binary is
$10^8 r^{-1}  (M_9/2.5)^{-3} (d_\mathrm{pc}/0.05)^4$~yr \citep{BBR80}. Both timescales
are highly uncertain. One might argue that the discovery of Cyg~A-2 makes it less
likely that there could be a further supermassive black hole in the system.
However, a rate of several 10~Gyr$^{-1}$ for 100:1 galaxy mergers  would be 
consistent with the expectations from cosmological simulations for a massive
galaxy like Cygnus~A \citep{Rodrea15}.}

\begin{table*}
	\centering
	\caption{\label{t:gw}
		Gravitational wave emission from nearby binary black hole candidates. 
		$d_\mathrm{L}$: luminosity distance. $M_9$: total black hole mass in units of $10^9 M_\odot$. 
		$P_\mathrm{gp}$: estimated geodetic precession period. 
		$h_0$: prediction for the gravitational wave strain amplitude. 
		For Cygnus A, \rva{we adopt a reference separation of $0.3\pm0.25$~\rvb{pc}.}
		Details for the individual sources are discussed in Appendix~\ref{ap:nearby_sources}.}
	\begin{tabular}{lccccc} 
		\hline
		Source & $d_\mathrm{L}$/Mpc & $M_9$ & $P_\mathrm{gp}$/Myr &  $h_0/10^{-17}$\\
		\hline 
	Centaurus A   & 11   & $0.055 \pm 0.01$ & $1 \pm 0.5$         & $0.6 \pm 0.5$ \\
	Cygnus A       & 237 & $  2.5 \pm 0.3 $    & $1.75 \pm 1.25$         & $4.1 \pm 4.7$\\
	Hydra A          & 240 & $0.5 \pm 0.4$      & $1 \pm 0.5$         & $0.5 \pm 0.5$\\
	Virgo A / M87 &22.2 & $6.6 \pm 0.4$      & $6\pm2$                 & $113 \pm 113$ \\
		\hline
	\end{tabular}
\end{table*}

\subsection{Orbital precession in parsec-scale jet sources}\label{sec:pc-scale-jets}

\rvb{The orbital period $P_\mathrm{orb}$ 
of a black hole binary can be related to the geodetic precession period as:
\begin{equation}
P_\mathrm{orb} = 164\, \mathrm{yr}\,\,
P_\mathrm{gp,Myr}^{3/5}
\left[\frac{(1+r)^2}{r(3r+4)}\right]^{-3/5} 
M_9^{-7/5}\, .
\end{equation}
This implies orbital timescales of the order of 10-1000~yr for geodetic precession periods
of the order of Myr and black hole masses of the order of $10^9 M_\odot$
as argued for in the sample of powerful jet sources above.
This should result in observable jet angle swings on the parsec scale, observable with VLBI,} 
at least in sources where the jet comes from 
the smaller, secondary black hole, particularly, where the jet axis is close to the line of sight (blazars). 
\rva{Swings of the jet angle in parsec-scale radio jets have been frequently reported, e.g., 
3C~273 \citep{Savolea06}, 3C~279 \citep{Jorstea04}, or 3C345 \citep{LR05}. In the latter case,
a subparsec black hole binary was suggested as the cause.}

\rva{From monitoring an overall sample of 259 blazar jets \citet{Listea13} 
conclude that jet angle variations are very frequent\footnote{\rvb{\citet{Lister16} discuss whether 
the whole jets are precessing or only 
an energised section of a wider, conical flow. They argue that the latter might be more likely,
because a precession of the whole jet would require enormous torques. 
Sub-parsec scale binary black holes, as suggested here, would, however, be able to provide
such torques. Therefore, precession of the entire jet is a viable interpretation of the data.}}.
In a subsample of the 60 most heavily observed blazar jets, they} find 
\rvb{potentially}
periodic reorientation of the jet
in 12 sources \rvb{where the periods would be}
5-12 years \rvb{(the time base is not long enough yet for a firm detection of periodicity),} 
5 more with a back-and-forth trend, 14 others with a 
monotonic trend and 29 with irregular changes. No source showed a constant jet angle. 
The sample was selected against obvious jet bends in the innermost jet structure.
\rvb{Thus, while the timebase is not long enough for a firm detection of periodicity,
the observed variations in the jet angles are consistent with the expectations
for binary black holes with orbital periods of 10-1000~yr as suggested from our analysis
of the 100~kpc-scale, powerful jets.}

\rva{From these 60 best observed sources, the majority were quasars (41 sources). Quasars are intrinsically
more powerful than the weaker BL~Lac objects, which constitute most of the rest of the sample, and are
thus comparable to our powerful jet sources. \citet{Listea13} show that for the quasars, the jet angle
varies even more than for the BL~Lac objects. These findings are in excellent agreement with the expectation,
if powerful radio sources frequently had close binary black holes, as suggested already by \citet{LR05}
in the case of 3C~345.}

\subsection{Complementary evidence for the case of close binary black 
holes associated with powerful jet sources}

Small-scale and large-scale jets are frequently misaligned with a secondary peak in the distribution 
function at $90\deg$, possibly related to binaries \citep{Applea96conf,Applea96,Kharbea10}. 
Abrupt changes of ejection direction are known, as expected if jets are produced alternately 
by one or the other black hole of a binary with misaligned spins \citep{Listea13,Luea13}. 
Binaries with orbital periods of the order of 10~yr are also suggested by Gamma ray light curves of blazars, where the luminosity changes due to the relativistic Doppler boosting \citep{Rieger07}.

{ While X-shaped radio sources probably include examples that originate from
peculiar jet-environment interactions \citep[e.g.,][]{Kraftea05,Rossiea17}, some may 
be related to binary black hole sources, where both black holes produced a jet at some point,
or spin flips after a black hole merger \citep{MeEk02}. A recent study of X-shaped sources 
has found that four~per cent of all radio sources need to re-orient black-hole spin during the radio source lifetime
\citep{SR18}.}

Observations of the host galaxies support this picture. 
Radio-loud and thus jet-producing supermassive black holes are frequently found in interacting 
galaxies \citep{Shabea12a,Ramea13,Sabea13}. When two black holes 
approach each other, they expel stars, thus imprinting a core profile on the stellar density structure. 
Galaxies with core profiles show enhanced radio loudness \citep{Richea11}. 
Near infrared observations with the upcoming James Webb Space Telescope might be able to 
address this connection for part of our sample of powerful radio galaxies.

X-ray cavities around radio sources in general are distributed almost isotropically which is consistent with the idea that binary supermassive black holes play an important role with precession and black hole mergers
re-orienting spins frequently \citep{Babulea13,Cielea18aph}.

\rva{One might expect kinematic signatures of supermassive black hole binaries to appear 
in their broad emission lines. 
We searched the literature for complementary data for the radio sources discussed
here. The only sources with double-peaked broad emission lines are 
3C~382 and 3C~390.3 \citep*[classified above as precessing,][]{EH03,LEH16}. 
3C~47, 3C~234, 3C~249.1 and 3C~334, two of which are classified as non-precessing
above, are single-peaked emitters \citep{EH03}. 
However, based on multi-epoch spectroscopy,
\citet{LEH16} and \citet {EH03} reject the binary black hole hypothesis as the explanation
for their double-peaked objects. \citet{LEH16} extend this conclusion
to double-peaked broad line emitters in general, which is confirmed by other studies \citep[e.g.,][]{WangLea17}.
3C~390.3 has, however been found to have a periodicity in its optical light
curve of about 10~years \citep{Kovaea18}, 
which could be related to the orbital period of a 
supermassive binary \citep{Charisea16}. 
Precession studies are sensitive to supermassive binaries with very unequal masses,
so that the secondary might frequently be too faint to appear spectroscopically.
Also, sub-parsec separations as suggested here are comparable to the size of the
broad line region \citep{Kaspiea00}. The kinematics of broad-line regions in general is not 
fully understood with radiation pressure and other factors likely playing a role
\citep{EH03,Marcea08,Net10,KBS11,KSB12,LEH16}. We would therefore not expect that two 
distinct broad line regions would be identifiable for our objects.
}


\subsection{Possible gravitational wave detections for nearby extragalactic radio sources}\label{sec:gw}
\rva{Final confirmation of our proposed binary supermassive black hole candidates would 
require a gravitational wave detection. Unfortunately, the signal will likely be too weak to expect detections in the 
near future. We exemplify this with the closest sources for which jet precession could be argued for from
their radio images.}

The characteristic strain amplitude for a source at a 
distance of $d_\mathrm{L,Gpc} 10^9$~pc precessing with a period $P_\mathrm{gp,Myr} 10^6$~yr 
is given by \citep{Zhuea15}:
\begin{equation}\label{eq:h}
h_0 = 1.12\times10^{-18} \, (f_r/0.0357) \,d_\mathrm{L,Gpc}^{-1} \, M_9^{7/5} \,P_\mathrm{gp,Myr}^{-2/5}\, ,
\end{equation}
where $f_r^5 = r^3 (1+r)^{-6} (3r+4)^{-2}$. The closest jet sources display ample signs of binary black holes, 
e.g., precession in Hydra A where a simulation study has constrained the period to 1 Myr
\citep{Nawea16a}. We give details of the best candidates for gravitational wave detections in 
Appendix~\ref{ap:nearby_sources} and summarise best estimates for $h_0$  in Table~\ref{t:gw}. 
They are generally of the order of $10^{-17}$ except for M87 \citep[a core elliptical][]{Cotea06} 
where $h_0$ could be as high as $10^{-15}$. 
Pulsar timing with the Square Kilometre Array radio telescope is expected to detect gravitational waves at relevant frequencies down to $h_0 = 6\times10^{-16}$ \citep{Lazio13}. 
The SKA pulsar timing array is currently the best observatory on the horizon for such observations.
Unfortunately, this means that we can hardly expect an independent confirmation of 
our binary supermassive black hole candidates by individual gravitational wave detections.
{This is in contrast to more optimistic predictions based on the ILLUSTRIS cosmological simulation \citep{Kelleyea18}.}
A detection for M87 appears possible, but we caution that this is perhaps the least secure of the
binary black hole candidates we discuss in Appendix~\ref{ap:nearby_sources}.
We also note that the black-hole mass of M87 may be lower than our presently adopted value 
based on stellar- dynamical models. 
Indeed the recent finding of a substantial increase in the stellar mass-to-light ratio towards the 
centre of M87 \citep[relating to variations in the stellar initial mass function;][]{Sarzea18} 
may eventually lead to  a downward revision of the black-hole mass in this galaxy,
which would also better agree with the gas-dynamical measurements \citep[e.g.,][]{Walshea13}.
Because the gravitational wave strain depends super-linearly on the total mass of a binary
(compare eq.~\ref{eq:h}), this would make the detection of a gravitational wave signal even less likely.

\section{Conclusions}\label{sec:conc}
\rva{Supermassive black hole binaries are predicted 
in hierarchical cosmological evolution.
Sophisticated galaxy merger simulations that treat the relativistic physics of the 
supermassive black holes predict subparsec separations for timescales comparable
to radio source lifetimes or longer \citep{Kahnea16,Mayer17b}. 
If such a source had a radio jet, two effects would be predicted
\citep[compare Sect.~\ref{sec:intro}]{BBR80}:
(1) a comparatively short-term precession likely observable with VLBI multi-epoch imaging, and
(2) a long-term precession, which can potentially be observed as morphological features
on VLA radio maps.}

\rva{For one of the best studied extragalactic radio sources,  Cygnus~A,
we find evidence for precession in the VLA radio map with a period of about 
0.5-3~Myr. A helical structure
exists on the parsec scale that could be produced by a supermassive
black hole binary with an orbital timescale of 18~yr. Both observations 
are consistent with a sub-parsec supermassive binary. The observations 
are, however, inconclusive regarding the mass ratio of the binary.}

\rva{We have shown that morphological features in kiloparsec-scale, VLA radio maps 
\rvb{of 3CRR radio sources}
are very frequently (73 per cent) consistent with jet precession with a timescale of the order of 
$10^6-10^7$~years. 
We argue that the likely cause for the jet precession in these radio sources is geodetic
precession in a close binary black hole system. Precession of the black hole spins
due to interaction with a massive accretion disc would predict a single precession cycle.
Radio morphologies suggest, however, multiple cycles. The single precession cycle would likely occur
at the beginning of an active phase, but we find precession frequently also for sources older
than 100~Myr. 
With reasonable assumptions about the black hole masses, the geodetic precession
interpretation requires supermassive black hole binaries with separations $\lesssim 1$~pc
in these powerful extragalactic jet sources. We support the precession interpretation
with additional pieces of circumstantial evidence.}

\rva{While the parsec-scale structure has not been systematically studied for our sample
of 100 kpc-scale radio sources, analogous samples of powerful parsec-scale jet sources
exist, for example the MOJAVE samples \citep{Listea13}. Essentially all of these sources
show evidence of significant variation of the innermost jet direction over the timescale 
covered by current multi-epoch observations over 12-16~yr, as shown by
\citet{Listea13}. Jet angle variations have been linked to orbital motion 
\rvb{of binary black holes} by \citet{LR05}.}

\rva{There is therefore evidence for both effects expected if the jets are produced
in close binary black holes, namely orbital and geodetic precession. Our results
for the 100-kpc scale radio sources showed that at least 73~per cent of these
sources show evidence for precessing jets. The MOJAVE results may hint 
at an even higher fraction. Indeed, since we conservatively required two precession features
for a firm assignment of precession, there could be more precessing sources which
are not picked up by our analysis.}

\rva{The high fraction of close binary black holes in powerful jet sources might suggest 
that supermassive black holes in general are very likely to occur as a binary system.
As the merger timescale due to gravitational wave emission and interaction with stars
is expected to be significantly shorter than the Hubble time 
\citep{BBR80,Kahnea16,Mayer17b}, this may imply a steady 
influx of, black holes probably from minor galaxy mergers. Recent cosmological 
simulations predict that a significant number of supermassive black holes will be present
in a given galaxy at any one time \citep{Tremmea18a}. How fast they approach each other
depends on the spatial distribution of stars within the galaxy \citep{Tremmea18b}.}

\rva{The alternative interpretation would be that binary \rvb{black hole} formation and jet formation 
are somehow linked. One possibility could be that galaxy mergers would cause both,
the close approach of the supermassive black holes and the accretion of gas into the centre
of the merged galaxy with subsequent jet formation.
}


Independent confirmation of our sub-parsec binary supermassive black hole 
candidates\rva{, and in fact any extragalactic jet source with some evidence for geodetic precession,}
via individual gravitational wave detections will be difficult. We estimate the gravitational wave
strain for the closest jet sources with \rva{some} evidence for binary black holes from precession. 
The only source where the SKA pulsar timing array could have some hope 
for detection is M87. Thus, improved modelling of the jet-environment interaction for
precessing jet sources is probably the best way to better constrain \rva{powerful objects}
in the coming years.

\section*{Acknowledgements}
We thank the anonymous referee for a very helpful review that has helped to improve this article. 
MGHK and SSS thank the Australian Research Council for an Early Career Fellowship, DE130101399. MJH acknowledges support from the UK Science and Technology Facilities Council [ST/M001008/1]. HB and GC acknowledge support from the DFG Transregio Program TR33 and the Munich Excellence Cluster ''Structure and Evolution of the Universe''. MAN acknowledges support from a grant of the Brazilian Agency FAPESP (2015/25126-2). The work of AYW has been supported in part by ERC Project No. 267117 (DARK) hosted by Université Pierre et Marie Curie (UPMC) - Paris 6, PI J. Silk. 
\rva{This work benefited from support by the International Space Science Institute, Bern, Switzerland,  through its International Team programme ref. no. 393 {\it The Evolution of Rich Stellar Populations \& BH Binaries} (2017-18).}
We thank \rva{Mitch Begelman and} Jim Lovell for helpful discussions and Chris Carilli for providing the 5~GHz radio map of Cygnus~A.
This work made use of data obtained with the National Radio Astronomy Observatory's Karl G. Jansky Very Large Array (VLA)
and from the Chandra Data Archive.
The National Radio Astronomy Observatory is a facility of the National Science Foundation operated under cooperative agreement by Associated Universities, Inc. 





\bibliographystyle{mnras}
\bibliography{/Users/mghkrause/texinput/references}

\begin{thebibliography}{}
\makeatletter
\relax
\def\mn@urlcharsother{\let\do\@makeother \do\$\do\&\do\#\do\^\do\_\do\%\do\~}
\def\mn@doi{\begingroup\mn@urlcharsother \@ifnextchar [ {\mn@doi@}
  {\mn@doi@[]}}
\def\mn@doi@[#1]#2{\def\@tempa{#1}\ifx\@tempa\@empty \href
  {http://dx.doi.org/#2} {doi:#2}\else \href {http://dx.doi.org/#2} {#1}\fi
  \endgroup}
\def\mn@eprint#1#2{\mn@eprint@#1:#2::\@nil}
\def\mn@eprint@arXiv#1{\href {http://arxiv.org/abs/#1} {{\tt arXiv:#1}}}
\def\mn@eprint@dblp#1{\href {http://dblp.uni-trier.de/rec/bibtex/#1.xml}
  {dblp:#1}}
\def\mn@eprint@#1:#2:#3:#4\@nil{\def\@tempa {#1}\def\@tempb {#2}\def\@tempc
  {#3}\ifx \@tempc \@empty \let \@tempc \@tempb \let \@tempb \@tempa \fi \ifx
  \@tempb \@empty \def\@tempb {arXiv}\fi \@ifundefined
  {mn@eprint@\@tempb}{\@tempb:\@tempc}{\expandafter \expandafter \csname
  mn@eprint@\@tempb\endcsname \expandafter{\@tempc}}}

\bibitem[\protect\citeauthoryear{{Abbott} et~al.,}{{Abbott}
  et~al.}{2016}]{Abbottea16a}
{Abbott} B.~P.,  et~al., 2016, \mn@doi [Physical Review Letters]
  {10.1103/PhysRevLett.116.061102}, \href
  {http://cdsads.u-strasbg.fr/abs/2016PhRvL.116f1102A} {116, 061102}

\bibitem[\protect\citeauthoryear{{Abraham} \& {Carrara}}{{Abraham} \&
  {Carrara}}{1998}]{AbrahamCar98}
{Abraham} Z.,  {Carrara} E.~A.,  1998, \mn@doi [\apj] {10.1086/305387}, \href
  {http://cdsads.u-strasbg.fr/abs/1998ApJ...496..172A} {496, 172}

\bibitem[\protect\citeauthoryear{{Alexander}}{{Alexander}}{1985}]{Alex85}
{Alexander} P.,  1985, \mn@doi [\mnras] {10.1093/mnras/213.4.743}, \href
  {http://cdsads.u-strasbg.fr/abs/1985MNRAS.213..743A} {213, 743}

\bibitem[\protect\citeauthoryear{{Alexander} \& {Pooley}}{{Alexander} \&
  {Pooley}}{1996}]{AP96}
{Alexander} P.,  {Pooley} G.~G.,  1996, in: Cygnus A -- Study of a Radio
  Galaxy, eds.: {Carilli}, C.~L. and {Harris}, D.~E..
Cambridge University Press, Cambridge, UK, p.~149

\bibitem[\protect\citeauthoryear{{Aloy}, {{Ib{\' a}{\~ n}ez}}, {Mart{\'\i}},
  {G{\' o}mez}  \& {M{\" u}ller}}{{Aloy} et~al.}{1999}]{Alea99}
{Aloy} M.~A.,  {{Ib{\' a}{\~ n}ez}} J.~M.,  {Mart{\'\i}} J.~M.,  {G{\' o}mez}
  J.-L.,   {M{\" u}ller} E.,  1999, \apjl, 523, L125

\bibitem[\protect\citeauthoryear{{Amaro-Seoane} et~al.,}{{Amaro-Seoane}
  et~al.}{2012}]{Amarea12}
{Amaro-Seoane} P.,  et~al., 2012, \mn@doi [Classical and Quantum Gravity]
  {10.1088/0264-9381/29/12/124016}, \href
  {http://cdsads.u-strasbg.fr/abs/2012CQGra..29l4016A} {29, 124016}

\bibitem[\protect\citeauthoryear{{Appl}}{{Appl}}{1996}]{Appl96}
{Appl} S.,  1996, \aap, 314, 995

\bibitem[\protect\citeauthoryear{{Appl}, {Sol}  \& {Vicente}}{{Appl}
  et~al.}{1996a}]{Applea96conf}
{Appl} S.,  {Sol} H.,   {Vicente} L.,  1996a, in {Hardee} P.~E.,  {Bridle}
  A.~H.,   {Zensus} J.~A.,  eds,  Astronomical Society of the Pacific
  Conference Series Vol. 100, Energy Transport in Radio Galaxies and Quasars.
  p.~181

\bibitem[\protect\citeauthoryear{{Appl}, {Sol}  \& {Vicente}}{{Appl}
  et~al.}{1996b}]{Applea96}
{Appl} S.,  {Sol} H.,   {Vicente} L.,  1996b, \aap, \href
  {http://cdsads.u-strasbg.fr/abs/1996A%26A...310..419A} {310, 419}

\bibitem[\protect\citeauthoryear{{Babul}, {Sharma}  \& {Reynolds}}{{Babul}
  et~al.}{2013}]{Babulea13}
{Babul} A.,  {Sharma} P.,   {Reynolds} C.~S.,  2013, \mn@doi [\apj]
  {10.1088/0004-637X/768/1/11}, \href
  {http://cdsads.u-strasbg.fr/abs/2013ApJ...768...11B} {768, 11}

\bibitem[\protect\citeauthoryear{{Balsara} \& {Norman}}{{Balsara} \&
  {Norman}}{1992}]{BalNorm92}
{Balsara} D.~S.,  {Norman} M.~L.,  1992, \mn@doi [\apj] {10.1086/171531}, \href
  {http://cdsads.u-strasbg.fr/abs/1992ApJ...393..631B} {393, 631}

\bibitem[\protect\citeauthoryear{{Bansal}, {Taylor}, {Peck}, {Zavala}  \&
  {Romani}}{{Bansal} et~al.}{2017}]{Bansea17}
{Bansal} K.,  {Taylor} G.~B.,  {Peck} A.~B.,  {Zavala} R.~T.,   {Romani} R.~W.,
   2017, \mn@doi [\apj] {10.3847/1538-4357/aa74e1}, \href
  {http://adsabs.harvard.edu/abs/2017ApJ...843...14B} {843, 14}

\bibitem[\protect\citeauthoryear{{Bardeen} \& {Petterson}}{{Bardeen} \&
  {Petterson}}{1975}]{BarPet75}
{Bardeen} J.~M.,  {Petterson} J.~A.,  1975, \mn@doi [\apjl] {10.1086/181711},
  \href {http://cdsads.u-strasbg.fr/abs/1975ApJ...195L..65B} {195, L65}

\bibitem[\protect\citeauthoryear{{Barker} \& {O'Connell}}{{Barker} \&
  {O'Connell}}{1975}]{BOC75}
{Barker} B.~M.,  {O'Connell} R.~F.,  1975, \mn@doi [\apjl] {10.1086/181840},
  \href {http://cdsads.u-strasbg.fr/abs/1975ApJ...199L..25B} {199, L25}

\bibitem[\protect\citeauthoryear{{Baryshev}}{{Baryshev}}{1983}]{Bary83}
{Baryshev} Y.~V.,  1983, Soviet Astronomy Letters, \href
  {http://cdsads.u-strasbg.fr/abs/1983SvAL....9..307B} {9, 307}

\bibitem[\protect\citeauthoryear{{Beckwith}, {Hawley}  \& {Krolik}}{{Beckwith}
  et~al.}{2008}]{BHK08}
{Beckwith} K.,  {Hawley} J.~F.,   {Krolik} J.~H.,  2008, \mn@doi [\apj]
  {10.1086/533492}, \href {http://adsabs.harvard.edu/abs/2008ApJ...678.1180B}
  {678, 1180}

\bibitem[\protect\citeauthoryear{{Begelman}, {Blandford}  \& {Rees}}{{Begelman}
  et~al.}{1980}]{BBR80}
{Begelman} M.~C.,  {Blandford} R.~D.,   {Rees} M.~J.,  1980, \mn@doi [\nat]
  {10.1038/287307a0}, \href {http://adsabs.harvard.edu/abs/1980Natur.287..307B}
  {287, 307}

\bibitem[\protect\citeauthoryear{{Begelman}, {Blandford}  \& {Rees}}{{Begelman}
  et~al.}{1984}]{BBR84}
{Begelman} M.~C.,  {Blandford} R.~D.,   {Rees} M.~J.,  1984, \mn@doi [Reviews
  of Modern Physics] {10.1103/RevModPhys.56.255}, \href
  {http://adsabs.harvard.edu/abs/1984RvMP...56..255B} {56, 255}

\bibitem[\protect\citeauthoryear{{Black}, {Baum}, {Leahy}, {Perley}, {Riley}
  \& {Scheuer}}{{Black} et~al.}{1992}]{Blackea92}
{Black} A.~R.~S.,  {Baum} S.~A.,  {Leahy} J.~P.,  {Perley} R.~A.,  {Riley}
  J.~M.,   {Scheuer} P.~A.~G.,  1992, \mn@doi [\mnras]
  {10.1093/mnras/256.2.186}, \href
  {http://cdsads.u-strasbg.fr/abs/1992MNRAS.256..186B} {256, 186}

\bibitem[\protect\citeauthoryear{{Boccardi}, {Krichbaum}, {Bach}, {Ros}  \&
  {Zensus}}{{Boccardi} et~al.}{2014}]{Boccardiea14}
{Boccardi} B.,  {Krichbaum} T.~P.,  {Bach} U.,  {Ros} E.,   {Zensus} A.~J.,
  2014, in Proceedings of the 12th European VLBI Network Symposium and Users
  Meeting (EVN 2014). 7-10 October 2014. Cagliari, Italy. Online at <A
  href=``http://pos.sissa.it/cgi-bin/reader/conf.cgi?confid=230''>http://pos.sissa.it/cgi-bin/reader/conf.cgi?confid=230</A>,
  id.16. p.~16

\bibitem[\protect\citeauthoryear{{Boccardi}, {Krichbaum}, {Bach}, {Mertens},
  {Ros}, {Alef}  \& {Zensus}}{{Boccardi} et~al.}{2016a}]{Boccardiea16a}
{Boccardi} B.,  {Krichbaum} T.~P.,  {Bach} U.,  {Mertens} F.,  {Ros} E.,
  {Alef} W.,   {Zensus} J.~A.,  2016a, \mn@doi [\aap]
  {10.1051/0004-6361/201526985}, \href
  {http://adsabs.harvard.edu/abs/2016A%26A...585A..33B} {585, A33}

\bibitem[\protect\citeauthoryear{{Boccardi}, {Krichbaum}, {Bach}, {Bremer}  \&
  {Zensus}}{{Boccardi} et~al.}{2016b}]{Boccardiea16b}
{Boccardi} B.,  {Krichbaum} T.~P.,  {Bach} U.,  {Bremer} M.,   {Zensus} J.~A.,
  2016b, \mn@doi [\aap] {10.1051/0004-6361/201628412}, \href
  {http://adsabs.harvard.edu/abs/2016A%26A...588L...9B} {588, L9}

\bibitem[\protect\citeauthoryear{{Bogd{\'a}n}, {Lovisari}, {Volonteri}  \&
  {Dubois}}{{Bogd{\'a}n} et~al.}{2018}]{Bogdea18}
{Bogd{\'a}n} {\'A}.,  {Lovisari} L.,  {Volonteri} M.,   {Dubois} Y.,  2018,
  \mn@doi [\apj] {10.3847/1538-4357/aa9ab5}, \href
  {http://cdsads.u-strasbg.fr/abs/2018ApJ...852..131B} {852, 131}

\bibitem[\protect\citeauthoryear{{Britzen} et~al.,}{{Britzen}
  et~al.}{2017}]{Britzea17}
{Britzen} S.,  et~al., 2017, \mn@doi [\aap] {10.1051/0004-6361/201629999},
  \href {http://cdsads.u-strasbg.fr/abs/2017A%26A...602A..29B} {602, A29}

\bibitem[\protect\citeauthoryear{{Cappellari}, {Neumayer}, {Reunanen}, {van der
  Werf}, {de Zeeuw}  \& {Rix}}{{Cappellari} et~al.}{2009}]{Cappea09}
{Cappellari} M.,  {Neumayer} N.,  {Reunanen} J.,  {van der Werf} P.~P.,  {de
  Zeeuw} P.~T.,   {Rix} H.-W.,  2009, \mn@doi [\mnras]
  {10.1111/j.1365-2966.2008.14377.x}, \href
  {http://cdsads.u-strasbg.fr/abs/2009MNRAS.394..660C} {394, 660}

\bibitem[\protect\citeauthoryear{{Capuzzo-Dolcetta} \& {Tosta e
  Melo}}{{Capuzzo-Dolcetta} \& {Tosta e Melo}}{2017}]{CaDoToMe17}
{Capuzzo-Dolcetta} R.,  {Tosta e Melo} I.,  2017, \mn@doi [\mnras]
  {10.1093/mnras/stx2246}, \href
  {http://adsabs.harvard.edu/abs/2017MNRAS.472.4013C} {472, 4013}

\bibitem[\protect\citeauthoryear{{Carilli} \& {Barthel}}{{Carilli} \&
  {Barthel}}{1996}]{CarBar96}
{Carilli} C.~L.,  {Barthel} P.~D.,  1996, \aap Review, 7, 1

\bibitem[\protect\citeauthoryear{{Carilli}, {Kurk}, {van der Werf}, {Perley}
  \& {Miley}}{{Carilli} et~al.}{1999}]{Cea99}
{Carilli} C.~L.,  {Kurk} J.~D.,  {van der Werf} P.~P.,  {Perley} R.~A.,
  {Miley} G.~K.,  1999, \mn@doi [\aj] {10.1086/301137}, \href
  {http://cdsads.u-strasbg.fr/abs/1999AJ....118.2581C} {118, 2581}

\bibitem[\protect\citeauthoryear{{Charisi}, {Bartos}, {Haiman}, {Price-Whelan},
  {Graham}, {Bellm}, {Laher}  \& {M{\'a}rka}}{{Charisi}
  et~al.}{2016}]{Charisea16}
{Charisi} M.,  {Bartos} I.,  {Haiman} Z.,  {Price-Whelan} A.~M.,  {Graham}
  M.~J.,  {Bellm} E.~C.,  {Laher} R.~R.,   {M{\'a}rka} S.,  2016, \mn@doi
  [\mnras] {10.1093/mnras/stw1838}, \href
  {http://cdsads.u-strasbg.fr/abs/2016MNRAS.463.2145C} {463, 2145}

\bibitem[\protect\citeauthoryear{{Chiaberge} et~al.,}{{Chiaberge}
  et~al.}{2017}]{Chiabea17}
{Chiaberge} M.,  et~al., 2017, \mn@doi [\aap] {10.1051/0004-6361/201629522},
  \href {http://cdsads.u-strasbg.fr/abs/2017A%26A...600A..57C} {600, A57}

\bibitem[\protect\citeauthoryear{{Choi}, {Wiita}  \& {Ryu}}{{Choi}
  et~al.}{2007}]{CWR07}
{Choi} E.,  {Wiita} P.~J.,   {Ryu} D.,  2007, \mn@doi [\apj] {10.1086/510120},
  \href
  {http://adsabs.harvard.edu/cgi-bin/nph-bib_query?bibcode=2007ApJ...655..769C&db_key=AST}
  {655, 769}

\bibitem[\protect\citeauthoryear{{Chon} \& {B{\"o}hringer}}{{Chon} \&
  {B{\"o}hringer}}{2017}]{ChonBo17}
{Chon} G.,  {B{\"o}hringer} H.,  2017, \mn@doi [\aap]
  {10.1051/0004-6361/201731854}, \href
  {http://cdsads.u-strasbg.fr/abs/2017A%26A...606L...4C} {606, L4}

\bibitem[\protect\citeauthoryear{{Chon}, {B{\"o}hringer}, {Krause}  \&
  {Tr{\"u}mper}}{{Chon} et~al.}{2012}]{Chonea12}
{Chon} G.,  {B{\"o}hringer} H.,  {Krause} M.,   {Tr{\"u}mper} J.,  2012,
  \mn@doi [\aap] {10.1051/0004-6361/201219538}, \href
  {http://adsabs.harvard.edu/abs/2012A%26A...545L...3C} {545, L3}

\bibitem[\protect\citeauthoryear{{Cielo}, {Babul}, {Antonuccio-Delogu}, {Silk}
  \& {Volonteri}}{{Cielo} et~al.}{2018}]{Cielea18aph}
{Cielo} S.,  {Babul} A.,  {Antonuccio-Delogu} V.,  {Silk} J.,   {Volonteri} M.,
   2018, preprint, \href {http://cdsads.u-strasbg.fr/abs/2018arXiv180104276C}
  {} (\mn@eprint {arXiv} {1801.04276})

\bibitem[\protect\citeauthoryear{{C{\^o}t{\'e}} et~al.,}{{C{\^o}t{\'e}}
  et~al.}{2006}]{Cotea06}
{C{\^o}t{\'e}} P.,  et~al., 2006, \mn@doi [\apjs] {10.1086/504042}, \href
  {http://cdsads.u-strasbg.fr/abs/2006ApJS..165...57C} {165, 57}

\bibitem[\protect\citeauthoryear{{Cox}, {Gull}  \& {Scheuer}}{{Cox}
  et~al.}{1991}]{CGS91}
{Cox} C.~I.,  {Gull} S.~F.,   {Scheuer} P.~A.~G.,  1991, \mnras, 252, 558

\bibitem[\protect\citeauthoryear{{Croston} et~al.,}{{Croston}
  et~al.}{2009}]{Crostea09}
{Croston} J.~H.,  et~al., 2009, \mn@doi [\mnras]
  {10.1111/j.1365-2966.2009.14715.x}, \href
  {http://adsabs.harvard.edu/abs/2009MNRAS.395.1999C} {395, 1999}

\bibitem[\protect\citeauthoryear{{Dabbech}, {Onose}, {Abdulaziz}, {Perley},
  {Smirnov}  \& {Wiaux}}{{Dabbech} et~al.}{2018}]{Dabbea18}
{Dabbech} A.,  {Onose} A.,  {Abdulaziz} A.,  {Perley} R.~A.,  {Smirnov} O.~M.,
   {Wiaux} Y.,  2018, \mn@doi [\mnras] {10.1093/mnras/sty372}, \href
  {http://cdsads.u-strasbg.fr/abs/2018MNRAS.476.2853D} {476, 2853}

\bibitem[\protect\citeauthoryear{{De Villiers}, {Hawley}  \& {Krolik}}{{De
  Villiers} et~al.}{2003}]{dVHK03}
{De Villiers} J.,  {Hawley} J.~F.,   {Krolik} J.~H.,  2003, \apj, 599, 1238

\bibitem[\protect\citeauthoryear{{Donohoe} \& {Smith}}{{Donohoe} \&
  {Smith}}{2016}]{DonSmi16}
{Donohoe} J.,  {Smith} M.~D.,  2016, \mn@doi [\mnras] {10.1093/mnras/stw335},
  \href {http://cdsads.u-strasbg.fr/abs/2016MNRAS.458..558D} {458, 558}

\bibitem[\protect\citeauthoryear{{Ekers}}{{Ekers}}{2016}]{Ekers16}
{Ekers} R.~D.,  2016, in {Meiron} Y.,  {Li} S.,  {Liu} F.-K.,   {Spurzem} R.,
  eds,  IAU Symposium Vol. 312, Star Clusters and Black Holes in Galaxies
  across Cosmic Time. pp 26--30, \mn@doi{10.1017/S1743921315007401}

\bibitem[\protect\citeauthoryear{{Ekers}, {Fanti}, {Lari}  \& {Parma}}{{Ekers}
  et~al.}{1978}]{Ekea78}
{Ekers} R.~D.,  {Fanti} R.,  {Lari} C.,   {Parma} P.,  1978, \mn@doi [\nat]
  {10.1038/276588a0}, \href
  {http://cdsads.u-strasbg.fr/abs/1978Natur.276..588E} {276, 588}

\bibitem[\protect\citeauthoryear{{English}, {Hardcastle}  \&
  {Krause}}{{English} et~al.}{2016}]{EHK2016}
{English} W.,  {Hardcastle} M.~J.,   {Krause} M.~G.~H.,  2016, \mn@doi [\mnras]
  {10.1093/mnras/stw1407}, \href
  {http://adsabs.harvard.edu/abs/2016MNRAS.461.2025E} {461, 2025}

\bibitem[\protect\citeauthoryear{{Eracleous} \& {Halpern}}{{Eracleous} \&
  {Halpern}}{2003}]{EH03}
{Eracleous} M.,  {Halpern} J.~P.,  2003, \mn@doi [\apj] {10.1086/379540}, \href
  {http://adsabs.harvard.edu/abs/2003ApJ...599..886E} {599, 886}

\bibitem[\protect\citeauthoryear{{Fanaroff} \& {Riley}}{{Fanaroff} \&
  {Riley}}{1974}]{FR74}
{Fanaroff} B.~L.,  {Riley} J.~M.,  1974, \mnras, 167, 31P

\bibitem[\protect\citeauthoryear{{Fender}, {Gallo}  \& {Russell}}{{Fender}
  et~al.}{2010}]{FenGR10}
{Fender} R.~P.,  {Gallo} E.,   {Russell} D.,  2010, \mn@doi [\mnras]
  {10.1111/j.1365-2966.2010.16754.x}, \href
  {http://adsabs.harvard.edu/abs/2010MNRAS.406.1425F} {406, 1425}

\bibitem[\protect\citeauthoryear{{Gaibler}, {Khochfar}  \& {Krause}}{{Gaibler}
  et~al.}{2011}]{GKK11}
{Gaibler} V.,  {Khochfar} S.,   {Krause} M.,  2011, \mn@doi [\mnras]
  {10.1111/j.1365-2966.2010.17674.x}, \href
  {http://adsabs.harvard.edu/abs/2010MNRAS.tmp.1669G} {411, 155}

\bibitem[\protect\citeauthoryear{{Gaibler}, {Khochfar}, {Krause}  \&
  {Silk}}{{Gaibler} et~al.}{2012}]{Gaiblea12}
{Gaibler} V.,  {Khochfar} S.,  {Krause} M.,   {Silk} J.,  2012, \mn@doi
  [\mnras] {10.1111/j.1365-2966.2012.21479.x}, \href
  {http://adsabs.harvard.edu/abs/2012MNRAS.425..438G} {425, 438}

\bibitem[\protect\citeauthoryear{{Gardner} \& {Done}}{{Gardner} \&
  {Done}}{2018}]{GarDon18}
{Gardner} E.,  {Done} C.,  2018, \mn@doi [\mnras] {10.1093/mnras/stx2516},
  \href {http://cdsads.u-strasbg.fr/abs/2018MNRAS.473.2639G} {473, 2639}

\bibitem[\protect\citeauthoryear{{Gebhardt}, {Adams}, {Richstone}, {Lauer},
  {Faber}, {G{\"u}ltekin}, {Murphy}  \& {Tremaine}}{{Gebhardt}
  et~al.}{2011}]{Gebhea11}
{Gebhardt} K.,  {Adams} J.,  {Richstone} D.,  {Lauer} T.~R.,  {Faber} S.~M.,
  {G{\"u}ltekin} K.,  {Murphy} J.,   {Tremaine} S.,  2011, \mn@doi [\apj]
  {10.1088/0004-637X/729/2/119}, \href
  {http://cdsads.u-strasbg.fr/abs/2011ApJ...729..119G} {729, 119}

\bibitem[\protect\citeauthoryear{{Ghisellini}, {Tavecchio}, {Maraschi},
  {Celotti}  \& {Sbarrato}}{{Ghisellini} et~al.}{2014}]{Ghiea14}
{Ghisellini} G.,  {Tavecchio} F.,  {Maraschi} L.,  {Celotti} A.,   {Sbarrato}
  T.,  2014, \mn@doi [\nat] {10.1038/nature13856}, \href
  {http://cdsads.u-strasbg.fr/abs/2014Natur.515..376G} {515, 376}

\bibitem[\protect\citeauthoryear{{Gilbert}, {Riley}, {Hardcastle}, {Croston},
  {Pooley}  \& {Alexander}}{{Gilbert} et~al.}{2004}]{Gilbea04}
{Gilbert} G.~M.,  {Riley} J.~M.,  {Hardcastle} M.~J.,  {Croston} J.~H.,
  {Pooley} G.~G.,   {Alexander} P.,  2004, \mn@doi [\mnras]
  {10.1111/j.1365-2966.2004.07824.x}, 351, 845

\bibitem[\protect\citeauthoryear{{Gower} \& {Hutchings}}{{Gower} \&
  {Hutchings}}{1982}]{GowHut82}
{Gower} A.~C.,  {Hutchings} J.~B.,  1982, \mn@doi [\apjl] {10.1086/183831},
  \href {http://cdsads.u-strasbg.fr/abs/1982ApJ...258L..63G} {258, L63}

\bibitem[\protect\citeauthoryear{{Gower}, {Gregory}, {Unruh}  \&
  {Hutchings}}{{Gower} et~al.}{1982}]{Gowea82}
{Gower} A.~C.,  {Gregory} P.~C.,  {Unruh} W.~G.,   {Hutchings} J.~B.,  1982,
  \mn@doi [\apj] {10.1086/160442}, \href
  {http://cdsads.u-strasbg.fr/abs/1982ApJ...262..478G} {262, 478}

\bibitem[\protect\citeauthoryear{{Graham} et~al.,}{{Graham}
  et~al.}{2015}]{Grahea15}
{Graham} M.~J.,  et~al., 2015, \mn@doi [\mnras] {10.1093/mnras/stv1726}, \href
  {http://cdsads.u-strasbg.fr/abs/2015MNRAS.453.1562G} {453, 1562}

\bibitem[\protect\citeauthoryear{{Hardcastle}, {Alexander}, {Pooley}  \&
  {Riley}}{{Hardcastle} et~al.}{1997}]{Hardea97}
{Hardcastle} M.~J.,  {Alexander} P.,  {Pooley} G.~G.,   {Riley} J.~M.,  1997,
  \mn@doi [\mnras] {10.1093/mnras/288.4.859}, \href
  {http://adsabs.harvard.edu/abs/1997MNRAS.288..859H} {288, 859}

\bibitem[\protect\citeauthoryear{{Hardcastle}, {Alexander}, {Pooley}  \&
  {Riley}}{{Hardcastle} et~al.}{1998}]{Hardea98}
{Hardcastle} M.~J.,  {Alexander} P.,  {Pooley} G.~G.,   {Riley} J.~M.,  1998,
  \mn@doi [\mnras] {10.1046/j.1365-8711.1998.01480.x}, \href
  {http://adsabs.harvard.edu/abs/1998MNRAS.296..445H} {296, 445}

\bibitem[\protect\citeauthoryear{{Hardcastle} et~al.,}{{Hardcastle}
  et~al.}{2007}]{Hardea07}
{Hardcastle} M.~J.,  et~al., 2007, \mn@doi [\apjl] {10.1086/524197}, \href
  {http://adsabs.harvard.edu/abs/2007ApJ...670L..81H} {670, L81}

\bibitem[\protect\citeauthoryear{{Hardee} \& {Norman}}{{Hardee} \&
  {Norman}}{1990}]{HN90}
{Hardee} P.~E.,  {Norman} M.~L.,  1990, \mn@doi [\apj] {10.1086/169464}, \href
  {http://cdsads.u-strasbg.fr/abs/1990ApJ...365..134H} {365, 134}

\bibitem[\protect\citeauthoryear{{H{\"a}ring} \& {Rix}}{{H{\"a}ring} \&
  {Rix}}{2004}]{HR04}
{H{\"a}ring} N.,  {Rix} H.-W.,  2004, \mn@doi [\apjl] {10.1086/383567}, \href
  {http://adsabs.harvard.edu/abs/2004ApJ...604L..89H} {604, L89}

\bibitem[\protect\citeauthoryear{{Harwood} et~al.,}{{Harwood}
  et~al.}{2017}]{Harwea17}
{Harwood} J.~J.,  et~al., 2017, \mn@doi [\mnras] {10.1093/mnras/stx820}, \href
  {http://cdsads.u-strasbg.fr/abs/2017MNRAS.469..639H} {469, 639}

\bibitem[\protect\citeauthoryear{{Hines}, {Owen}  \& {Eilek}}{{Hines}
  et~al.}{1989}]{HOE89}
{Hines} D.~C.,  {Owen} F.~N.,   {Eilek} J.~A.,  1989, \mn@doi [\apj]
  {10.1086/168163}, \href {http://adsabs.harvard.edu/abs/1989ApJ...347..713H}
  {347, 713}

\bibitem[\protect\citeauthoryear{{Hutchings}, {Price}  \& {Gower}}{{Hutchings}
  et~al.}{1988}]{HutPriGow88}
{Hutchings} J.~B.,  {Price} R.,   {Gower} A.~C.,  1988, \mn@doi [\apj]
  {10.1086/166363}, \href {http://cdsads.u-strasbg.fr/abs/1988ApJ...329..122H}
  {329, 122}

\bibitem[\protect\citeauthoryear{{Jorstad}, {Marscher}, {Lister}, {Stirling},
  {Cawthorne}, {G{\'o}mez}  \& {Gear}}{{Jorstad} et~al.}{2004}]{Jorstea04}
{Jorstad} S.~G.,  {Marscher} A.~P.,  {Lister} M.~L.,  {Stirling} A.~M.,
  {Cawthorne} T.~V.,  {G{\'o}mez} J.-L.,   {Gear} W.~K.,  2004, \mn@doi [\aj]
  {10.1086/420996}, \href {http://cdsads.u-strasbg.fr/abs/2004AJ....127.3115J}
  {127, 3115}

\bibitem[\protect\citeauthoryear{{Kaiser} \& {Alexander}}{{Kaiser} \&
  {Alexander}}{1999}]{KA99}
{Kaiser} C.~R.,  {Alexander} P.,  1999, \mnras, 305, 707

\bibitem[\protect\citeauthoryear{{Kaspi}, {Smith}, {Netzer}, {Maoz}, {Jannuzi}
  \& {Giveon}}{{Kaspi} et~al.}{2000}]{Kaspiea00}
{Kaspi} S.,  {Smith} P.~S.,  {Netzer} H.,  {Maoz} D.,  {Jannuzi} B.~T.,
  {Giveon} U.,  2000, \mn@doi [\apj] {10.1086/308704}, \href
  {http://cdsads.u-strasbg.fr/abs/2000ApJ...533..631K} {533, 631}

\bibitem[\protect\citeauthoryear{{Kelley}, {Blecha}, {Hernquist}, {Sesana}  \&
  {Taylor}}{{Kelley} et~al.}{2018}]{Kelleyea18}
{Kelley} L.~Z.,  {Blecha} L.,  {Hernquist} L.,  {Sesana} A.,   {Taylor} S.~R.,
  2018, \mn@doi [\mnras] {10.1093/mnras/sty689}, \href
  {http://cdsads.u-strasbg.fr/abs/2018MNRAS.477..964K} {477, 964}

\bibitem[\protect\citeauthoryear{{Khan}, {Fiacconi}, {Mayer}, {Berczik}  \&
  {Just}}{{Khan} et~al.}{2016}]{Kahnea16}
{Khan} F.~M.,  {Fiacconi} D.,  {Mayer} L.,  {Berczik} P.,   {Just} A.,  2016,
  \mn@doi [\apj] {10.3847/0004-637X/828/2/73}, \href
  {http://adsabs.harvard.edu/abs/2016ApJ...828...73K} {828, 73}

\bibitem[\protect\citeauthoryear{{Kharb}, {Lister}  \& {Cooper}}{{Kharb}
  et~al.}{2010}]{Kharbea10}
{Kharb} P.,  {Lister} M.~L.,   {Cooper} N.~J.,  2010, \mn@doi [\apj]
  {10.1088/0004-637X/710/1/764}, \href
  {http://cdsads.u-strasbg.fr/abs/2010ApJ...710..764K} {710, 764}

\bibitem[\protect\citeauthoryear{{Kharb}, {O'Dea}, {Baum}, {Hardcastle},
  {Dicken}, {Croston}, {Mingo}  \& {Noel-Storr}}{{Kharb}
  et~al.}{2014}]{Kharbea14}
{Kharb} P.,  {O'Dea} C.~P.,  {Baum} S.~A.,  {Hardcastle} M.~J.,  {Dicken} D.,
  {Croston} J.~H.,  {Mingo} B.,   {Noel-Storr} J.,  2014, \mn@doi [\mnras]
  {10.1093/mnras/stu421}, \href
  {http://cdsads.u-strasbg.fr/abs/2014MNRAS.440.2976K} {440, 2976}

\bibitem[\protect\citeauthoryear{{Kharb}, {Lal}  \& {Merritt}}{{Kharb}
  et~al.}{2017}]{KLM17}
{Kharb} P.,  {Lal} D.~V.,   {Merritt} D.,  2017, \mn@doi [Nature Astronomy]
  {10.1038/s41550-017-0256-4}, \href
  {http://cdsads.u-strasbg.fr/abs/2017NatAs...1..727K} {1, 727}

\bibitem[\protect\citeauthoryear{{Koide}, {Meier}, {Shibata}  \&
  {Kudoh}}{{Koide} et~al.}{2000}]{Koidea00}
{Koide} S.,  {Meier} D.~L.,  {Shibata} K.,   {Kudoh} T.,  2000, \apj, 536, 668

\bibitem[\protect\citeauthoryear{{Komissarov} \& {McKinney}}{{Komissarov} \&
  {McKinney}}{2007}]{KomMcK07}
{Komissarov} S.~S.,  {McKinney} J.~C.,  2007, \mn@doi [\mnras]
  {10.1111/j.1745-3933.2007.00301.x}, \href
  {http://cdsads.u-strasbg.fr/abs/2007MNRAS.377L..49K} {377, L49}

\bibitem[\protect\citeauthoryear{{Komissarov}, {Barkov}, {Vlahakis}  \&
  {K{\"o}nigl}}{{Komissarov} et~al.}{2007}]{Komea07}
{Komissarov} S.~S.,  {Barkov} M.~V.,  {Vlahakis} N.,   {K{\"o}nigl} A.,  2007,
  \mn@doi [\mnras] {10.1111/j.1365-2966.2007.12050.x}, \href
  {http://adsabs.harvard.edu/abs/2007MNRAS.380...51K} {380, 51}

\bibitem[\protect\citeauthoryear{{Komossa}, {Burwitz}, {Hasinger}, {Predehl},
  {Kaastra}  \& {Ikebe}}{{Komossa} et~al.}{2003}]{Kmsea03}
{Komossa} S.,  {Burwitz} V.,  {Hasinger} G.,  {Predehl} P.,  {Kaastra} J.~S.,
  {Ikebe} Y.,  2003, \mn@doi [\apjl] {10.1086/346145}, \href
  {http://cdsads.u-strasbg.fr/abs/2003ApJ...582L..15K} {582, L15}

\bibitem[\protect\citeauthoryear{{Kova{\v c}evi{\'c}},
  {P{\'e}rez-Hern{\'a}ndez}, {Popovi{\'c}}, {Shapovalova}, {Kollatschny}  \&
  {Ili{\'c}}}{{Kova{\v c}evi{\'c}} et~al.}{2018}]{Kovaea18}
{Kova{\v c}evi{\'c}} A.~B.,  {P{\'e}rez-Hern{\'a}ndez} E.,  {Popovi{\'c}}
  L.~{\v C}.,  {Shapovalova} A.~I.,  {Kollatschny} W.,   {Ili{\'c}} D.,  2018,
  \mn@doi [\mnras] {10.1093/mnras/stx3137}, \href
  {http://cdsads.u-strasbg.fr/abs/2018MNRAS.475.2051K} {475, 2051}

\bibitem[\protect\citeauthoryear{{Kraft}, {Hardcastle}, {Worrall}  \&
  {Murray}}{{Kraft} et~al.}{2005}]{Kraftea05}
{Kraft} R.~P.,  {Hardcastle} M.~J.,  {Worrall} D.~M.,   {Murray} S.~S.,  2005,
  \mn@doi [\apj] {10.1086/427822}, \href
  {http://cdsads.u-strasbg.fr/abs/2005ApJ...622..149K} {622, 149}

\bibitem[\protect\citeauthoryear{{Krause}}{{Krause}}{2003}]{mypap03a}
{Krause} M.,  2003, \aap, 398, 113

\bibitem[\protect\citeauthoryear{{Krause}}{{Krause}}{2005}]{Krause2005a}
{Krause} M.,  2005, \aap, 431, 45

\bibitem[\protect\citeauthoryear{{Krause} \& {Camenzind}}{{Krause} \&
  {Camenzind}}{2001}]{mypap01a}
{Krause} M.,  {Camenzind} M.,  2001, \aap, 380, 789

\bibitem[\protect\citeauthoryear{{Krause}, {Burkert}  \& {Schartmann}}{{Krause}
  et~al.}{2011}]{KBS11}
{Krause} M.,  {Burkert} A.,   {Schartmann} M.,  2011, \mn@doi [\mnras]
  {10.1111/j.1365-2966.2010.17698.x}, \href
  {http://adsabs.harvard.edu/abs/2011MNRAS.411..550K} {411, 550}

\bibitem[\protect\citeauthoryear{{Krause}, {Schartmann}  \& {Burkert}}{{Krause}
  et~al.}{2012a}]{KSB12}
{Krause} M.,  {Schartmann} M.,   {Burkert} A.,  2012a, \mn@doi [\mnras]
  {10.1111/j.1365-2966.2012.21642.x}, \href
  {http://adsabs.harvard.edu/abs/2012MNRAS.425.3172K} {425, 3172}

\bibitem[\protect\citeauthoryear{{Krause}, {Alexander}, {Riley}  \&
  {Hopton}}{{Krause} et~al.}{2012b}]{Krausea12b}
{Krause} M.,  {Alexander} P.,  {Riley} J.,   {Hopton} D.,  2012b, \mn@doi
  [\mnras] {10.1111/j.1365-2966.2012.21645.x}, \href
  {http://adsabs.harvard.edu/abs/2012MNRAS.427.3196K} {427, 3196}

\bibitem[\protect\citeauthoryear{{Kun}, {Gab{\'a}nyi}, {Karouzos}, {Britzen}
  \& {Gergely}}{{Kun} et~al.}{2014}]{Kunea14}
{Kun} E.,  {Gab{\'a}nyi} K.~{\'E}.,  {Karouzos} M.,  {Britzen} S.,   {Gergely}
  L.~{\'A}.,  2014, \mn@doi [\mnras] {10.1093/mnras/stu1813}, \href
  {http://cdsads.u-strasbg.fr/abs/2014MNRAS.445.1370K} {445, 1370}

\bibitem[\protect\citeauthoryear{{Kun}, {Frey}, {Gab{\'a}nyi}, {Britzen},
  {Cseh}  \& {Gergely}}{{Kun} et~al.}{2015}]{Kunea15}
{Kun} E.,  {Frey} S.,  {Gab{\'a}nyi} K.~{\'E}.,  {Britzen} S.,  {Cseh} D.,
  {Gergely} L.~{\'A}.,  2015, \mn@doi [\mnras] {10.1093/mnras/stv2049}, \href
  {http://cdsads.u-strasbg.fr/abs/2015MNRAS.454.1290K} {454, 1290}

\bibitem[\protect\citeauthoryear{{Laing}, {Riley}  \& {Longair}}{{Laing}
  et~al.}{1983}]{Laingea83}
{Laing} R.~A.,  {Riley} J.~M.,   {Longair} M.~S.,  1983, \mnras, \href
  {http://adsabs.harvard.edu/abs/1983MNRAS.204..151L} {204, 151}

\bibitem[\protect\citeauthoryear{{Lazio}}{{Lazio}}{2013}]{Lazio13}
{Lazio} T.~J.~W.,  2013, \mn@doi [Classical and Quantum Gravity]
  {10.1088/0264-9381/30/22/224011}, \href
  {http://cdsads.u-strasbg.fr/abs/2013CQGra..30v4011L} {30, 224011}

\bibitem[\protect\citeauthoryear{{Lazio}, {Cohen}, {Kassim}, {Perley},
  {Erickson}, {Carilli}  \& {Crane}}{{Lazio} et~al.}{2006}]{Lazioea06}
{Lazio} T.~J.~W.,  {Cohen} A.~S.,  {Kassim} N.~E.,  {Perley} R.~A.,  {Erickson}
  W.~C.,  {Carilli} C.~L.,   {Crane} P.~C.,  2006, \mn@doi [\apjl]
  {10.1086/504408}, \href {http://cdsads.u-strasbg.fr/abs/2006ApJ...642L..33L}
  {642, L33}

\bibitem[\protect\citeauthoryear{{Leahy}, {Black}, {Dennett-Thorpe},
  {Hardcastle}, {Komissarov}, {Perley}, {Riley}  \& {Scheuer}}{{Leahy}
  et~al.}{1997}]{Lea97}
{Leahy} J.~P.,  {Black} A.~R.~S.,  {Dennett-Thorpe} J.,  {Hardcastle} M.~J.,
  {Komissarov} S.,  {Perley} R.~A.,  {Riley} J.~M.,   {Scheuer} P.~A.~G.,
  1997, \mnras, \href {http://adsabs.harvard.edu/abs/1997MNRAS.291...20L} {291,
  20}

\bibitem[\protect\citeauthoryear{{Li} et~al.,}{{Li} et~al.}{2016}]{Liea16}
{Li} Y.-R.,  et~al., 2016, \mn@doi [\apj] {10.3847/0004-637X/822/1/4}, \href
  {http://cdsads.u-strasbg.fr/abs/2016ApJ...822....4L} {822, 4}

\bibitem[\protect\citeauthoryear{{Lister}}{{Lister}}{2016}]{Lister16}
{Lister} M.,  2016, \mn@doi [Galaxies] {10.3390/galaxies4030029}, \href
  {http://cdsads.u-strasbg.fr/abs/2016Galax...4...29L} {4, 29}

\bibitem[\protect\citeauthoryear{{Lister} et~al.,}{{Lister}
  et~al.}{2013}]{Listea13}
{Lister} M.~L.,  et~al., 2013, \mn@doi [\aj] {10.1088/0004-6256/146/5/120},
  \href {http://cdsads.u-strasbg.fr/abs/2013AJ....146..120L} {146, 120}

\bibitem[\protect\citeauthoryear{{Liu}, {Eracleous}  \& {Halpern}}{{Liu}
  et~al.}{2016}]{LEH16}
{Liu} J.,  {Eracleous} M.,   {Halpern} J.~P.,  2016, \mn@doi [\apj]
  {10.3847/0004-637X/817/1/42}, \href
  {http://cdsads.u-strasbg.fr/abs/2016ApJ...817...42L} {817, 42}

\bibitem[\protect\citeauthoryear{{Lobanov} \& {Roland}}{{Lobanov} \&
  {Roland}}{2005}]{LR05}
{Lobanov} A.~P.,  {Roland} J.,  2005, \mn@doi [\aap]
  {10.1051/0004-6361:20041831}, \href
  {http://cdsads.u-strasbg.fr/abs/2005A%26A...431..831L} {431, 831}

\bibitem[\protect\citeauthoryear{{Lodato} \& {Pringle}}{{Lodato} \&
  {Pringle}}{2006}]{LoPri06}
{Lodato} G.,  {Pringle} J.~E.,  2006, \mn@doi [\mnras]
  {10.1111/j.1365-2966.2006.10194.x}, \href
  {http://cdsads.u-strasbg.fr/abs/2006MNRAS.368.1196L} {368, 1196}

\bibitem[\protect\citeauthoryear{{Lu} et~al.,}{{Lu} et~al.}{2013}]{Luea13}
{Lu} R.-S.,  et~al., 2013, \mn@doi [\apj] {10.1088/0004-637X/772/1/13}, \href
  {http://cdsads.u-strasbg.fr/abs/2013ApJ...772...13L} {772, 13}

\bibitem[\protect\citeauthoryear{{Magorrian} et~al.,}{{Magorrian}
  et~al.}{1998}]{Magea98}
{Magorrian} J.,  et~al., 1998, \aj, 115, 2285

\bibitem[\protect\citeauthoryear{{Marconi}, {Schreier}, {Koekemoer}, {Capetti},
  {Axon}, {Macchetto}  \& {Caon}}{{Marconi} et~al.}{2000}]{Marcea00}
{Marconi} A.,  {Schreier} E.~J.,  {Koekemoer} A.,  {Capetti} A.,  {Axon} D.,
  {Macchetto} D.,   {Caon} N.,  2000, \mn@doi [\apj] {10.1086/308168}, \href
  {http://cdsads.u-strasbg.fr/abs/2000ApJ...528..276M} {528, 276}

\bibitem[\protect\citeauthoryear{{Marconi}, {Axon}, {Maiolino}, {Nagao},
  {Pastorini}, {Pietrini}, {Robinson}  \& {Torricelli}}{{Marconi}
  et~al.}{2008}]{Marcea08}
{Marconi} A.,  {Axon} D.~J.,  {Maiolino} R.,  {Nagao} T.,  {Pastorini} G.,
  {Pietrini} P.,  {Robinson} A.,   {Torricelli} G.,  2008, \mn@doi [\apj]
  {10.1086/529360}, \href {http://adsabs.harvard.edu/abs/2008ApJ...678..693M}
  {678, 693}

\bibitem[\protect\citeauthoryear{{Mayer}}{{Mayer}}{2017}]{Mayer17b}
{Mayer} L.,  2017, in Journal of Physics Conference Series. p. 012025
  (\mn@eprint {arXiv} {1703.00661}), \mn@doi{10.1088/1742-6596/840/1/012025}

\bibitem[\protect\citeauthoryear{{McKean} et~al.,}{{McKean}
  et~al.}{2016}]{McKeanea16}
{McKean} J.~P.,  et~al., 2016, \mn@doi [\mnras] {10.1093/mnras/stw2105}, \href
  {http://cdsads.u-strasbg.fr/abs/2016MNRAS.463.3143M} {463, 3143}

\bibitem[\protect\citeauthoryear{{Merritt} \& {Ekers}}{{Merritt} \&
  {Ekers}}{2002}]{MeEk02}
{Merritt} D.,  {Ekers} R.~D.,  2002, \mn@doi [Science]
  {10.1126/science.1074688}, \href
  {http://cdsads.u-strasbg.fr/abs/2002Sci...297.1310M} {297, 1310}

\bibitem[\protect\citeauthoryear{{Moderski}, {Sikora}  \& {Lasota}}{{Moderski}
  et~al.}{1998}]{MSL98}
{Moderski} R.,  {Sikora} M.,   {Lasota} J.,  1998, \mn@doi [\mnras]
  {10.1046/j.1365-8711.1998.02009.x}, \href
  {http://adsabs.harvard.edu/abs/1998MNRAS.301..142M} {301, 142}

\bibitem[\protect\citeauthoryear{{Monceau-Baroux}, {Porth}, {Meliani}  \&
  {Keppens}}{{Monceau-Baroux} et~al.}{2014}]{Moncea14}
{Monceau-Baroux} R.,  {Porth} O.,  {Meliani} Z.,   {Keppens} R.,  2014, \mn@doi
  [\aap] {10.1051/0004-6361/201322682}, \href
  {http://cdsads.u-strasbg.fr/abs/2014A%26A...561A..30M} {561, A30}

\bibitem[\protect\citeauthoryear{{Morganti}, {Oosterloo}, {Tadhunter}, {Aiudi},
  {Jones}  \& {Villar-Martin}}{{Morganti} et~al.}{1999a}]{Morgea99b}
{Morganti} R.,  {Oosterloo} T.,  {Tadhunter} C.~N.,  {Aiudi} R.,  {Jones} P.,
  {Villar-Martin} M.,  1999a, \mn@doi [\aaps] {10.1051/aas:1999427}, \href
  {http://cdsads.u-strasbg.fr/abs/1999A%26AS..140..355M} {140, 355}

\bibitem[\protect\citeauthoryear{{Morganti}, {Killeen}, {Ekers}  \&
  {Oosterloo}}{{Morganti} et~al.}{1999b}]{Morgea99}
{Morganti} R.,  {Killeen} N.~E.~B.,  {Ekers} R.~D.,   {Oosterloo} T.~A.,
  1999b, \mn@doi [\mnras] {10.1046/j.1365-8711.1999.02622.x}, \href
  {http://adsabs.harvard.edu/abs/1999MNRAS.307..750M} {307, 750}

\bibitem[\protect\citeauthoryear{{Mukherjee}, {Bicknell}, {Sutherland}  \&
  {Wagner}}{{Mukherjee} et~al.}{2016}]{Mukhea16}
{Mukherjee} D.,  {Bicknell} G.~V.,  {Sutherland} R.,   {Wagner} A.,  2016,
  \mn@doi [\mnras] {10.1093/mnras/stw1368}, \href
  {http://cdsads.u-strasbg.fr/abs/2016MNRAS.461..967M} {461, 967}

\bibitem[\protect\citeauthoryear{{M{\"u}ller} et~al.,}{{M{\"u}ller}
  et~al.}{2014}]{Muellea14}
{M{\"u}ller} C.,  et~al., 2014, \mn@doi [\aap] {10.1051/0004-6361/201423948},
  \href {http://cdsads.u-strasbg.fr/abs/2014A%26A...569A.115M} {569, A115}

\bibitem[\protect\citeauthoryear{{Mullin}, {Hardcastle}  \& {Riley}}{{Mullin}
  et~al.}{2006}]{Mulea06}
{Mullin} L.~M.,  {Hardcastle} M.~J.,   {Riley} J.~M.,  2006, \mn@doi [\mnras]
  {10.1111/j.1365-2966.2006.10763.x}, \href
  {http://adsabs.harvard.edu/abs/2006MNRAS.372..113M} {372, 113}

\bibitem[\protect\citeauthoryear{{Mullin}, {Riley}  \& {Hardcastle}}{{Mullin}
  et~al.}{2008}]{Mulea08}
{Mullin} L.~M.,  {Riley} J.~M.,   {Hardcastle} M.~J.,  2008, \mn@doi [\mnras]
  {10.1111/j.1365-2966.2008.13534.x}, \href
  {http://adsabs.harvard.edu/abs/2008MNRAS.390..595M} {390, 595}

\bibitem[\protect\citeauthoryear{{Nawaz}, {Bicknell}, {Wagner}, {Sutherland}
  \& {McNamara}}{{Nawaz} et~al.}{2016}]{Nawea16a}
{Nawaz} M.~A.,  {Bicknell} G.~V.,  {Wagner} A.~Y.,  {Sutherland} R.~S.,
  {McNamara} B.~R.,  2016, \mn@doi [\mnras] {10.1093/mnras/stw330}, \href
  {http://cdsads.u-strasbg.fr/abs/2016MNRAS.458..802N} {458, 802}

\bibitem[\protect\citeauthoryear{{Netzer} \& {Marziani}}{{Netzer} \&
  {Marziani}}{2010}]{Net10}
{Netzer} H.,  {Marziani} P.,  2010, \mn@doi [\apj]
  {10.1088/0004-637X/724/1/318}, \href
  {http://adsabs.harvard.edu/abs/2010ApJ...724..318N} {724, 318}

\bibitem[\protect\citeauthoryear{{O'Neill}, {Beckwith}  \&
  {Begelman}}{{O'Neill} et~al.}{2012}]{OBB12}
{O'Neill} S.~M.,  {Beckwith} K.,   {Begelman} M.~C.,  2012, \mn@doi [\mnras]
  {10.1111/j.1365-2966.2012.20721.x}, \href
  {http://adsabs.harvard.edu/abs/2012MNRAS.422.1436O} {422, 1436}

\bibitem[\protect\citeauthoryear{{Owen}, {Eilek}  \& {Kassim}}{{Owen}
  et~al.}{2000}]{Owea00}
{Owen} F.~N.,  {Eilek} J.~A.,   {Kassim} N.~E.,  2000, \mn@doi [\apj]
  {10.1086/317151}, \href {http://cdsads.u-strasbg.fr/abs/2000ApJ...543..611O}
  {543, 611}

\bibitem[\protect\citeauthoryear{{Perley}, {Perley}, {Dhawan}  \&
  {Carilli}}{{Perley} et~al.}{2017}]{Perlea17}
{Perley} D.~A.,  {Perley} R.~A.,  {Dhawan} V.,   {Carilli} C.~L.,  2017,
  \mn@doi [\apj] {10.3847/1538-4357/aa725b}, \href
  {http://cdsads.u-strasbg.fr/abs/2017ApJ...841..117P} {841, 117}

\bibitem[\protect\citeauthoryear{{Perucho}, {Mart{\'{\i}}}  \&
  {Hanasz}}{{Perucho} et~al.}{2004}]{Peruea04b}
{Perucho} M.,  {Mart{\'{\i}}} J.~M.,   {Hanasz} M.,  2004, \mn@doi [\aap]
  {10.1051/0004-6361:20040350}, \href
  {http://adsabs.harvard.edu/abs/2004A%26A...427..431P} {427, 431}

\bibitem[\protect\citeauthoryear{{Perucho}, {Mart{\'{\i}}}, {Cela}, {Hanasz},
  {de La Cruz}  \& {Rubio}}{{Perucho} et~al.}{2010}]{Peruea10}
{Perucho} M.,  {Mart{\'{\i}}} J.~M.,  {Cela} J.~M.,  {Hanasz} M.,  {de La Cruz}
  R.,   {Rubio} F.,  2010, \mn@doi [\aap] {10.1051/0004-6361/200913012}, \href
  {http://adsabs.harvard.edu/abs/2010A%26A...519A..41P} {519, A41}

\bibitem[\protect\citeauthoryear{{Peterson} et~al.,}{{Peterson}
  et~al.}{2004}]{Peterea04a}
{Peterson} B.~M.,  et~al., 2004, \mn@doi [\apj] {10.1086/423269}, \href
  {http://adsabs.harvard.edu/abs/2004ApJ...613..682P} {613, 682}

\bibitem[\protect\citeauthoryear{{Porth} \& {Fendt}}{{Porth} \&
  {Fendt}}{2010}]{PF10}
{Porth} O.,  {Fendt} C.,  2010, \mn@doi [\apj] {10.1088/0004-637X/709/2/1100},
  \href {http://adsabs.harvard.edu/abs/2010ApJ...709.1100P} {709, 1100}

\bibitem[\protect\citeauthoryear{{Rafferty}, {McNamara}, {Nulsen}  \&
  {Wise}}{{Rafferty} et~al.}{2006}]{Rafea06}
{Rafferty} D.~A.,  {McNamara} B.~R.,  {Nulsen} P.~E.~J.,   {Wise} M.~W.,  2006,
  \mn@doi [\apj] {10.1086/507672}, \href
  {http://adsabs.harvard.edu/abs/2006ApJ...652..216R} {652, 216}

\bibitem[\protect\citeauthoryear{{Ramos Almeida}, {Bessiere}, {Tadhunter},
  {Inskip}, {Morganti}, {Dicken}, {Gonz{\'a}lez-Serrano}  \& {Holt}}{{Ramos
  Almeida} et~al.}{2013}]{Ramea13}
{Ramos Almeida} C.,  {Bessiere} P.~S.,  {Tadhunter} C.~N.,  {Inskip} K.~J.,
  {Morganti} R.,  {Dicken} D.,  {Gonz{\'a}lez-Serrano} J.~I.,   {Holt} J.,
  2013, \mn@doi [\mnras] {10.1093/mnras/stt1595}, \href
  {http://cdsads.u-strasbg.fr/abs/2013MNRAS.436..997R} {436, 997}

\bibitem[\protect\citeauthoryear{{Richings}, {Uttley}  \&
  {K{\"o}rding}}{{Richings} et~al.}{2011}]{Richea11}
{Richings} A.~J.,  {Uttley} P.,   {K{\"o}rding} E.,  2011, \mn@doi [\mnras]
  {10.1111/j.1365-2966.2011.18845.x}, \href
  {http://cdsads.u-strasbg.fr/abs/2011MNRAS.415.2158R} {415, 2158}

\bibitem[\protect\citeauthoryear{{Rieger}}{{Rieger}}{2007}]{Rieger07}
{Rieger} F.~M.,  2007, \mn@doi [\apss] {10.1007/s10509-007-9467-y}, \href
  {http://cdsads.u-strasbg.fr/abs/2007Ap%26SS.309..271R} {309, 271}

\bibitem[\protect\citeauthoryear{{Rodriguez-Gomez} et~al.,}{{Rodriguez-Gomez}
  et~al.}{2015}]{Rodrea15}
{Rodriguez-Gomez} V.,  et~al., 2015, \mn@doi [\mnras] {10.1093/mnras/stv264},
  \href {http://cdsads.u-strasbg.fr/abs/2015MNRAS.449...49R} {449, 49}

\bibitem[\protect\citeauthoryear{{Rodr{\'{\i}}guez-Mart{\'{\i}}nez},
  {Vel{\'a}zquez}, {Binette}  \& {Raga}}{{Rodr{\'{\i}}guez-Mart{\'{\i}}nez}
  et~al.}{2006}]{Rodrea06}
{Rodr{\'{\i}}guez-Mart{\'{\i}}nez} M.,  {Vel{\'a}zquez} P.~F.,  {Binette} L.,
  {Raga} A.~C.,  2006, \mn@doi [\aap] {10.1051/0004-6361:20053382}, \href
  {http://cdsads.u-strasbg.fr/abs/2006A%26A...448...15R} {448, 15}

\bibitem[\protect\citeauthoryear{{Rosen}, {Hardee}, {Clarke}  \&
  {Johnson}}{{Rosen} et~al.}{1999a}]{RHC99}
{Rosen} A.,  {Hardee} P.~E.,  {Clarke} D.~A.,   {Johnson} A.,  1999a, \apj,
  510, 136

\bibitem[\protect\citeauthoryear{{Rosen}, {Hughes}, {Duncan}  \&
  {Hardee}}{{Rosen} et~al.}{1999b}]{Rosea99}
{Rosen} A.,  {Hughes} P.~A.,  {Duncan} G.~C.,   {Hardee} P.~E.,  1999b, \apj,
  516, 729

\bibitem[\protect\citeauthoryear{{Rossi}, {Bodo}, {Capetti}  \&
  {Massaglia}}{{Rossi} et~al.}{2017}]{Rossiea17}
{Rossi} P.,  {Bodo} G.,  {Capetti} A.,   {Massaglia} S.,  2017, \mn@doi [\aap]
  {10.1051/0004-6361/201730594}, \href
  {http://cdsads.u-strasbg.fr/abs/2017A%26A...606A..57R} {606, A57}

\bibitem[\protect\citeauthoryear{{Sabater}, {Best}  \&
  {Argudo-Fern{\'a}ndez}}{{Sabater} et~al.}{2013}]{Sabea13}
{Sabater} J.,  {Best} P.~N.,   {Argudo-Fern{\'a}ndez} M.,  2013, \mn@doi
  [\mnras] {10.1093/mnras/sts675}, \href
  {http://cdsads.u-strasbg.fr/abs/2013MNRAS.430..638S} {430, 638}

\bibitem[\protect\citeauthoryear{{Saripalli} \& {Roberts}}{{Saripalli} \&
  {Roberts}}{2018}]{SR18}
{Saripalli} L.,  {Roberts} D.~H.,  2018, \mn@doi [\apj]
  {10.3847/1538-4357/aa9c4b}, \href
  {http://cdsads.u-strasbg.fr/abs/2018ApJ...852...48S} {852, 48}

\bibitem[\protect\citeauthoryear{{Sarzi}, {Spiniello}, {La Barbera},
  {Krajnovi{\'c}}  \& {van den Bosch}}{{Sarzi} et~al.}{2018}]{Sarzea18}
{Sarzi} M.,  {Spiniello} C.,  {La Barbera} F.,  {Krajnovi{\'c}} D.,   {van den
  Bosch} R.,  2018, \mn@doi [\mnras] {10.1093/mnras/sty1092}, \href
  {http://cdsads.u-strasbg.fr/abs/2018MNRAS.478.4084S} {478, 4084}

\bibitem[\protect\citeauthoryear{{Savolainen}, {Wiik}, {Valtaoja}  \&
  {Tornikoski}}{{Savolainen} et~al.}{2006}]{Savolea06}
{Savolainen} T.,  {Wiik} K.,  {Valtaoja} E.,   {Tornikoski} M.,  2006, \mn@doi
  [\aap] {10.1051/0004-6361:20053753}, \href
  {http://cdsads.u-strasbg.fr/abs/2006A%26A...446...71S} {446, 71}

\bibitem[\protect\citeauthoryear{{Scheuer}}{{Scheuer}}{1982}]{Scheuer82}
{Scheuer} P.~A.~G.,  1982, in {Heeschen} D.~S.,  {Wade} C.~M.,  eds,  IAU
  Symposium Vol. 97, Extragalactic Radio Sources. pp 163--165

\bibitem[\protect\citeauthoryear{{Scheuer} \& {Feiler}}{{Scheuer} \&
  {Feiler}}{1996}]{ScheuerF96}
{Scheuer} P.~A.~G.,  {Feiler} R.,  1996, \mn@doi [\mnras]
  {10.1093/mnras/282.1.291}, \href
  {http://cdsads.u-strasbg.fr/abs/1996MNRAS.282..291S} {282, 291}

\bibitem[\protect\citeauthoryear{{Shabala} et~al.,}{{Shabala}
  et~al.}{2012}]{Shabea12a}
{Shabala} S.~S.,  et~al., 2012, \mn@doi [\mnras]
  {10.1111/j.1365-2966.2012.20598.x}, \href
  {http://cdsads.u-strasbg.fr/abs/2012MNRAS.423...59S} {423, 59}

\bibitem[\protect\citeauthoryear{{Smith}, {Wilson}, {Arnaud}, {Terashima}  \&
  {Young}}{{Smith} et~al.}{2002a}]{Sea01}
{Smith} D.~A.,  {Wilson} A.~S.,  {Arnaud} K.~A.,  {Terashima} Y.,   {Young}
  A.~J.,  2002a, \apj, 565

\bibitem[\protect\citeauthoryear{{Smith}, {Wilson}, {Arnaud}, {Terashima}  \&
  {Young}}{{Smith} et~al.}{2002b}]{Smithea02}
{Smith} D.~A.,  {Wilson} A.~S.,  {Arnaud} K.~A.,  {Terashima} Y.,   {Young}
  A.~J.,  2002b, \apj, 565

\bibitem[\protect\citeauthoryear{{Stairs}}{{Stairs}}{2003}]{Stairs03}
{Stairs} I.~H.,  2003, \mn@doi [Living Reviews in Relativity]
  {10.12942/lrr-2003-5}, \href
  {http://cdsads.u-strasbg.fr/abs/2003LRR.....6....5S} {6, 5}

\bibitem[\protect\citeauthoryear{{Steenbrugge} \& {Blundell}}{{Steenbrugge} \&
  {Blundell}}{2008}]{SteenBlun08}
{Steenbrugge} K.~C.,  {Blundell} K.~M.,  2008, \mn@doi [\mnras]
  {10.1111/j.1365-2966.2007.12665.x}, \href
  {http://cdsads.u-strasbg.fr/abs/2008MNRAS.388.1457S} {388, 1457}

\bibitem[\protect\citeauthoryear{{Tadhunter}, {Marconi}, {Axon}, {Wills},
  {Robinson}  \& {Jackson}}{{Tadhunter} et~al.}{2003}]{Tadea03}
{Tadhunter} C.,  {Marconi} A.,  {Axon} D.,  {Wills} K.,  {Robinson} T.~G.,
  {Jackson} N.,  2003, \mnras, 342, 861

\bibitem[\protect\citeauthoryear{{Taylor}}{{Taylor}}{1996}]{Taylor96}
{Taylor} G.~B.,  1996, \mn@doi [\apj] {10.1086/177874}, \href
  {http://cdsads.u-strasbg.fr/abs/1996ApJ...470..394T} {470, 394}

\bibitem[\protect\citeauthoryear{{Taylor}, {Perley}, {Inoue}, {Kato}, {Tabara}
  \& {Aizu}}{{Taylor} et~al.}{1990}]{Taylea90}
{Taylor} G.~B.,  {Perley} R.~A.,  {Inoue} M.,  {Kato} T.,  {Tabara} H.,
  {Aizu} K.,  1990, \mn@doi [\apj] {10.1086/169094}, \href
  {http://cdsads.u-strasbg.fr/abs/1990ApJ...360...41T} {360, 41}

\bibitem[\protect\citeauthoryear{{Thorne} \& {Blandford}}{{Thorne} \&
  {Blandford}}{2017}]{TP17}
{Thorne} K.~S.,  {Blandford} R.~D.,  2017, {Modern Classical Physics: Optics,
  Fluids, Plasmas, Elasticity, Relativity, and Statistical Physics}

\bibitem[\protect\citeauthoryear{{Tremmel}, {Governato}, {Volonteri}, {Quinn}
  \& {Pontzen}}{{Tremmel} et~al.}{2018a}]{Tremmea18b}
{Tremmel} M.,  {Governato} F.,  {Volonteri} M.,  {Quinn} T.~R.,   {Pontzen} A.,
   2018a, \mn@doi [\mnras] {10.1093/mnras/sty139}, \href
  {http://cdsads.u-strasbg.fr/abs/2018MNRAS.475.4967T} {475, 4967}

\bibitem[\protect\citeauthoryear{{Tremmel}, {Governato}, {Volonteri}, {Pontzen}
   \& {Quinn}}{{Tremmel} et~al.}{2018b}]{Tremmea18a}
{Tremmel} M.,  {Governato} F.,  {Volonteri} M.,  {Pontzen} A.,   {Quinn} T.~R.,
   2018b, \mn@doi [\apjl] {10.3847/2041-8213/aabc0a}, \href
  {http://cdsads.u-strasbg.fr/abs/2018ApJ...857L..22T} {857, L22}

\bibitem[\protect\citeauthoryear{{Turner}, {Shabala}  \& {Krause}}{{Turner}
  et~al.}{2018}]{Turnea18b}
{Turner} R.~J.,  {Shabala} S.~S.,   {Krause} M.~G.~H.,  2018, \mn@doi [\mnras]
  {10.1093/mnras/stx2947}, \href
  {http://adsabs.harvard.edu/abs/2018MNRAS.474.3361T} {474, 3361}

\bibitem[\protect\citeauthoryear{{Wagner} \& {Bicknell}}{{Wagner} \&
  {Bicknell}}{2011}]{WB11}
{Wagner} A.~Y.,  {Bicknell} G.~V.,  2011, \mn@doi [\apj]
  {10.1088/0004-637X/728/1/29}, \href
  {http://adsabs.harvard.edu/abs/2011ApJ...728...29W} {728, 29}

\bibitem[\protect\citeauthoryear{{Wagner}, {Bicknell}  \& {Umemura}}{{Wagner}
  et~al.}{2012}]{WBU12}
{Wagner} A.~Y.,  {Bicknell} G.~V.,   {Umemura} M.,  2012, \mn@doi [\apj]
  {10.1088/0004-637X/757/2/136}, \href
  {http://adsabs.harvard.edu/abs/2012ApJ...757..136W} {757, 136}

\bibitem[\protect\citeauthoryear{{Wall} \& {Peacock}}{{Wall} \&
  {Peacock}}{1985}]{WaPe85}
{Wall} J.~V.,  {Peacock} J.~A.,  1985, \mn@doi [\mnras]
  {10.1093/mnras/216.2.173}, \href
  {http://cdsads.u-strasbg.fr/abs/1985MNRAS.216..173W} {216, 173}

\bibitem[\protect\citeauthoryear{{Walsh}, {Barth}, {Ho}  \& {Sarzi}}{{Walsh}
  et~al.}{2013}]{Walshea13}
{Walsh} J.~L.,  {Barth} A.~J.,  {Ho} L.~C.,   {Sarzi} M.,  2013, \mn@doi [\apj]
  {10.1088/0004-637X/770/2/86}, \href
  {http://cdsads.u-strasbg.fr/abs/2013ApJ...770...86W} {770, 86}

\bibitem[\protect\citeauthoryear{{Wang}, {Greene}, {Ju}, {Rafikov}, {Ruan}  \&
  {Schneider}}{{Wang} et~al.}{2017}]{WangLea17}
{Wang} L.,  {Greene} J.~E.,  {Ju} W.,  {Rafikov} R.~R.,  {Ruan} J.~J.,
  {Schneider} D.~P.,  2017, \mn@doi [\apj] {10.3847/1538-4357/834/2/129}, \href
  {http://cdsads.u-strasbg.fr/abs/2017ApJ...834..129W} {834, 129}

\bibitem[\protect\citeauthoryear{{Weijmans} et~al.,}{{Weijmans}
  et~al.}{2014}]{Weijmea14}
{Weijmans} A.-M.,  et~al., 2014, \mn@doi [\mnras] {10.1093/mnras/stu1603},
  \href {http://cdsads.u-strasbg.fr/abs/2014MNRAS.444.3340W} {444, 3340}

\bibitem[\protect\citeauthoryear{{Williams} \& {Gull}}{{Williams} \&
  {Gull}}{1985}]{WG85}
{Williams} A.~G.,  {Gull} S.~F.,  1985, \mn@doi [\nat] {10.1038/313034a0},
  \href {http://cdsads.u-strasbg.fr/abs/1985Natur.313...34W} {313, 34}

\bibitem[\protect\citeauthoryear{{Wilson}, {Young}  \& {Shopbell}}{{Wilson}
  et~al.}{2000}]{WYS00}
{Wilson} A.~S.,  {Young} A.~J.,   {Shopbell} P.~L.,  2000, \apjl, 544, L27

\bibitem[\protect\citeauthoryear{{Zhu} et~al.,}{{Zhu} et~al.}{2015}]{Zhuea15}
{Zhu} X.-J.,  et~al., 2015, \mn@doi [\mnras] {10.1093/mnras/stv381}, \href
  {http://cdsads.u-strasbg.fr/abs/2015MNRAS.449.1650Z} {449, 1650}

\makeatother
\end{thebibliography}



\begin{figure*}
	\includegraphics[width=\textwidth]{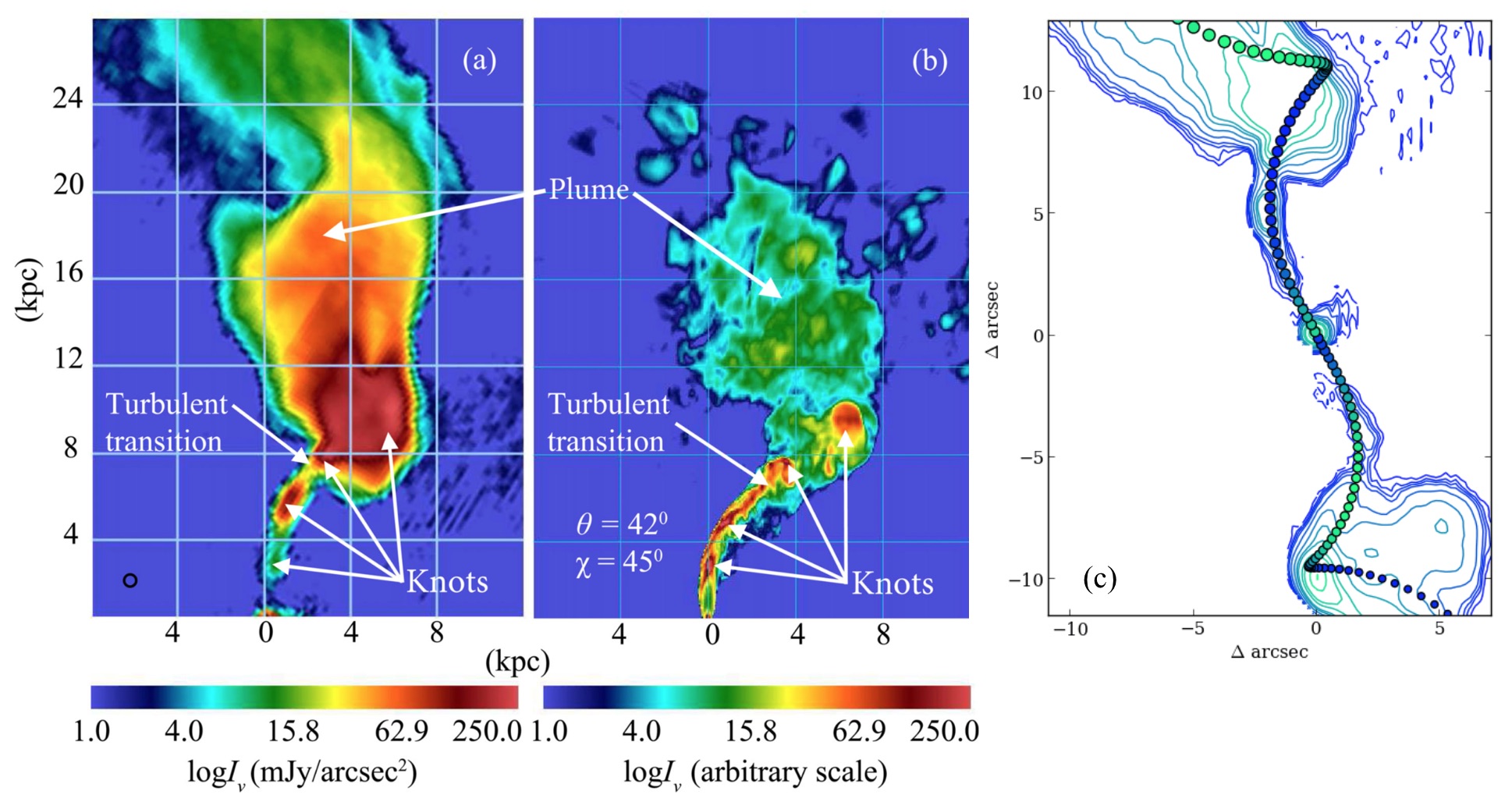}
	\caption{ Different methods of modelling the precessing jet in Hydra~A. 
	(a) VLA image of Hydra A. 
	(b) Hydrodynamic simulation of a jet with parameters empirically matched to Hydra~A.
	(c) Ballistic jet model \citep{Gowea82} for Hydra~A. 
	Even though Hydra~A has a weaker jet than the powerful jet sources discussed
	elsewhere in this article, the jet path is over most of its length well modelled 
	by the hydrodynamic simulation as well as the ballistic model.
	This suggests that precession effects are indeed observable in real 
	radio sources.
	Panels (a) and (b) were adopted from \citet{Nawea16a}.
}
    \label{fig:Hydracomp}
\end{figure*}

\appendix
{
\section{Comparison between ballistic jet models and hydrodynamic simulations}\label{ap:hydra}
Hydrodynamic effects in radio lobes have been shown to be able to bend also
non-precessing jets \citep{HN90}. This is especially important close to the terminal hotspots,
where lobe pressures are high. To investigate the role of the hydrodynamic effects,
we compare a 3D hydrodynamic simulation, published in \citet{Nawea16a}, to a ballistic
jet model \citep{Gowea82} for the precessing jet source Hydra~A in Fig.~\ref{fig:Hydracomp}.
Model parameters are given in Table~\ref{t:hyapars}.
It is clear that even for this comparatively low-power jet, the curvature induced by precession dominates
the observational appearance for most of the region where the jet is observed.
It is therefore reasonable to take curvature as signature for precessing jets.

}
\begin{table}
	\centering
	\caption{Parameters for the jet models compared in Fig.~\ref{fig:Hydracomp}\label{t:hyapars}}
	\begin{tabular}{lll} 
		\hline
		Parameter & Simulation & Ballistic model\\
		\hline 
		Jet velocity & $0.8c$ & $0.1c$\\
		Precession period & 1~Myr & 0.85 Myr \\
		Inclination from line of sight  & $42\deg$  & $42\deg$ \\  
		Starting angle of precession\footnote{$\Omega t_\mathrm{ref}$ in \citet{Gowea82}} &$-45\deg $
			     & $-45\deg $\\
		Prec. cone half opening angle &   $20\deg $  & $10\deg $\\  
		\hline
	\end{tabular}
\end{table}

\newlength{\swidth} 
\setlength{\swidth}{0.000784\columnwidth} 

\section{Radio maps for the complete sample}\label{ap:data}
We provide here all the radio maps of our complete sample. Shown is Stokes I in Jy/beam. 
\rvb{The mapping of the flux levels onto the colour map is non-linear
and has been adjusted carefully
for each map to ensure lobe and jet structures are well visible.}
\rva{Equatorial coordinates are given on the maps.} Observing frequency, resolution,
redshift and precession indicators are given in the individual figure captions. The features 
of precessing jets are indicated on each map if applicable.
\begin{figure}\centering
	\includegraphics[width=1160\swidth]{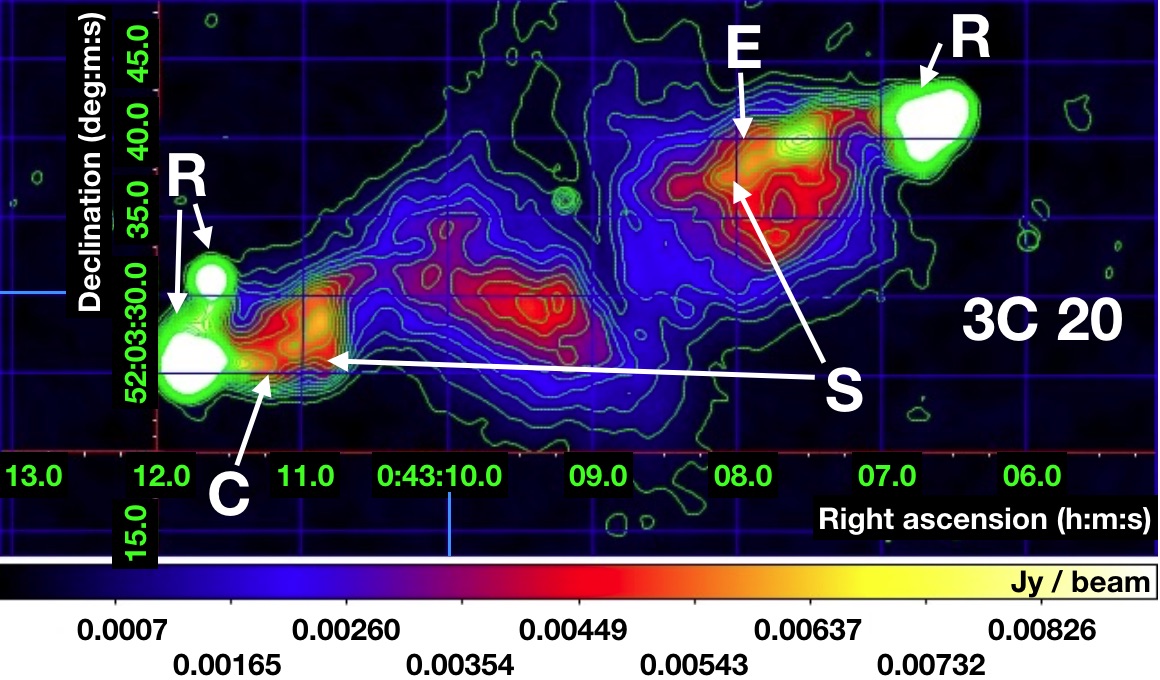}
	\caption{3C~20, 8.41~GHz, resolution: $1.1\arcsec$, redshift: 0.174.
	C: Curved jet structure on the eastern side. E: Lobe brightness asymmetry / jet at edge of lobe
	on the western side. R: Multiple terminal hotspots on the eastern side; 
	wide terminal hotspot on the western side. S: S-symmetry. Additionally, interaction with ambient medium.
}
    \label{fig:3c20}
\end{figure}
\begin{figure}\centering
	\includegraphics[width=1268\swidth]{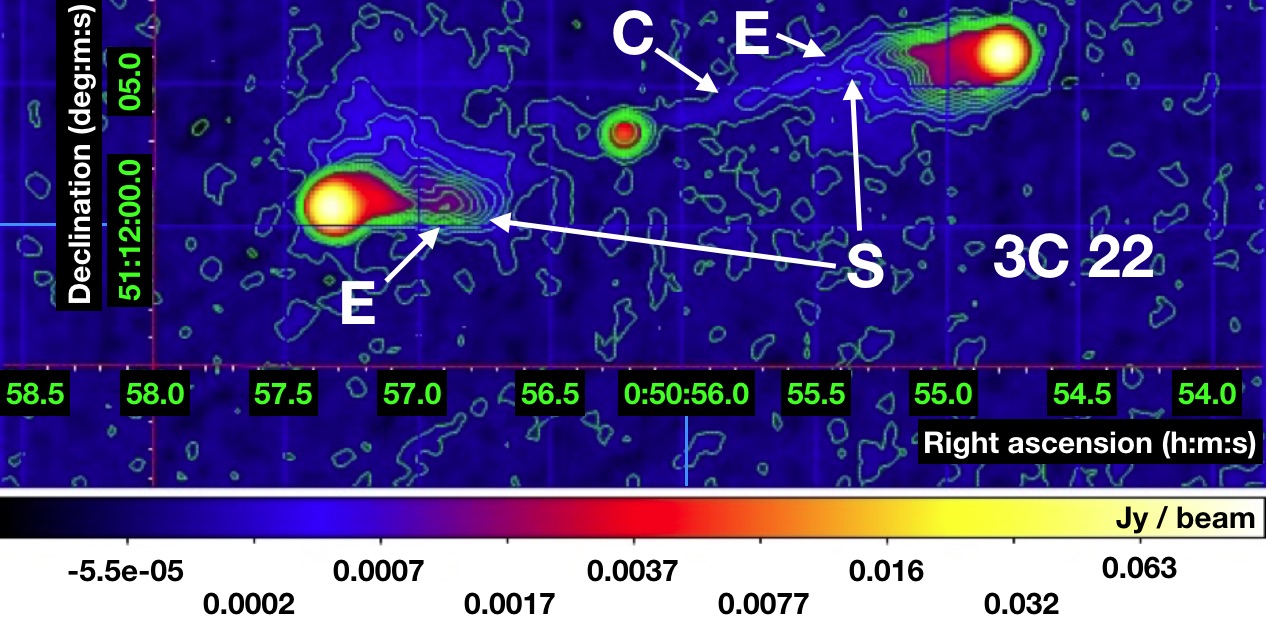}
	\caption{3C~22, 8.56 GHz, resolution: $0.75\arcsec$, redshift: 0.938. 
	C: Curved jet. E: Misaligned lobe axis on the eastern side. 
	Lobe brightness asymmetry on the western side.   S: S symmetric jets.}
    \label{fig:3c22}
\end{figure}
\begin{figure}\centering
	\includegraphics[width=865\swidth]{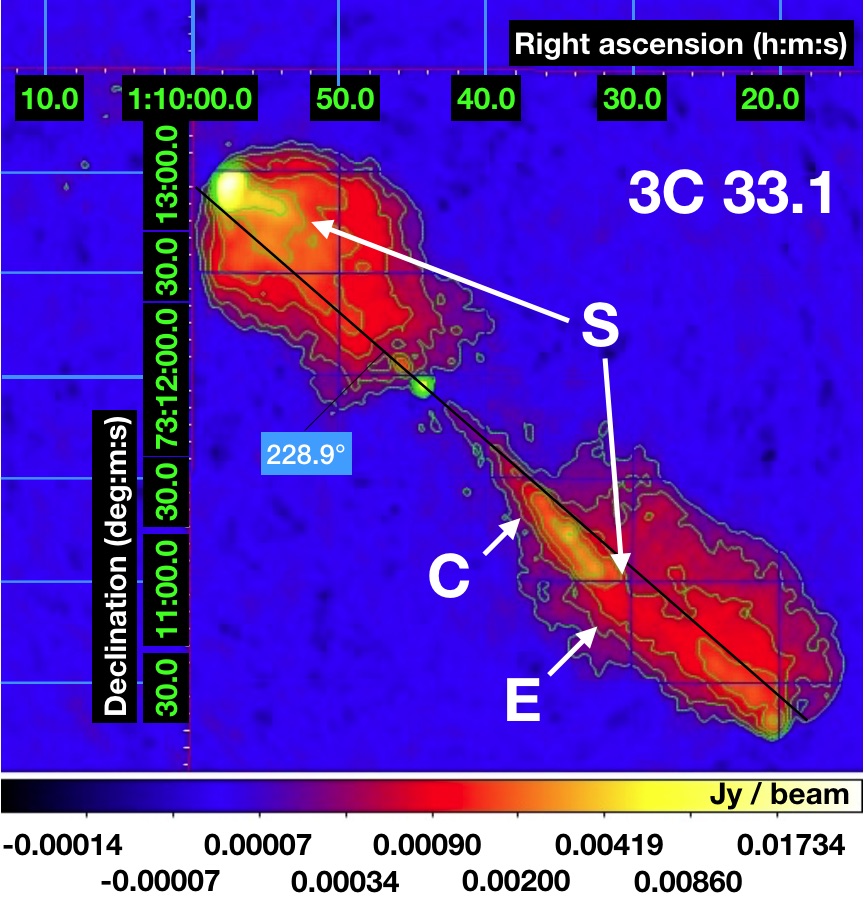}
	\caption{3C~33.1, 4.85 GHz, resolution: $2.948\arcsec\times2.472\arcsec$, redshift: 0.18.
	C: Curved eastern jet.
	E: Misaligned lobe axis on the eastern side.  
	S: S-symmetric jets.
	Jet enters lobe on southern side. Jet bend with brightening, possibly related to 
	jet-cloud interaction. 
	Southern tip-of-lobe hotspot lies along the continuation of inner southern jet. 
	\rva{The innermost part of the jets is aligned with the lobe axis (solid black line, position angle (jet side) indicated).}}
    \label{fig:3c33P1}
\end{figure}
\begin{figure}\centering
	\includegraphics[width=665\swidth]{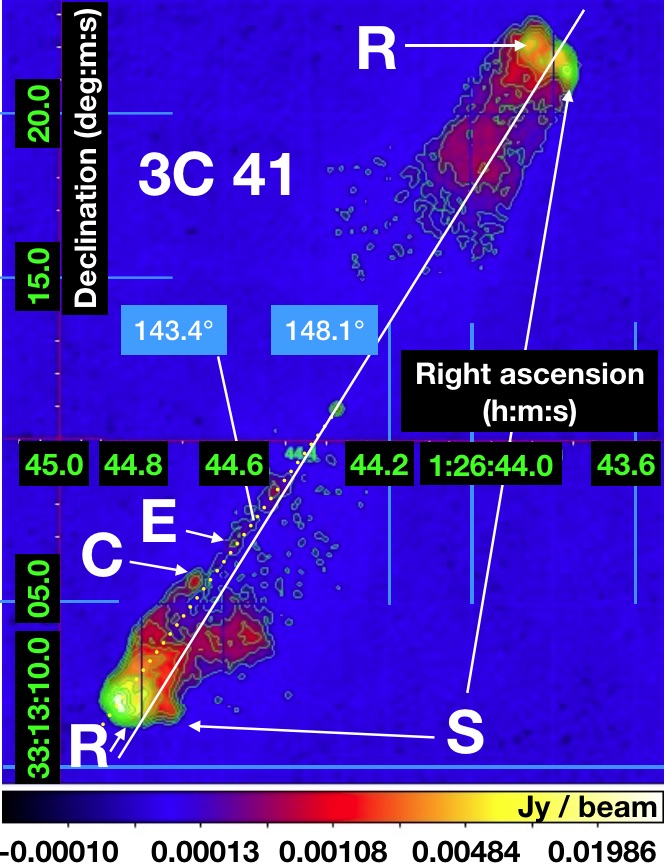}
	\caption{3C~41, 8.46 GHz, resolution: $0.2\arcsec$, redshift: 0.795.
	C: Curved southern jet. 
	E: Misaligned lobe.
	R: Wide terminal hotspots on both sides.  
	S: S-symmetric jet-hotspots.
	Southern jet \rva{(yellow dotted line) misaligned} with \rva{lobe axis (solid white line) by $5\deg$}.}
    	\label{fig:3c41}
\end{figure}
\begin{figure}\centering
	\includegraphics[width=656\swidth]{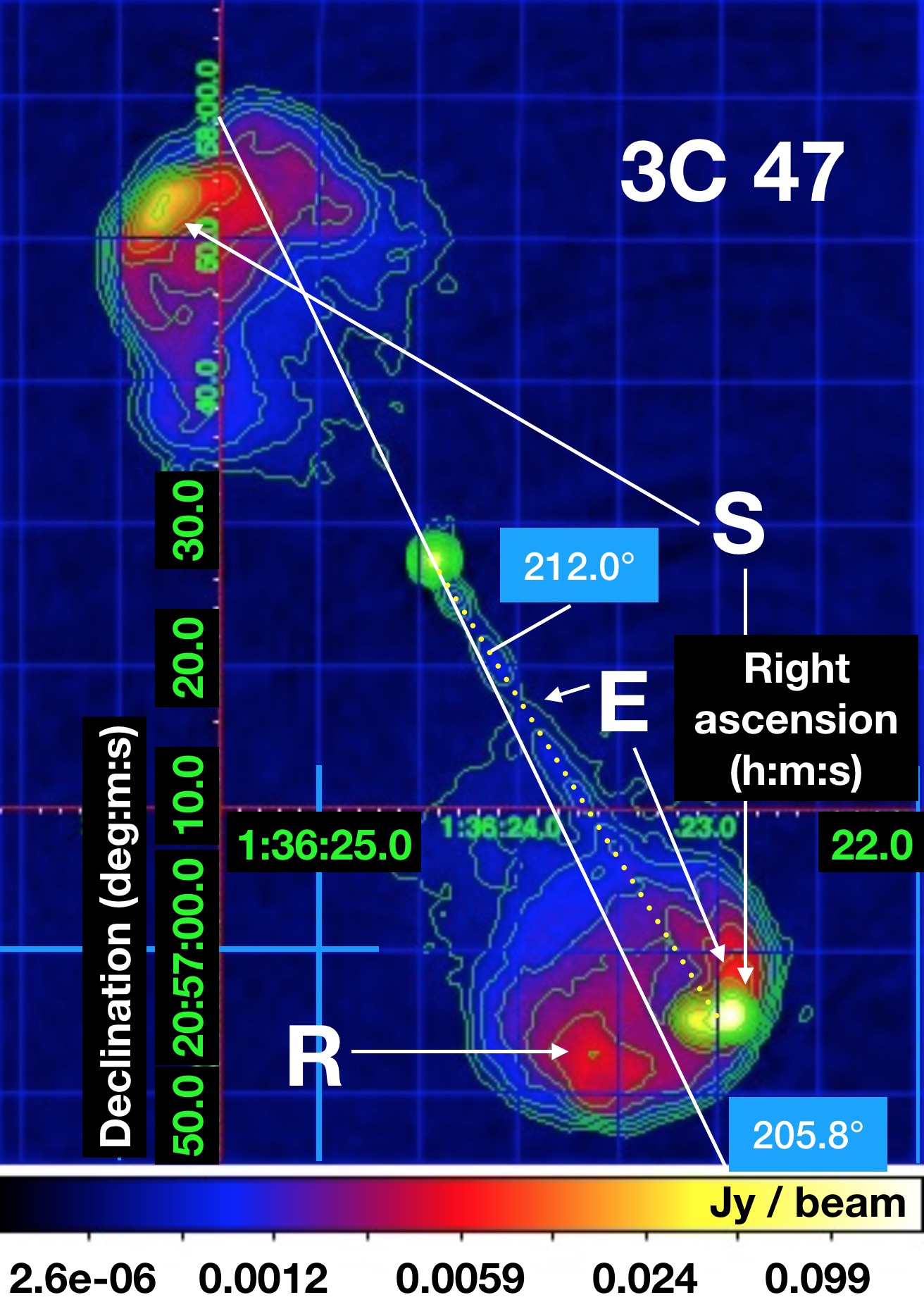}
	\caption{3C~47, 4.89 GHz, resolution: $1.3\arcsec$, redshift: 0.43. 
	E: Jet and hotspots at edge of lobe.
	R: Ring structure near hot spot.
	S: S-symmetry. \rva{Jet (yellow dotted line) misaligned with lobe axis by $6\deg$.}}
    	\label{fig:3c47}
\end{figure}
\begin{figure}\centering
	\includegraphics[width=540\swidth]{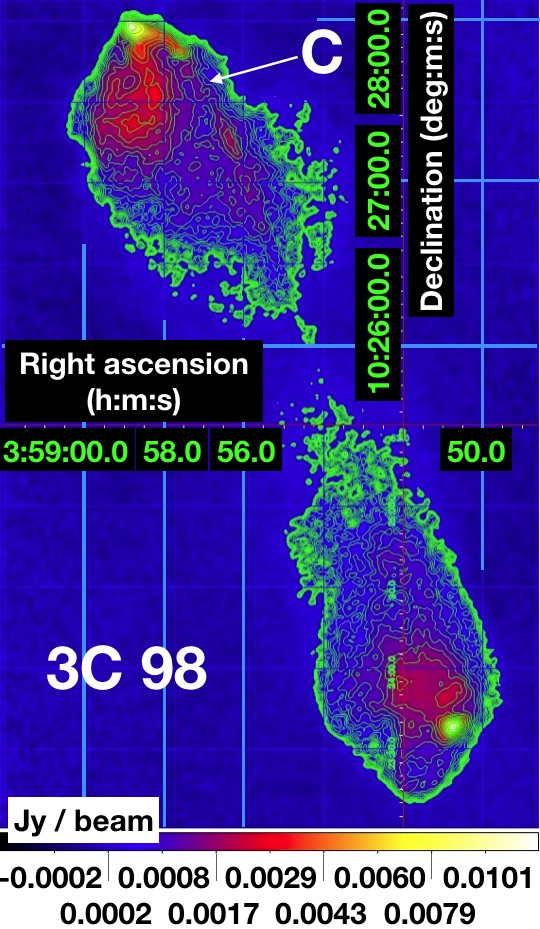}
	\caption{3C~98, 8.35 GHz, resolution: $2~\arcsec$, redshift: 0.03.
	C: The outer part of the northern jet is curved, possibly related to precession.
	Jet and hotspot at one side of lobe, but possibly due to interaction with ambient medium.}
    	\label{fig:3c98}
\end{figure}
\begin{figure}\centering
	\includegraphics[width=1150\swidth]{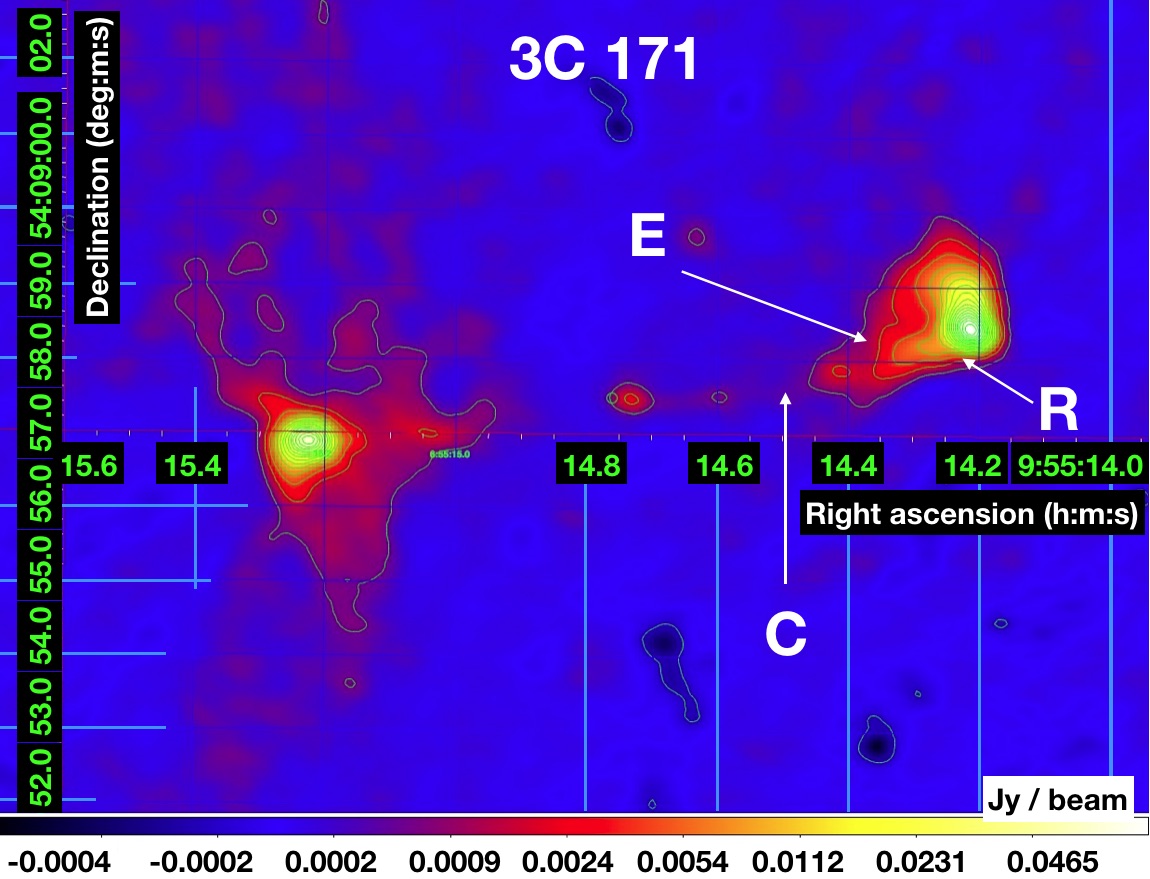}
	\caption{3C~171, 8.06 GHz, resolution: $0.35 \times 0.25\arcsec$, redshift: 0.24.
	C: Curved eastern jet.
	E: Misaligned lobe axis on the eastern side. 
	R: Wide terminal hotspot on the eastern side.
	Image quality is not very good and the source has complex structure.}
    	\label{fig:3c171}
\end{figure}
\begin{figure}\centering
	\includegraphics[width=492\swidth]{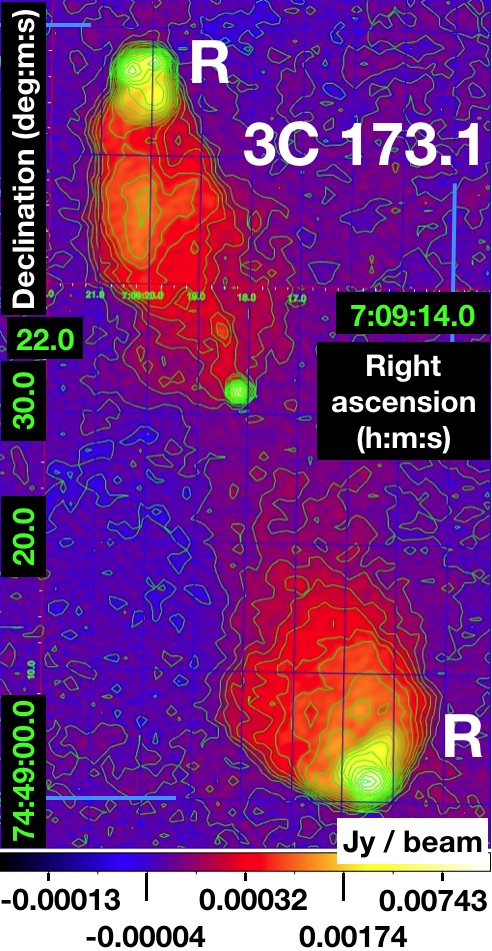}
	\caption{3C~173.1, 8.44 GHz, resolution: $0.75\arcsec$, redshift: 0.292.
	R: Double hotspot in the north, extended hotspot in the south.
	Moderate bend in northern lobe structure. Otherwise the source appears simple and symmetric.}
    	\label{fig:3c173P1}
\end{figure}
\begin{figure}\centering
	\includegraphics[width=1219\swidth]{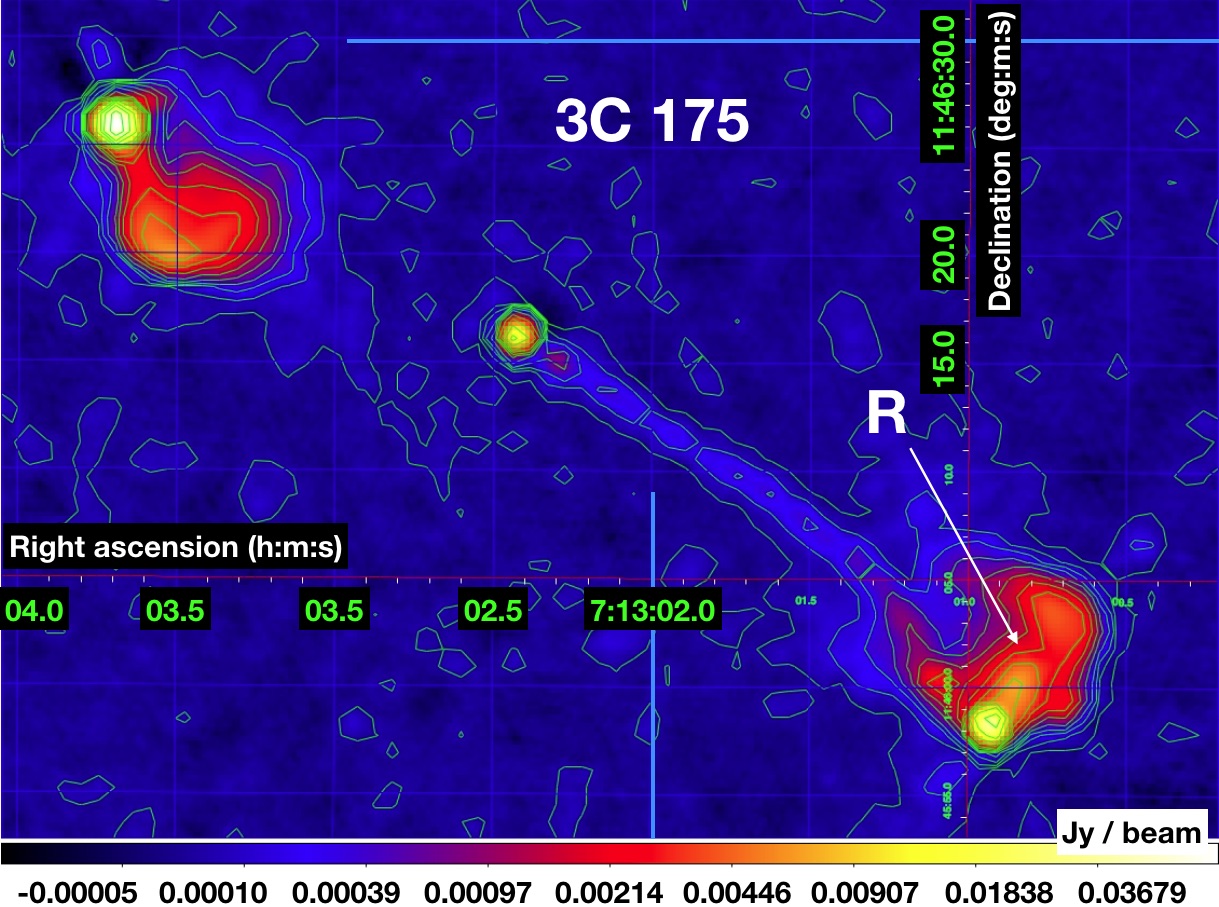}
	\caption{3C~175, 8.45 GHz, resolution: $1.0\arcsec$, redshift: 0.768.
	R: Some evidence of ring structure, clearer in southern hotspot.
	Jet mainly in the middle of the lobe, bends south towards hotspot at edge of lobe. 
	Since the bend is at the tip of the lobe, this could be due to hydrodynamic effects.}
    	\label{fig:3c175}
\end{figure}
\begin{figure}\centering
	\includegraphics[width=579\swidth]{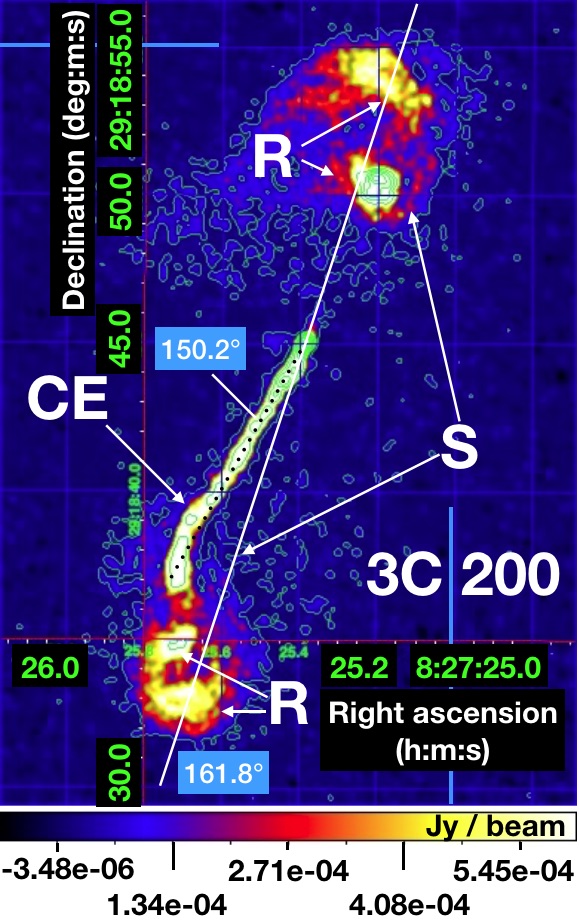}
	\caption{3C~200, 8.46 GHz, resolution: $0.25\arcsec$, redshift: 0.46. 
	CE: Curved jet at the edge of southern lobe.
	R: Ring-like structure for both hotspots.
	S: Clear S-symmetry. \rva{Jet misaligned with lobe axis by $12\deg$.}}
    	\label{fig:3c200}
\end{figure}
\begin{figure}\centering
	\includegraphics[width=1241\swidth]{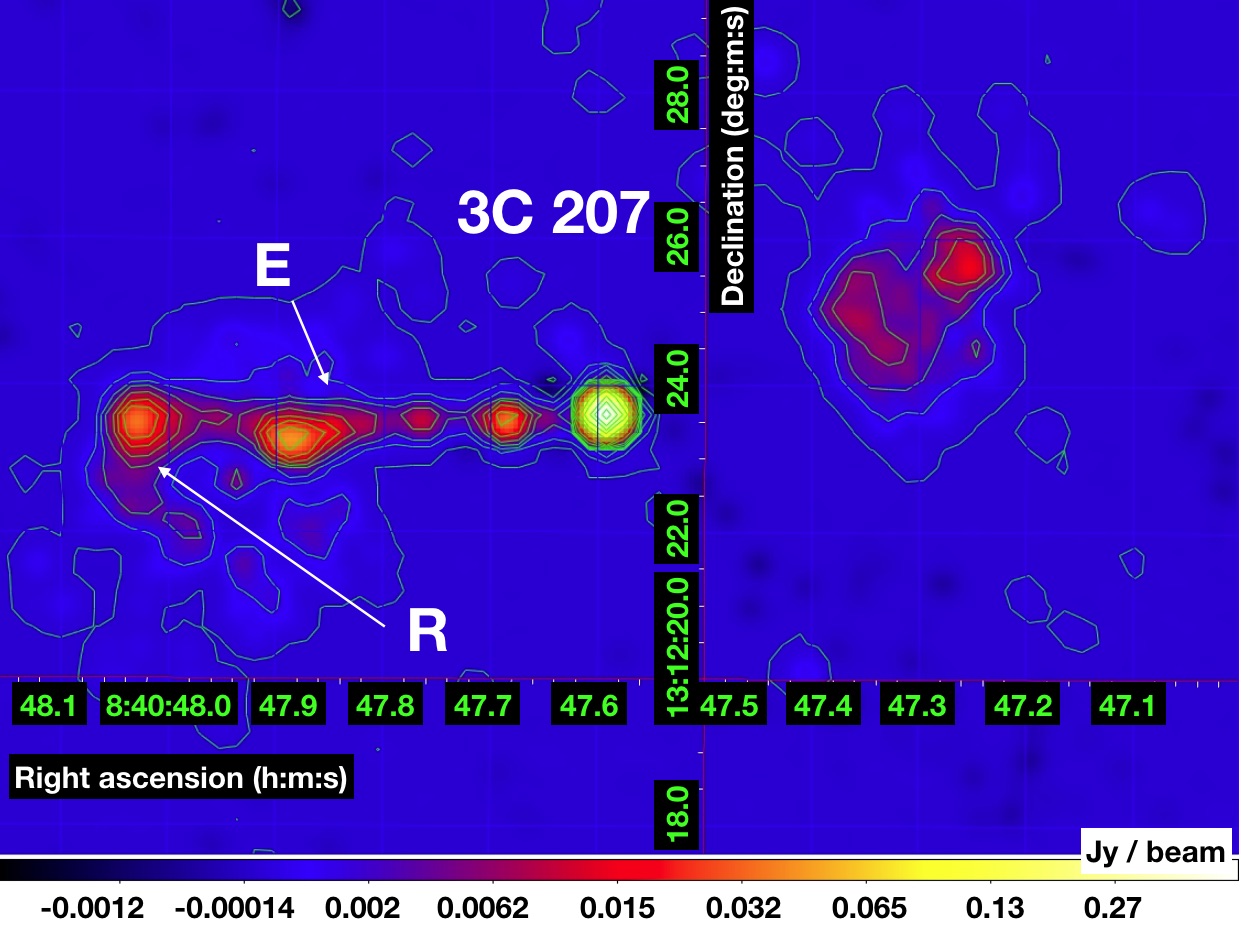}
	\caption{3C~207, 4.86 GHz, resolution: $0.35\arcsec$, redshift: 0.90. 
	E: Jet/lobe misalignment on eastern side.
	R: Wide hotspot on eastern side.
	Faint lobe.}
    	\label{fig:3c207}
\end{figure}
\begin{figure}\centering
	\includegraphics[width=1233\swidth]{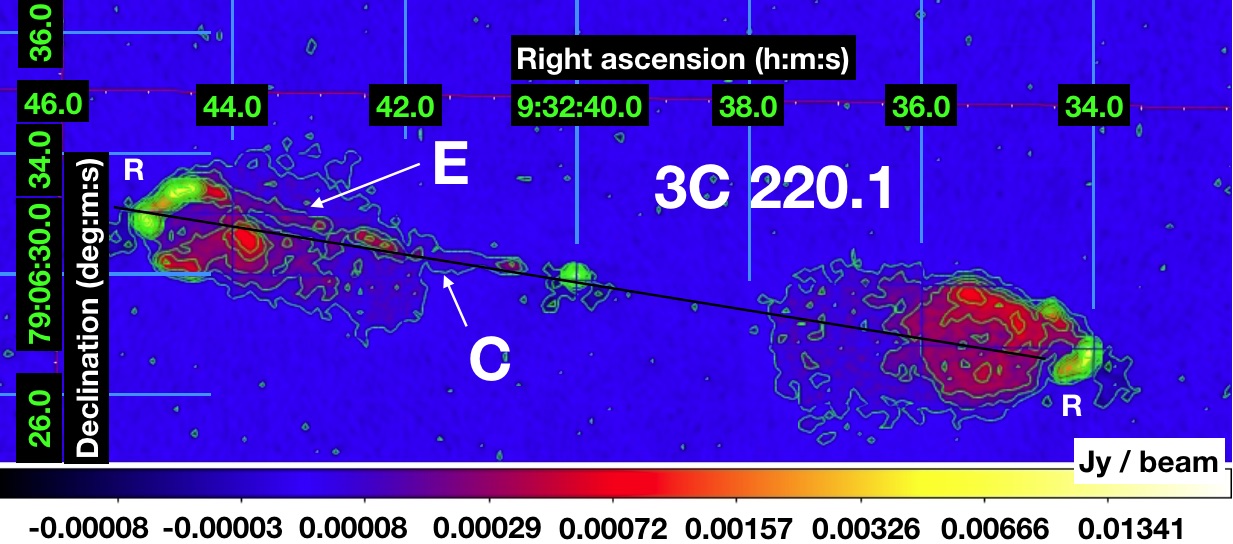}
	\caption{3C~220.1, 8.44 GHz, resolution: $0.25\arcsec$, redshift: 0.61.
	C: Curvature of jet on the western side,
	E: Jets at edge of lobes,
	R: Wide terminal hotspots on both sides; ring-like feature at eastern hotspot.}
    	\label{fig:3c220P1}
\end{figure}
\begin{figure}\centering
	\includegraphics[width=507\swidth]{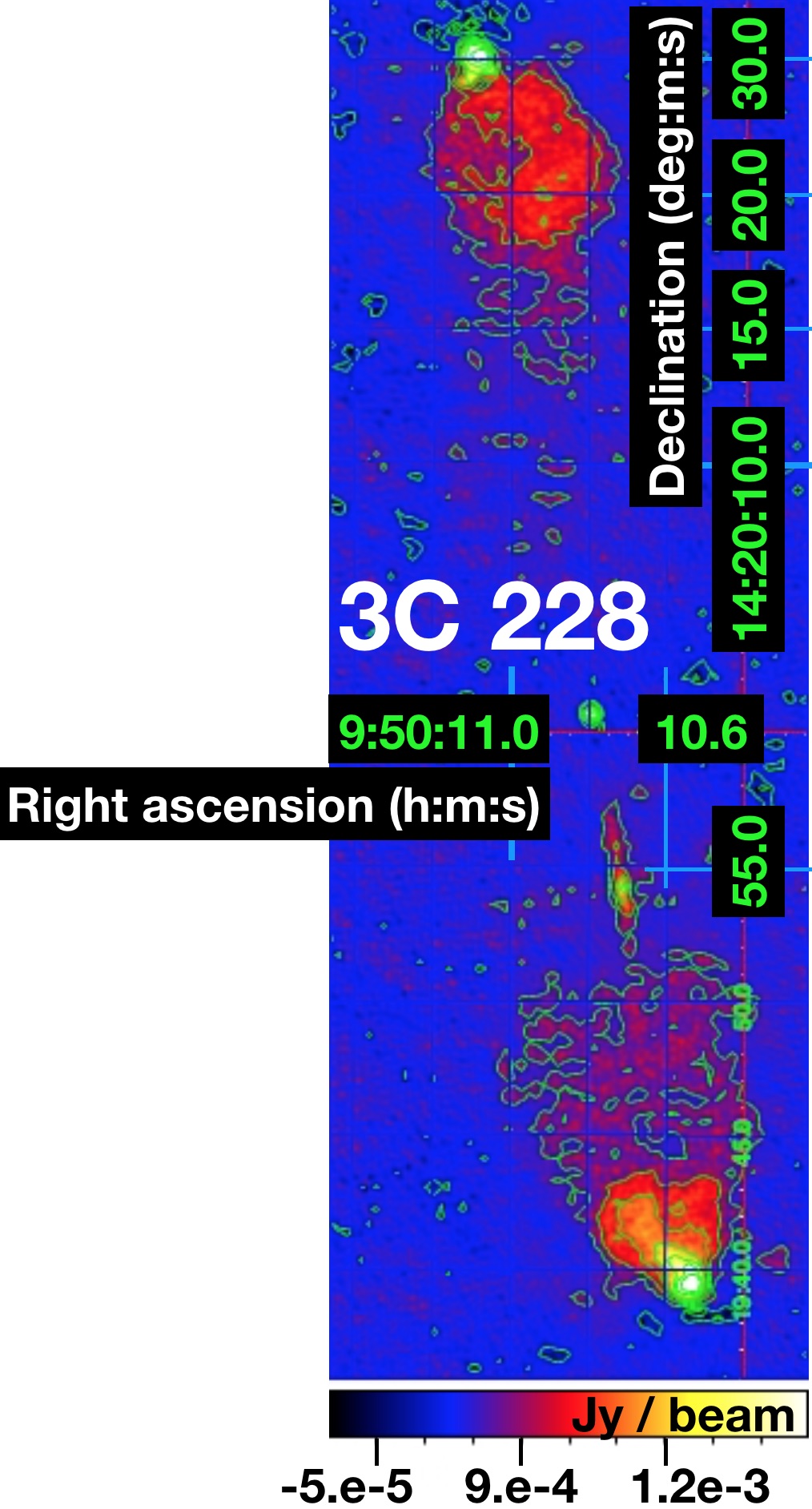}
	\caption{3C~228, 8.47 GHz, resolution: $0.225\arcsec$, redshift: 0.55. 
	Hotspots, jet and core perfectly aligned. }
    	\label{fig:3c228}
\end{figure}
\begin{figure}\centering
	\includegraphics[width=1258\swidth]{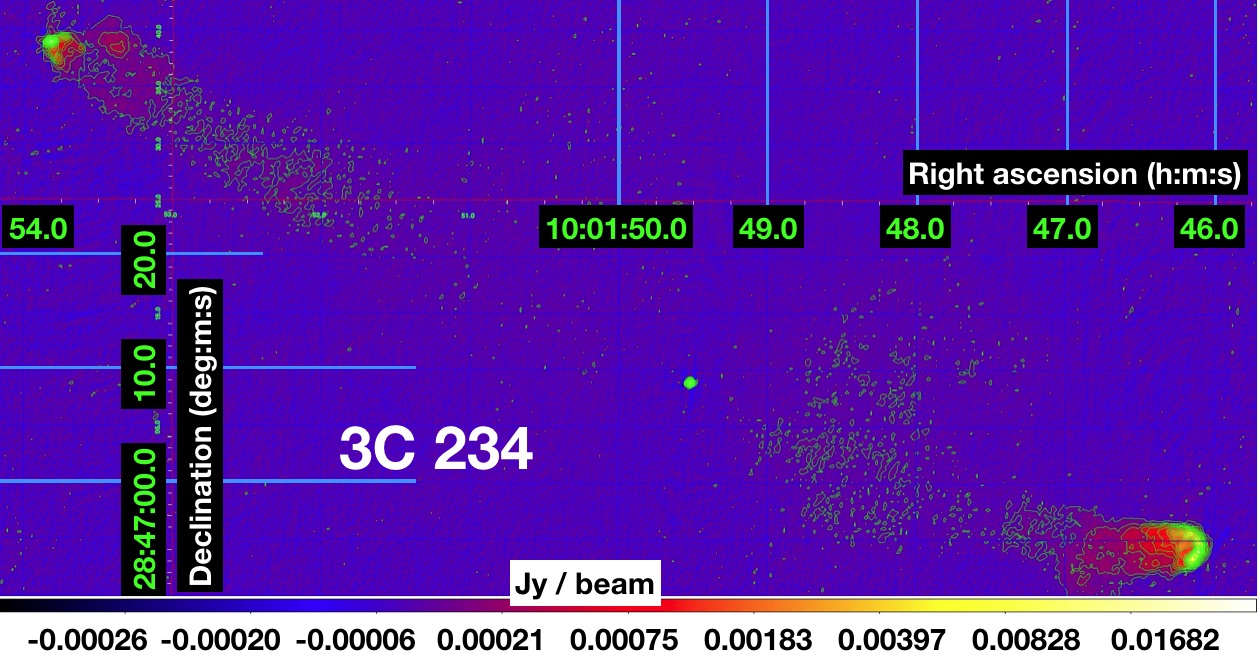}
	\caption{3C~234, 8.44 GHz, resolution: $0.3\arcsec$, redshift: 0.18.
	Wide central, low surface brightness, structure could indicate interaction 
	with dense host ISM. Deviation in lobe direction could point to interaction 
	with neighbouring galaxy. Complex structure, no evidence for precession.}
    	\label{fig:3c234}
\end{figure}
\begin{figure}\centering
	\includegraphics[width=465\swidth]{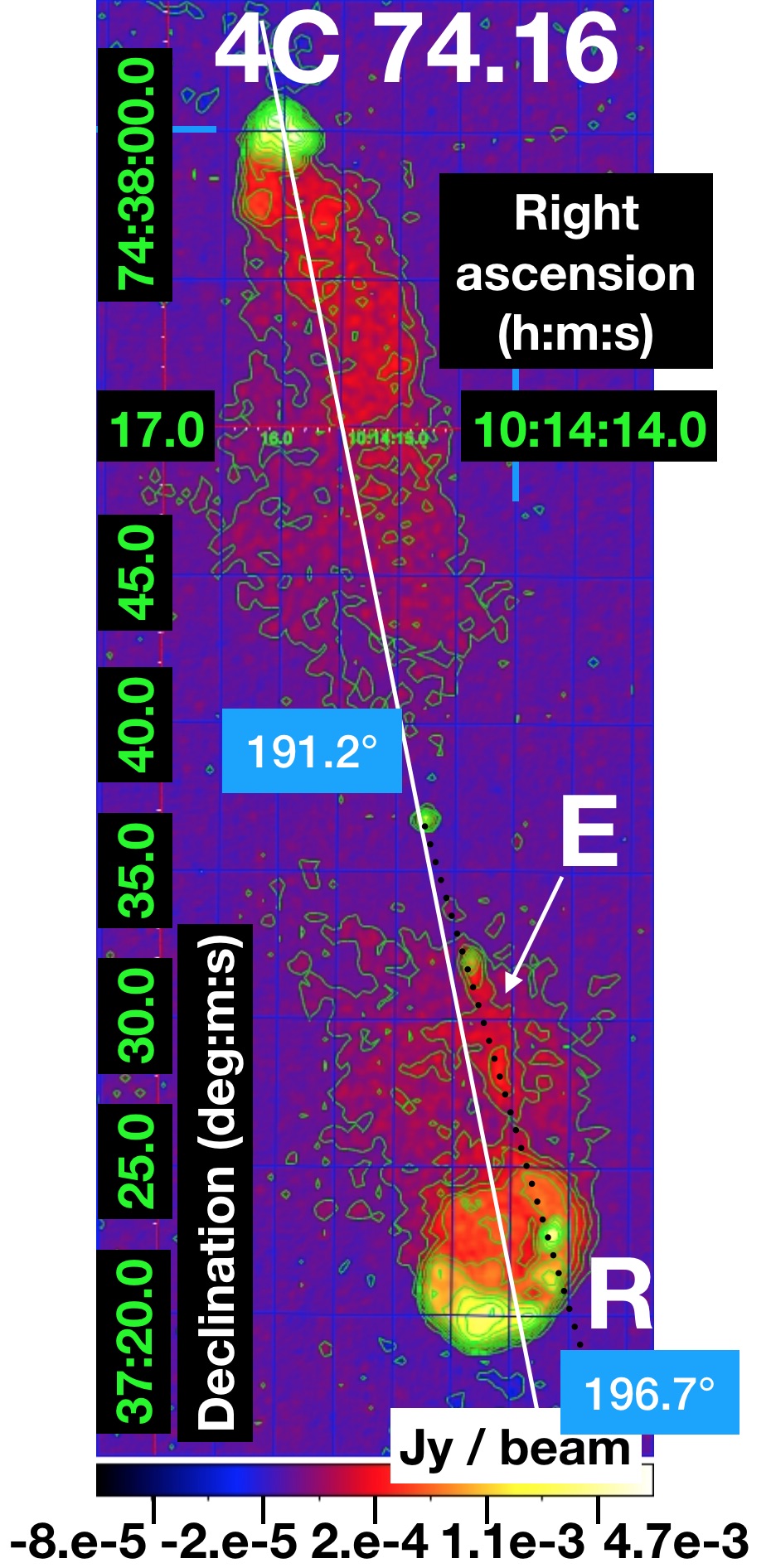}
	\caption{4C~74.16, 8.47 GHz, resolution: $0.3\arcsec$, redshift: 0.81.
	E: Jet runs into lobe along its edge,
	R: Partial ring structure in the southern lobe,
	Wiggles in the northern lobes which is not part of our classification scheme.
	\rva{The angle between jet (black dotted line) and lobe axis (white solid line) is $6\deg$.}}
    	\label{fig:4c74P16}
\end{figure}
\begin{figure}\centering
	\includegraphics[width=1266\swidth]{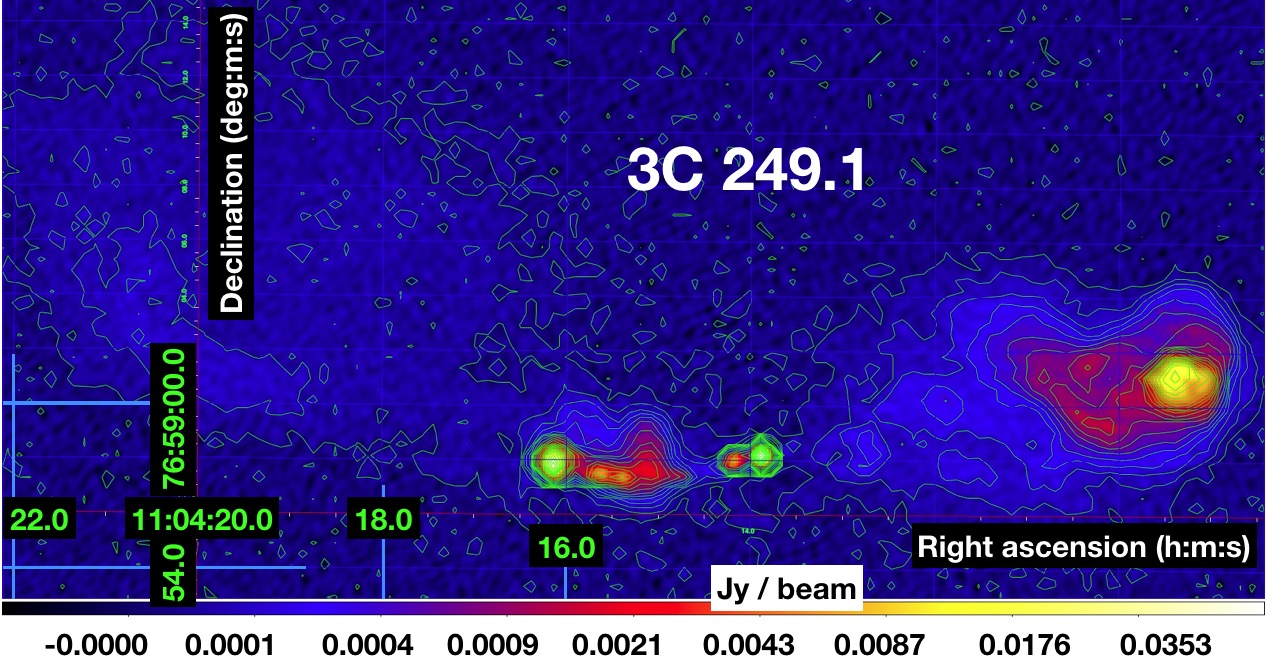}
	\caption{3C~249.1, 4.89 GHz, resolution: $0.35\arcsec$, redshift: 0.31.
	Complex structure. Bent at a bright spot, pointing to persistent jet-cloud 
	(jet-galaxy) interaction. Crosswind from south. The source structure is too 
	complex for firm diagnosis of precession signatures.}
    	\label{fig:3c249P1}
\end{figure}
\begin{figure}\centering
	\includegraphics[width=1270\swidth]{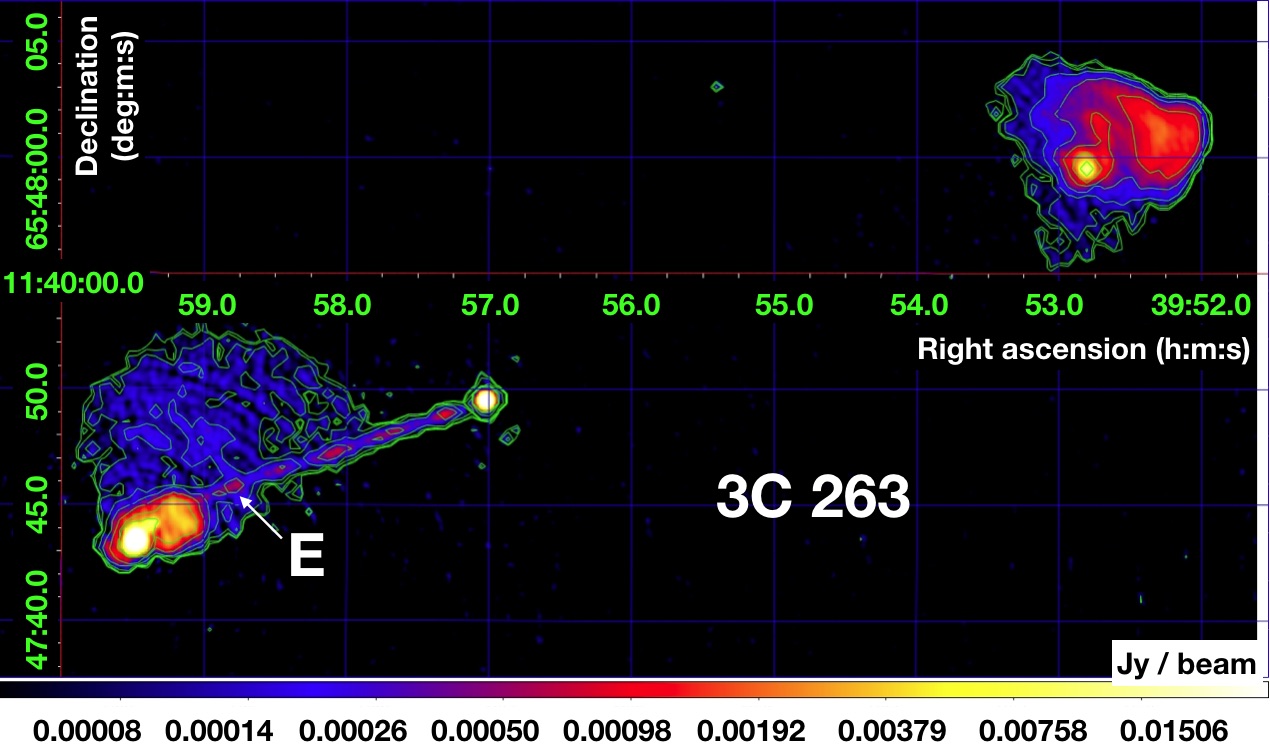}
	\caption{3C~263, 4.86 GHz, resolution: $0.35\arcsec$, redshift: 0.65. 
	E: Misaligned lobe axis on the eastern side.}
    	\label{fig:3c263} 
\end{figure}
\begin{figure}\centering
	\includegraphics[width=614\swidth]{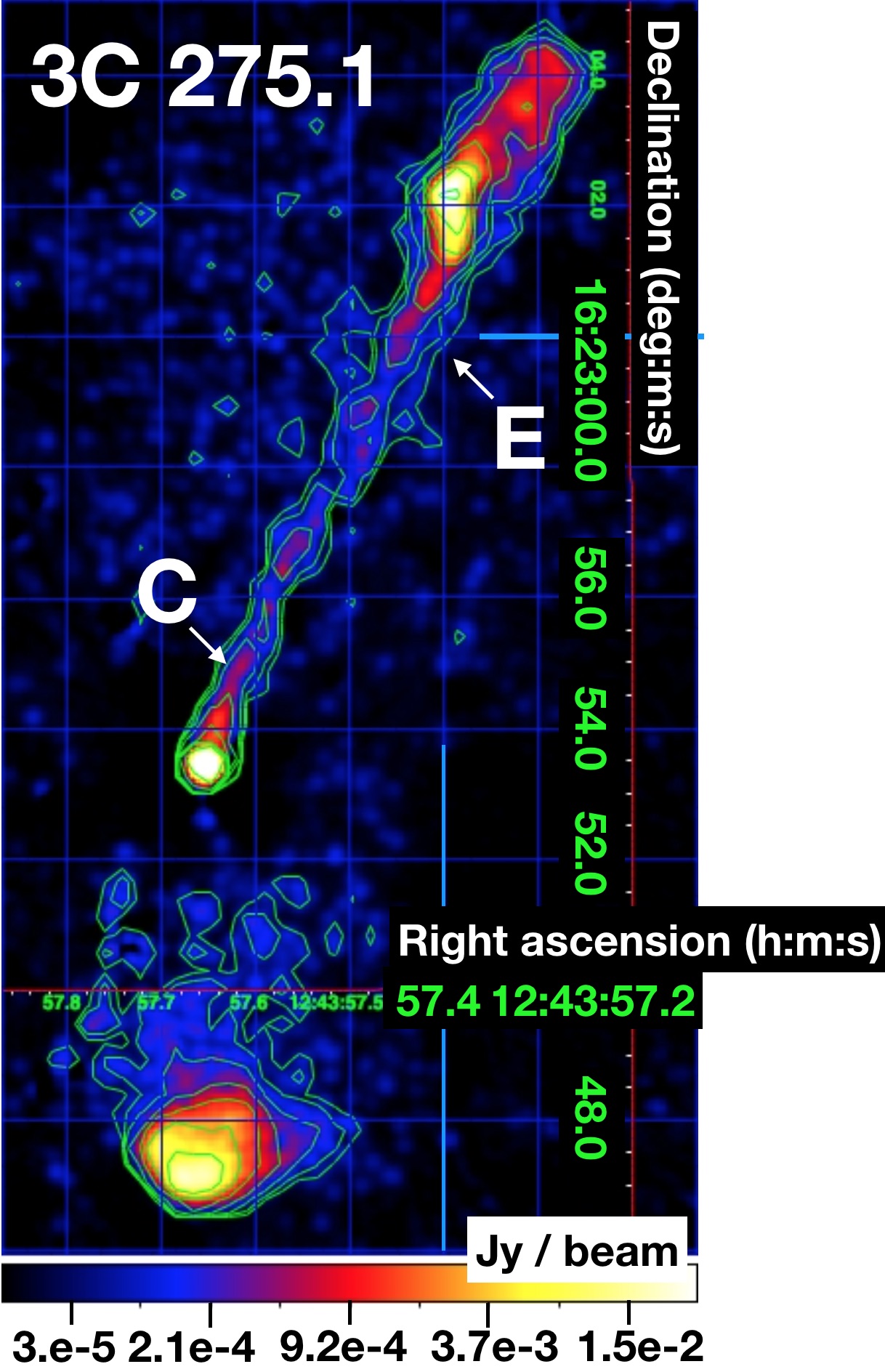}
	\caption{3C~275.1, 8.46 GHz, resolution: $0.225\arcsec$, redshift: 0.42.
	C: Jet is curved,
	E: Jet is not symmetric with respect to lobe axis 
	\rva{(either northern or southern jet as implied by the southern hotspot position).}
	Northern lobe almost lost in noise. 
	\rva{We can therfore not assign a misalignment angle with confidence.}}
    	\label{fig:3c275P1}
\end{figure}
\begin{figure}\centering
	\includegraphics[width=1276\swidth]{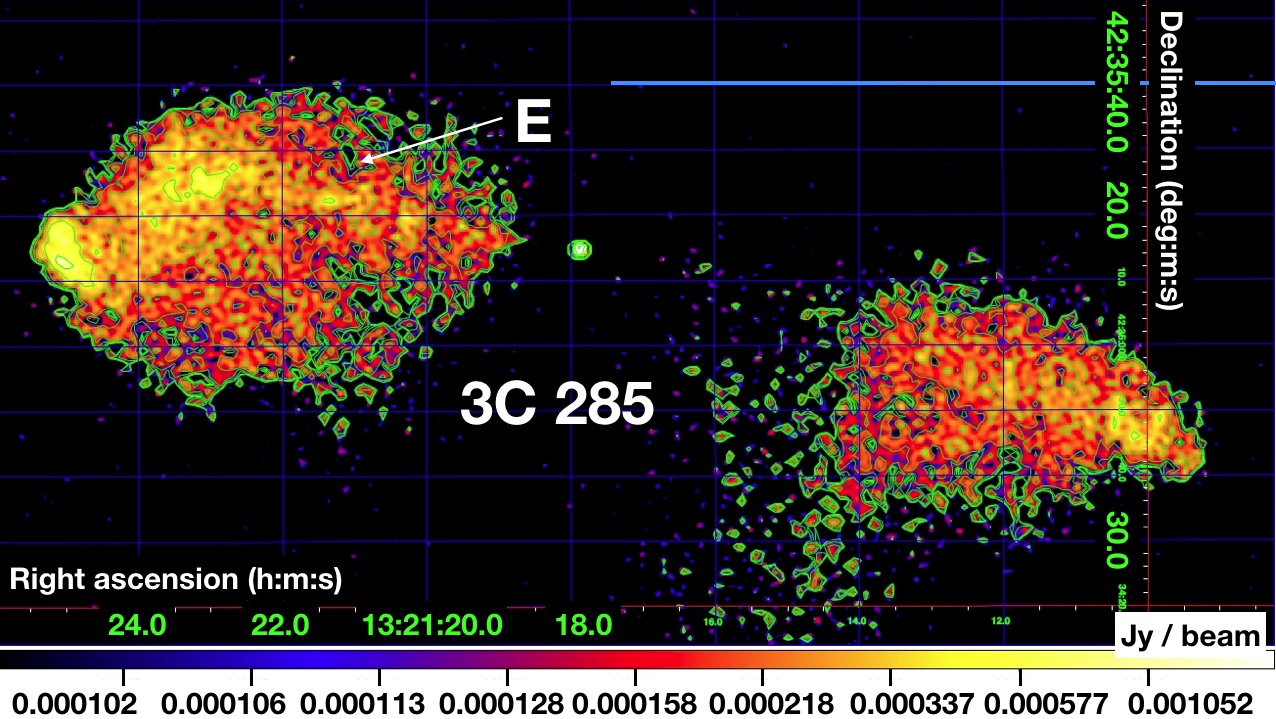}
	\caption{3C~285, 4.87 GHz, resolution: $1\arcsec$, redshift: 0.079.
	E: Jet at edge of lobes.
	Indication for motion through ICM roughly perpendicular to jet axis. 
	Northern jet axis differs from southern jet axis. Possible interaction near tip of north lobe.}
    	\label{fig:3c285}
\end{figure}
\begin{figure}\centering
	\includegraphics[width=844\swidth]{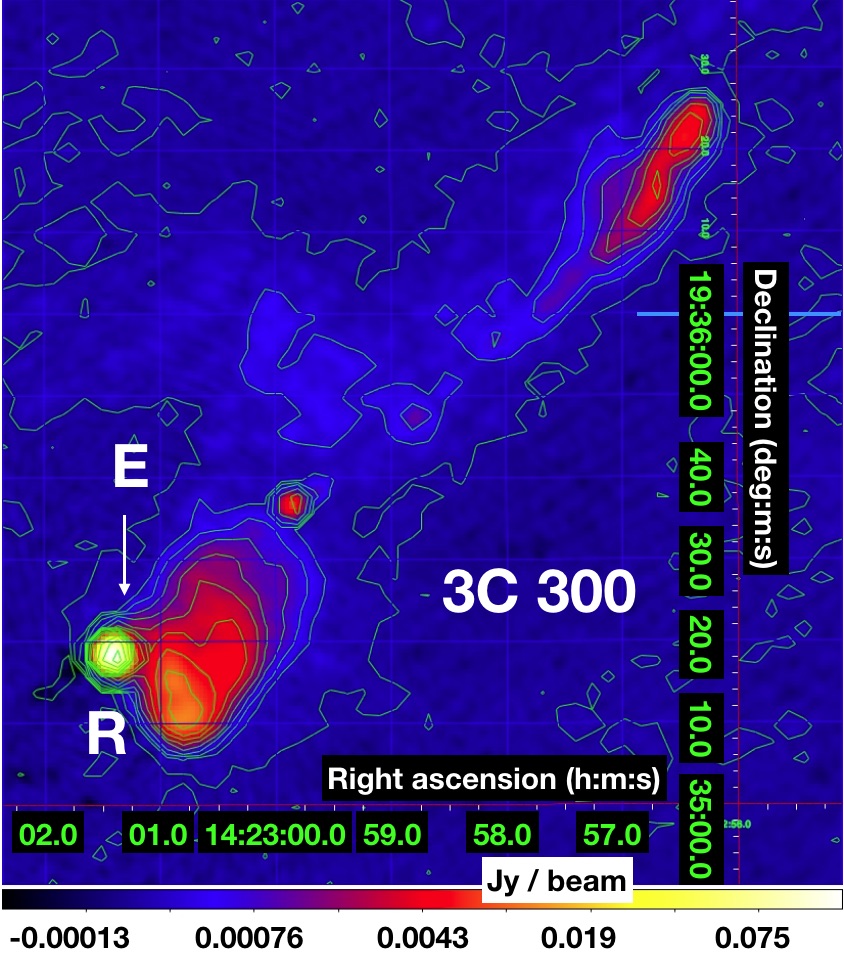}
	\caption{3C~300, 8.06 GHz, resolution: $2.1\arcsec$, redshift: 0.27. 
	E: Hotspot / lobe misalignment southern lobe.
	R: Wide / multiple southern hotspot.}
    	\label{fig:3c300}
\end{figure}
\begin{figure}\centering
	\includegraphics[width=1148\swidth]{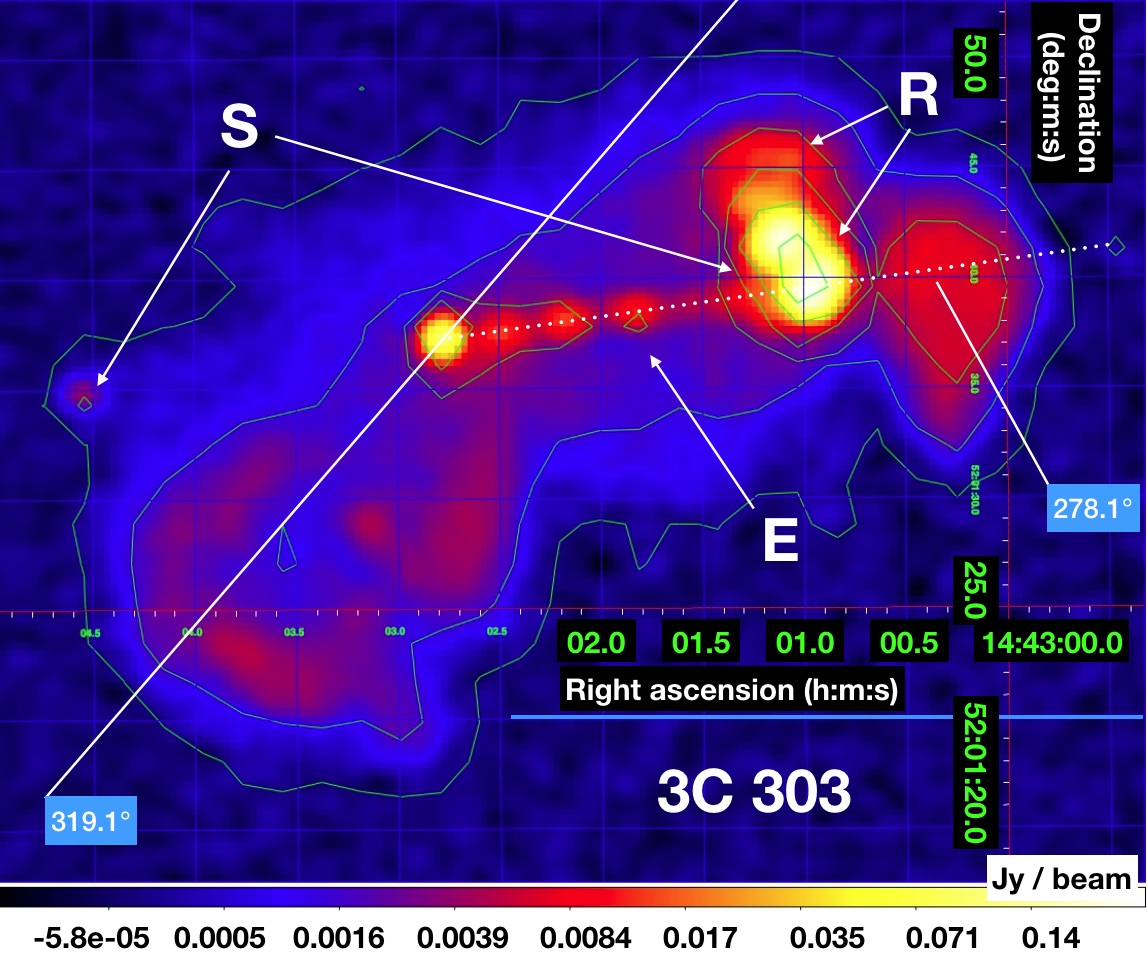}
	\caption{3C~3C 303, 1.45 GHz, resolution: $1.2\arcsec$, redshift: 0.141.
	 E: Misaligned lobe axis on the western side.
	R: Multiple terminal hotspots on the western side.
	S: S-symmetric hotspots.}
    	\label{fig:3c303}
\end{figure}
\begin{figure}\centering
	\includegraphics[width=1253\swidth]{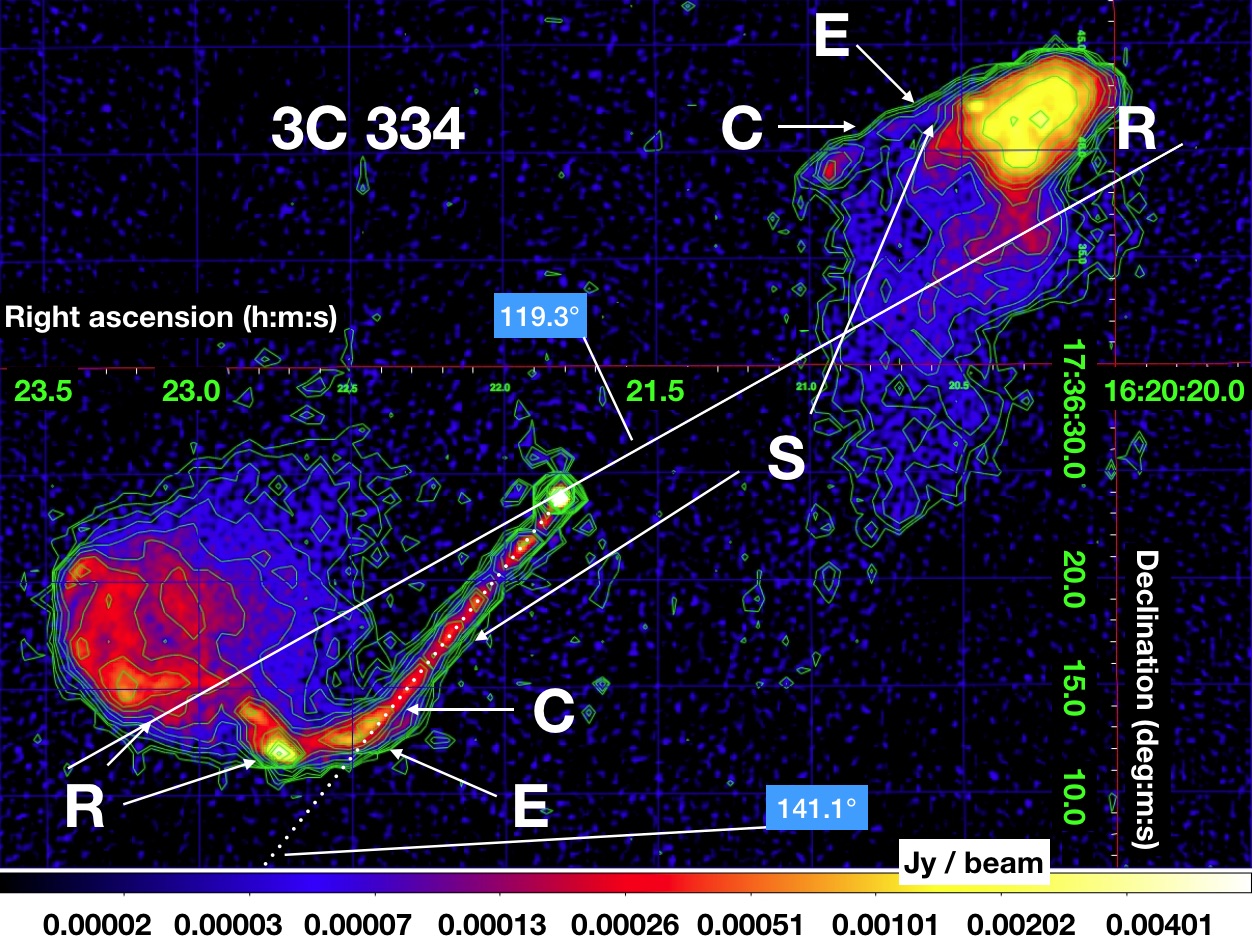}
	\caption{3C~334, 4.89 GHz, resolution: $0.35\arcsec$, redshift: 0.555.
	C: Curved jets. 
	E: Misaligned lobe axis on both sides.  
	R: Wide terminal hotspots on western side; Ring-shaped feature at hotspot on eastern side.   
	S: S-symmetric jet-lobe structure.
}
    	\label{fig:3c334}
\end{figure}
\begin{figure}\centering
	\includegraphics[width=638\swidth]{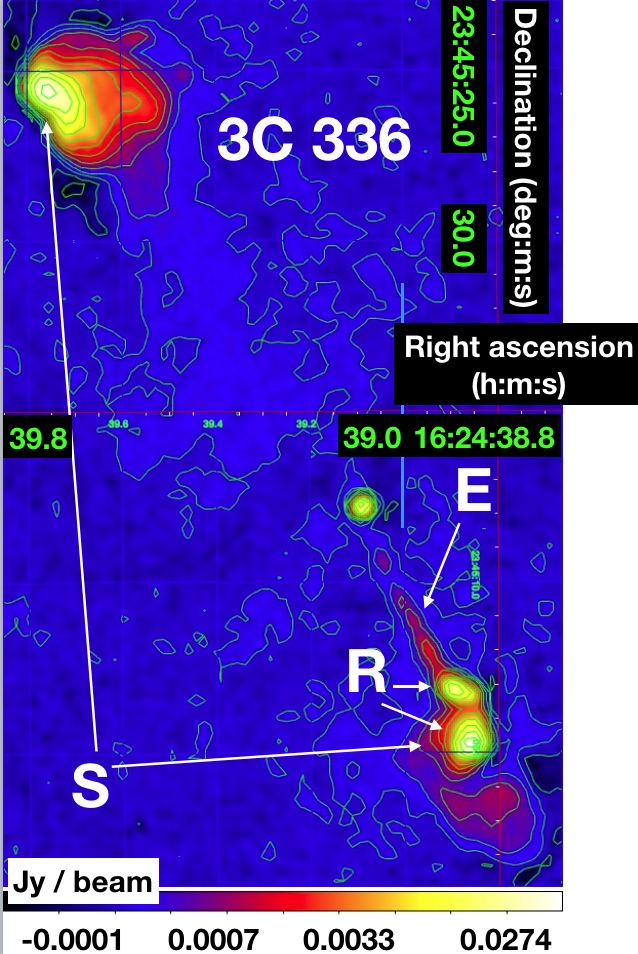}
	\caption{3C~336, 4.86 GHz, resolution: $0.35\arcsec$, redshift: 0.93.
	E: Misaligned lobe axis on both sides,
	R: Multiple terminal hotspots on the southern side,  
	S: S-symmetric jet-hotspots.}
    	\label{fig:3c336}
\end{figure}
\begin{figure}\centering
	\includegraphics[width=455\swidth]{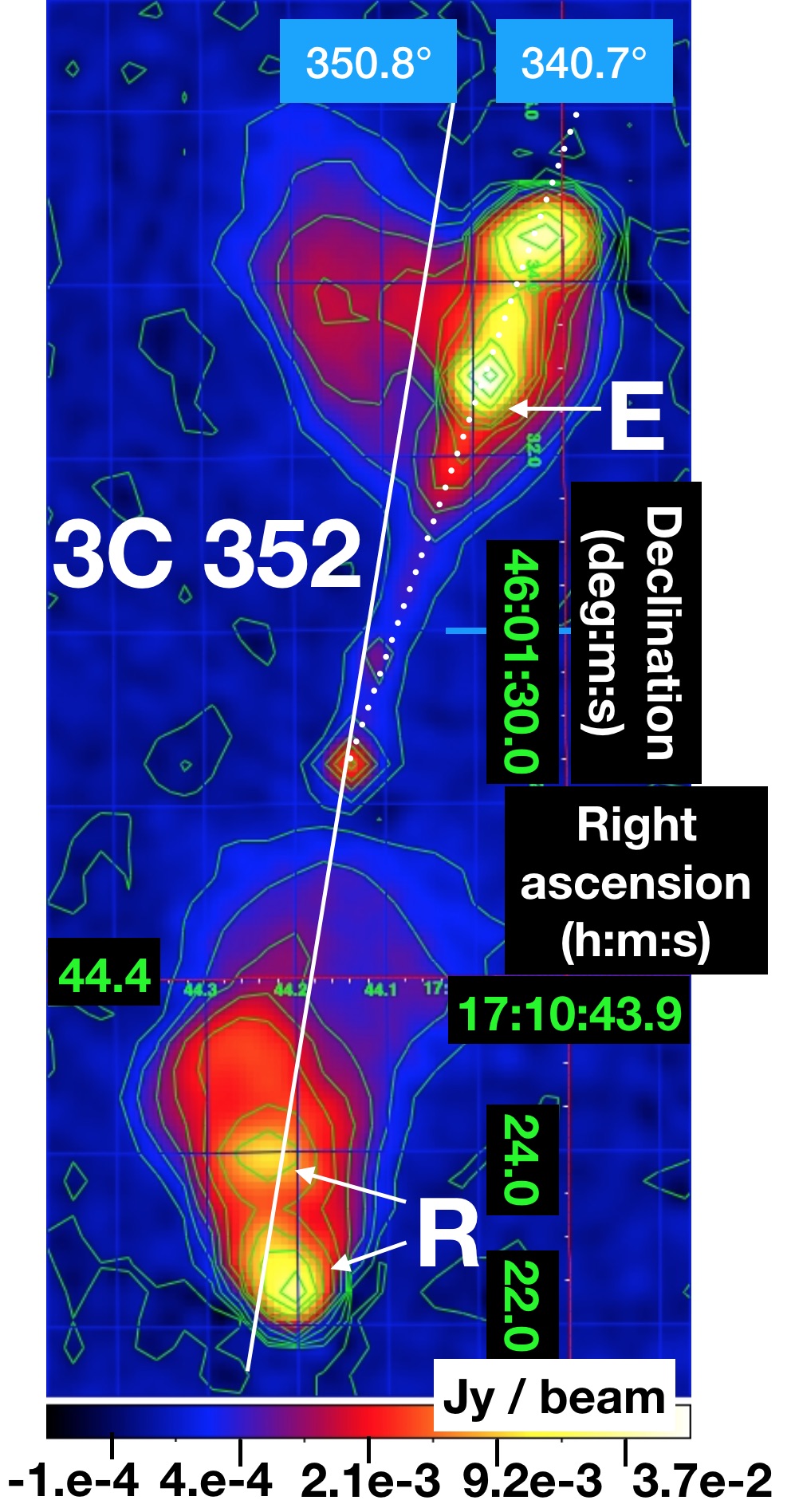}
	\caption{3C~352, 4.71 GHz, resolution: $0.35\arcsec$, redshift: 0.37.
	E: Jets towards edge of lobes; Lobe brightness asymmetry on the southern side,
	R: Multiple hotspots on southern side.
	\rva{Lobe extension on the northern side as expected for precession with long} 
	period and wide precession cone \citep{DonSmi16}.}
    	\label{fig:3c352}
\end{figure}
\begin{figure}\centering
	\includegraphics[width=960\swidth]{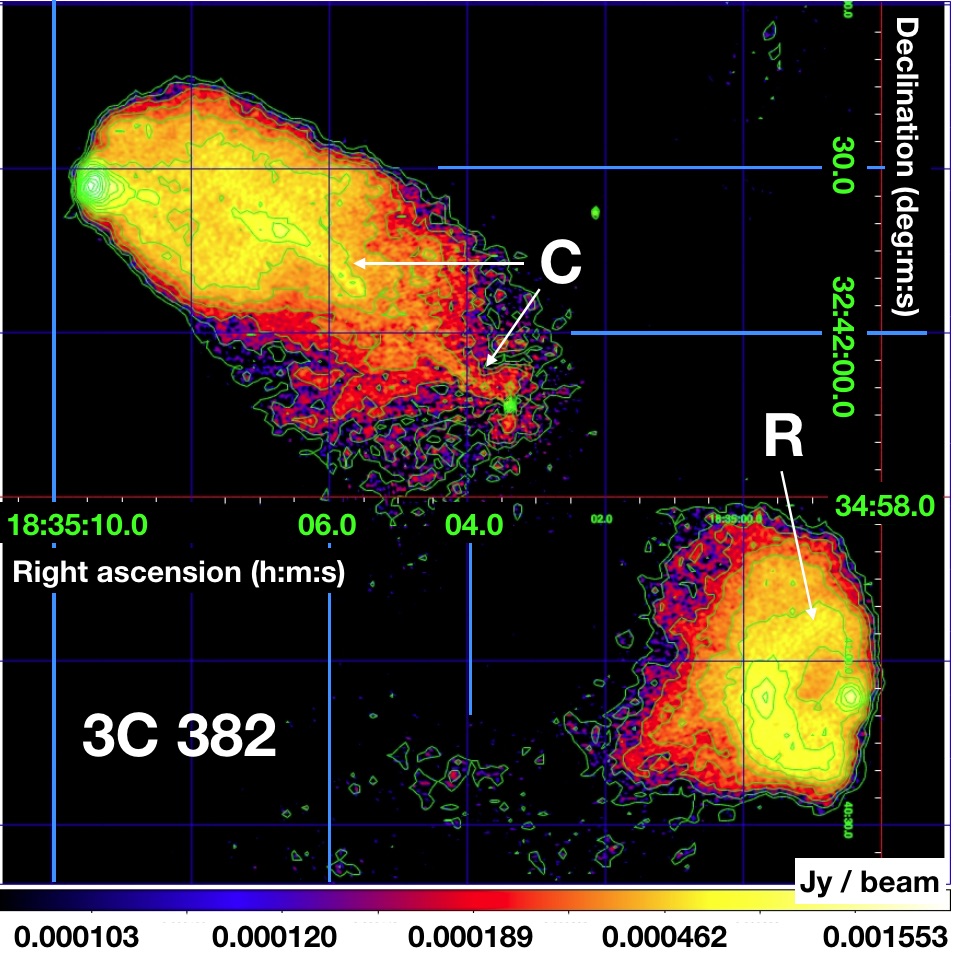}
	\caption{3C~382, 8.47 GHz, resolution: $0.75\arcsec$, redshift: 0.058.
	C: Jet bent in northern lobe over whole length, possibly additionally a jet-cloud interaction,
	R: Ring-shaped southern hotspot.
	Jet slightly off lobe axis. Crosswind.}
    	\label{fig:3c382}
\end{figure}
\begin{figure}\centering
	\includegraphics[width=1081\swidth]{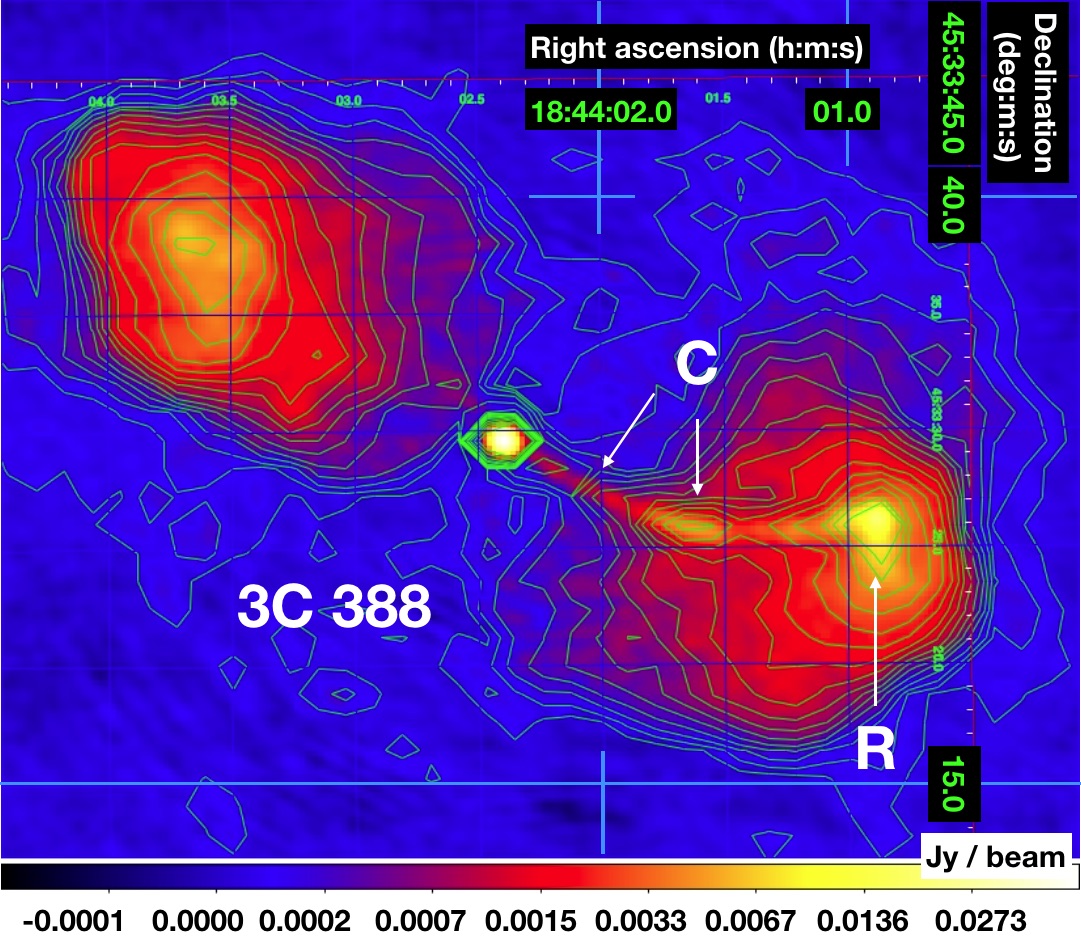}
	\caption{3C~388, 8.46 GHz, resolution: $0.783 \times 0.693 \arcsec$, redshift: 0.091.
	C: Obvious curvature in jet. Additionally, bending at bright spot, possibly a jet-cloud interaction. 
	R: Wide western hotspot.
	Lobe symmetry indicates that jet path would have moved towards the cloud/galaxy recently.}
    	\label{fig:3c388}
\end{figure}
\begin{figure}\centering
	\includegraphics[width=793\swidth]{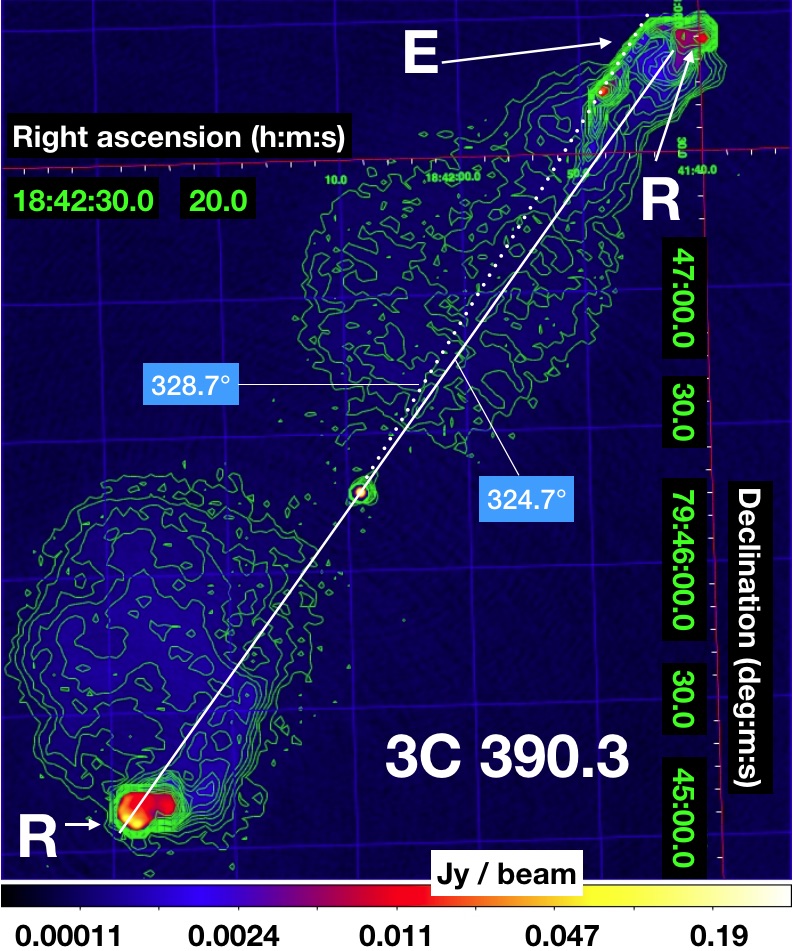}
	\caption{3C~390.3, 4.99 GHz, resolution: $1 \arcsec$, redshift: 0.057.
	E: Jet at leeward (see below) edge of lobe,
	R: Wide terminal hotspots on both sides.
	Likely crosswind pushing lobes towards east. For the southern lobe, 
	there is an indication for a jet on the western (windward) side of the lobe.}
    	\label{fig:3c390P3}
\end{figure}
\begin{figure}\centering
	\includegraphics[width=681\swidth]{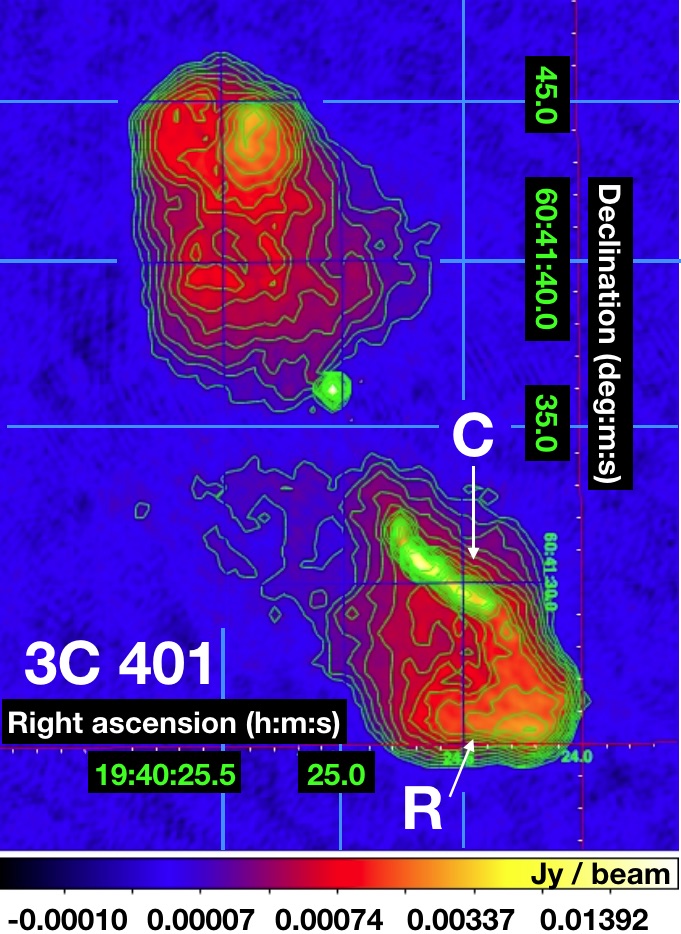}
	\caption{3C~401, 8.44 GHz, resolution: $0.27 \arcsec$, redshift: 0.20.
	C: Curved jet on the southern side. E: Jets / hotspots at edge of lobes.
	Lobe brightness asymmetry on the northern side. 
	 R: Wide hotspot / trail of hot back flow on the southern side.}
    	\label{fig:3c401}
\end{figure}
\begin{figure}\centering
	\includegraphics[width=854\swidth]{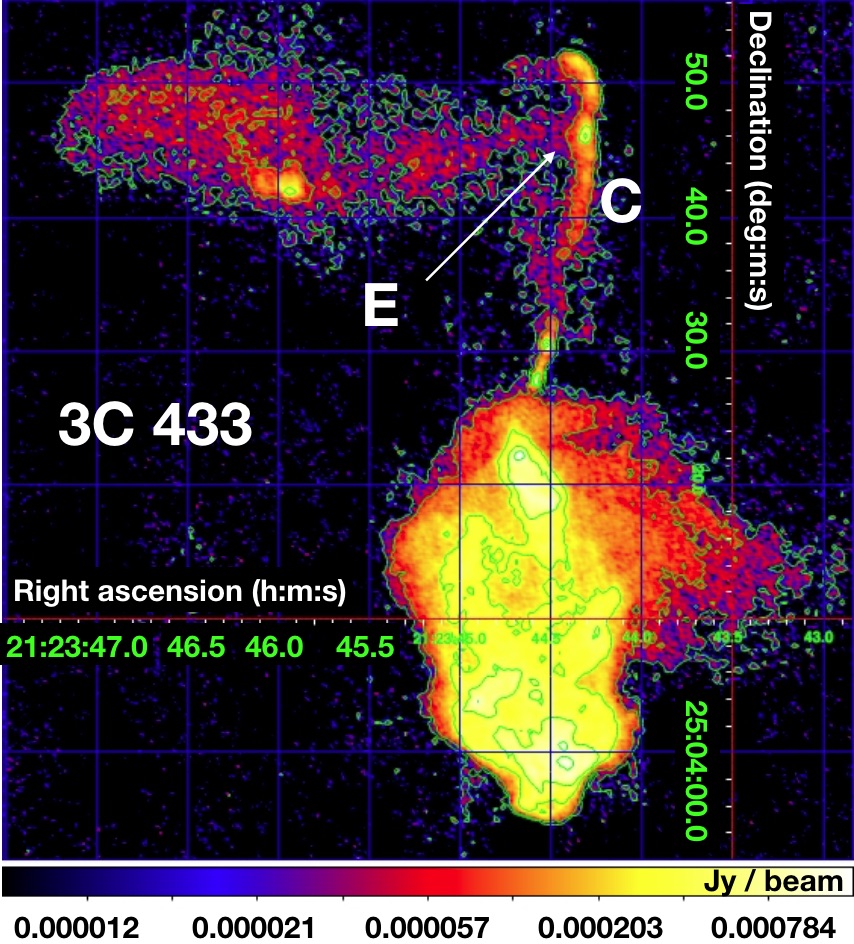}
	\caption{3C~433, 8.47 GHz, resolution: $0.25 \arcsec$, redshift: 0.10.
	C: Curved jet on the northern side. 
	E: Misaligned lobe axis on the northern side; Lobe brightness asymmetry on the southern side. 
	Additional hotspot in the northeast lobe extension.}
    	\label{fig:3c433}
\end{figure}
\begin{figure}\centering
	\includegraphics[width=439\swidth]{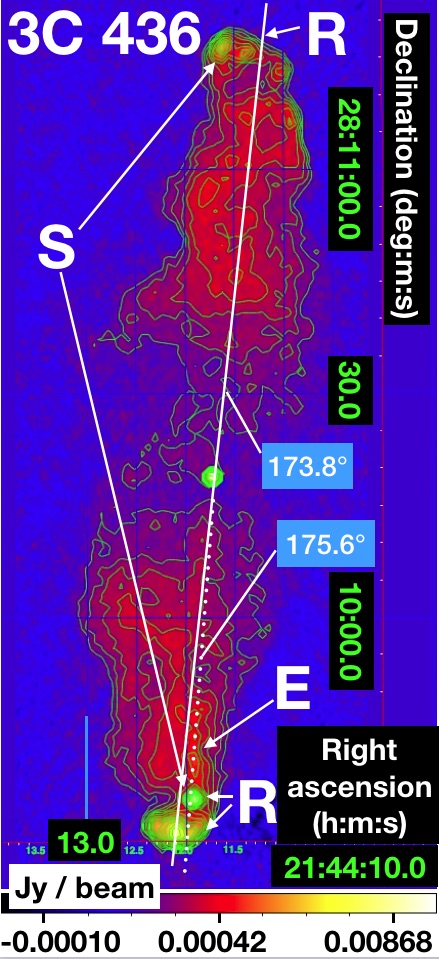}
	\caption{3C~436, 8.44 GHz, resolution: $0.75 \arcsec$, redshift: 0.21.
	E: Southern jet at edge of lobe.
	R: Multiple hotspots.
	S: Northern hotspots on opposite side of lobe as southern hotspot. 
	\citet{Hardea97} note a jet in the southern lobe, on the western side.}
    	\label{fig:3c436}
\end{figure}
\begin{figure}\centering
	\includegraphics[width=944\swidth]{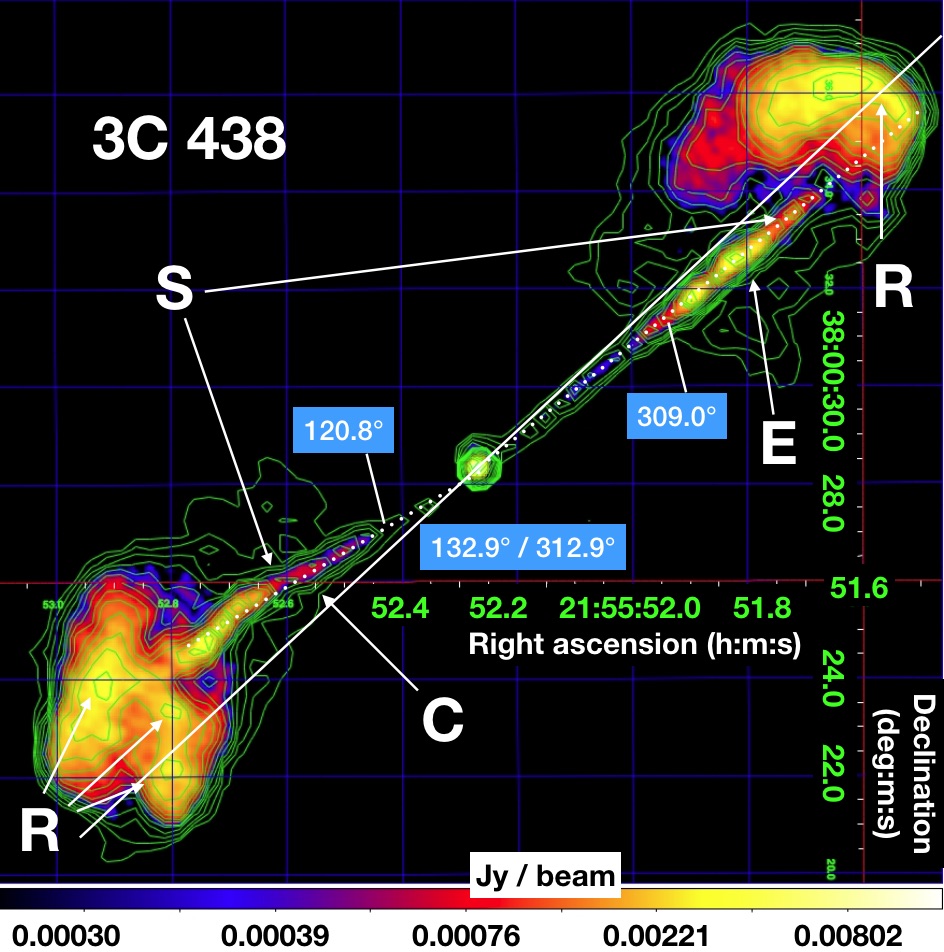}
	\caption{3C~438, 8.44 GHz, resolution: $0.23 \arcsec$, redshift: 0.29.
	C: Curved jet on the eastern side.  More curvature in counterjet (East).
	E: Misaligned lobe axis on the western side.  
	R: Wide terminal hotspot on the eastern side. Trail of hot back flow on the western side.  
	S: S symmetric jets.
	Angle between northern and southern jet $\ne 180\deg$}
    	\label{fig:3c438}
\end{figure}
\begin{figure}\centering
	\includegraphics[width=713\swidth]{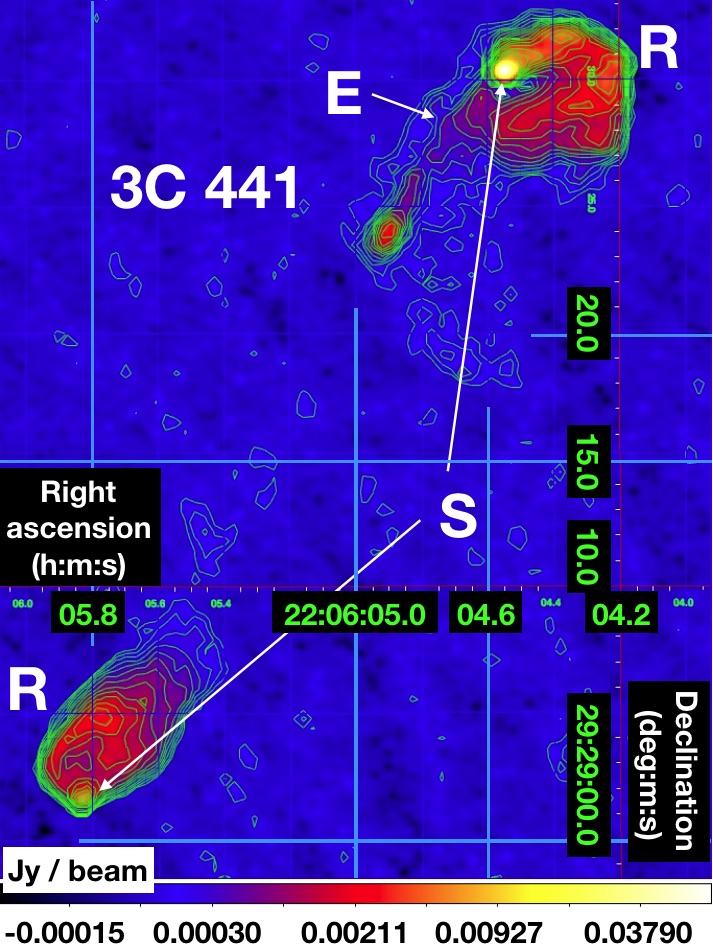}
	\caption{3C~441, 4.86 GHz, resolution: $0.35 \arcsec$, redshift: 0.71.
	E: Jet and hotspot at one edge of lobe.
	R: Wide hotspot / ring-like feature in both lobes. S: Hotspot S-symmetry.}
    	\label{fig:3c441}
\end{figure}
\begin{figure}\centering
	\includegraphics[width=1268\swidth]{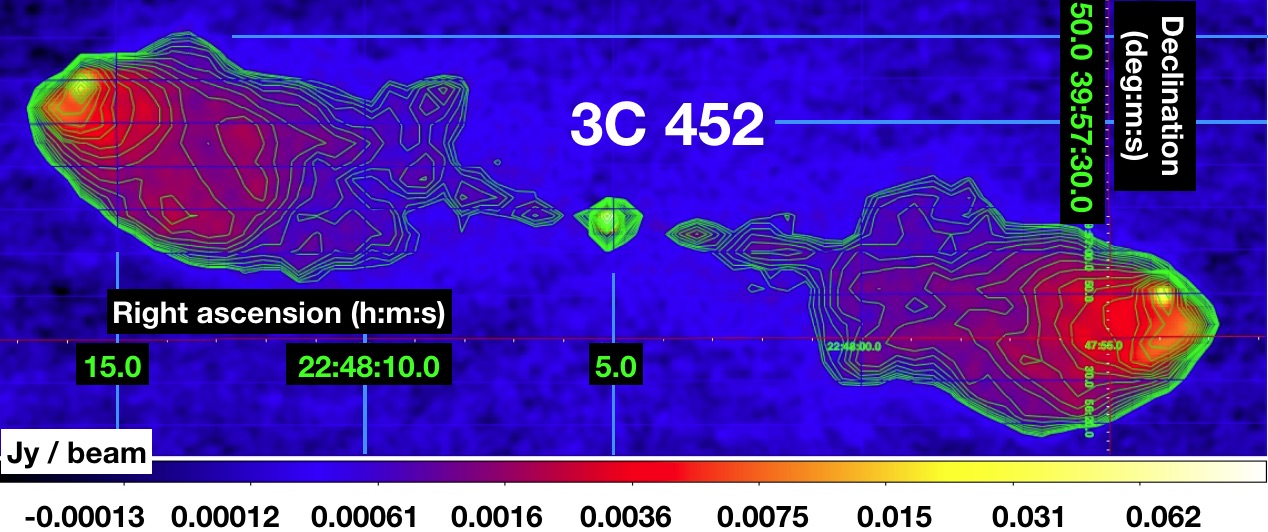}
	\caption{3C~452, 8.06 GHz, resolution: $2.1 \arcsec$, redshift: 0.081.
	A crosswind from north can probably not fully explain the hotspot asymmetry, 
	because a straight line connecting the hotspots would leave the core south. 
	However, since the situation appears doubtful we conservatively assign no precession indicator.}
    	\label{fig:3c452}
\end{figure}


\section{Nearby binary black hole candidates}\label{ap:nearby_sources}
The closest sources are the best candidates for a direct detection of gravitational waves if they indeed 
contain close binary black holes. We summarise here important data, give an estimate for the precession period 
and predict the gravitational wave strain (Table~\ref{t:gw}). 
We also comment on the presence of a core structure in the stellar profiles where available, 
which is expected as a result of stars being lost by interaction with the supermassive black hole 
binary during their mutual approach.

\subsection{Hydra A}
This well studied radio source has a large-scale S-symmetry \citep{Taylea90}. 
An extended 3D simulation study of the jet-environment interaction for this particular source, where source 
geometry and precession period has been varied, has recently determined the precession period to 1~Myr 
\citep{Nawea16a}. \rva{This is also compatible with a \citet{Gowea82} model of precessing jets
with relativistic aberration (Fig~\ref{fig:Hydracomp} and Table~\ref{t:hyapars}).} As argued above for similar precession periods, this precession is most likely related to 
geodetic precession due to a binary supermassive black hole. The precession period together with the 
black hole mass of $(5\pm4)\times10^8 M_\odot$ \citep{Rafea06} limits the binary separation to a maximum of 
0.2~pc (eq.~\ref{eq:geoprecsep}). This is accessible to high-resolution radio imaging. 
A helical pattern would hence be expected for the parsec-scale jet 
{ unless the jet is produced by the primary in a very unequal mass binary.}
The parsec-scale image shows a straight jet \citep{Taylor96}. 
This might mean that in this source the jet is { indeed produced by a dominant primary.} 
The parsec-scale jet is obviously misaligned with the large-scale jet, as expected for a precessing source.

\begin{figure}\label{fig:cenA}
	\includegraphics[width=\columnwidth]{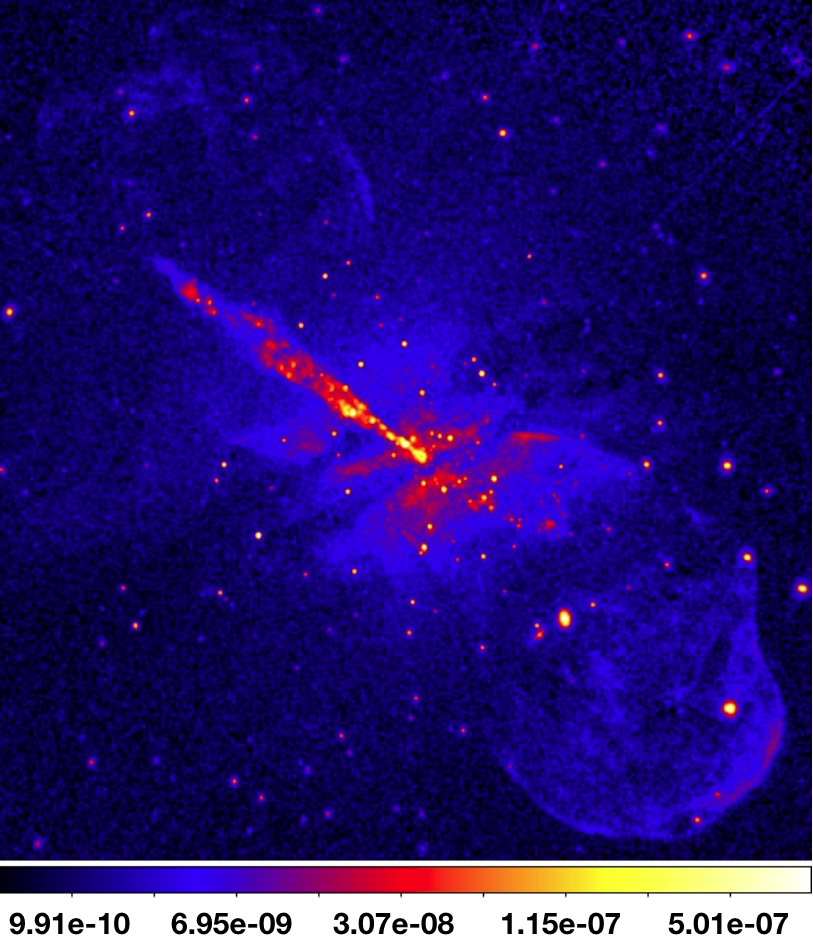}
	\caption{{ Chandra} X-ray image of Centaurus A, exposure-corrected 
	image in the 0.4-2.5 keV band 
	taken from the dataset presented by \citet{Hardea07}. 
	The bright X-ray structure emphasises the current jet, whereas the extended X-ray emission 
	identifies the long-term average of the jet direction. The extent of this part of the source 
	is about 11~kpc. S-symmetry and a misalignment of the jet with the lobe axis
	suggest precession. There is a weak indication of a curved counterjet.}
\end{figure}

\subsection{Centaurus A}
Similar to Hydra~A, Cen~A is also a source with S-symmetric structure \citep[Fig.~\ref{fig:cenA}]{Morgea99}. 
The inner structure has a size and geometry very similar to Hydra A, 
suggesting also a precession with period of approximately 1~Myr. 
\rva{We can also estimate the precession period from the X-ray image (Fig.~\ref{fig:cenA}). 
The bow shock in the ambient gas is clearly detected. \citet{Crostea09} analyse the shock and derive  
a source age of 2~Myr. The southern bow shock has an extension which could be related
to the current impact of the jet. If the precession period was comparable to the source age, the extension
would be comparable in size to the overall bow shock region. The fact that a small extension is observed
requires that the precession period is significantly less than 2 Myr, consistent with the above estimate. The argument
becomes stronger, if even the observed bow shock extension was not related to the current impact of the jet.   
The jet is straight for about 4.5~kpc \citep{Hardea07}. This requires a precession period significantly longer than
0.02~Myr. We adopt $1\pm0.5$~Myr as an estimate.}
The faint X-ray counterjet \citep{Hardea07} appears more strongly curved than the jet, consistent with expectations 
for relativistic aberration from a precessing jet \citep{Gowea82}. 
Assuming a binary black hole with a total mass of $(5.5\pm1)\times10^7 M_\odot$ \citep{Cappea09}, 
a separation of less than about 0.05~parsec is derived (eq.~\ref{eq:geoprecsep}). 
\rva{\citet{Muellea14} present high-resolution VLBI images of Cen~A with similar resolution. No sign of a secondary
radio core is seen. This could indicate that the separation is smaller than 0.05~pc, or that the secondary core
was not luminous enough at the time of observation, possibly because the secondary black hole is smaller.}
The stellar density as a function of radius shows a core structure, 
as expected for a binary supermassive black hole, and a dust lane signifies a 
recent galaxy merging event \citep{Marcea00}.

\begin{figure}\label{fig:M87}
	\includegraphics[width=\columnwidth]{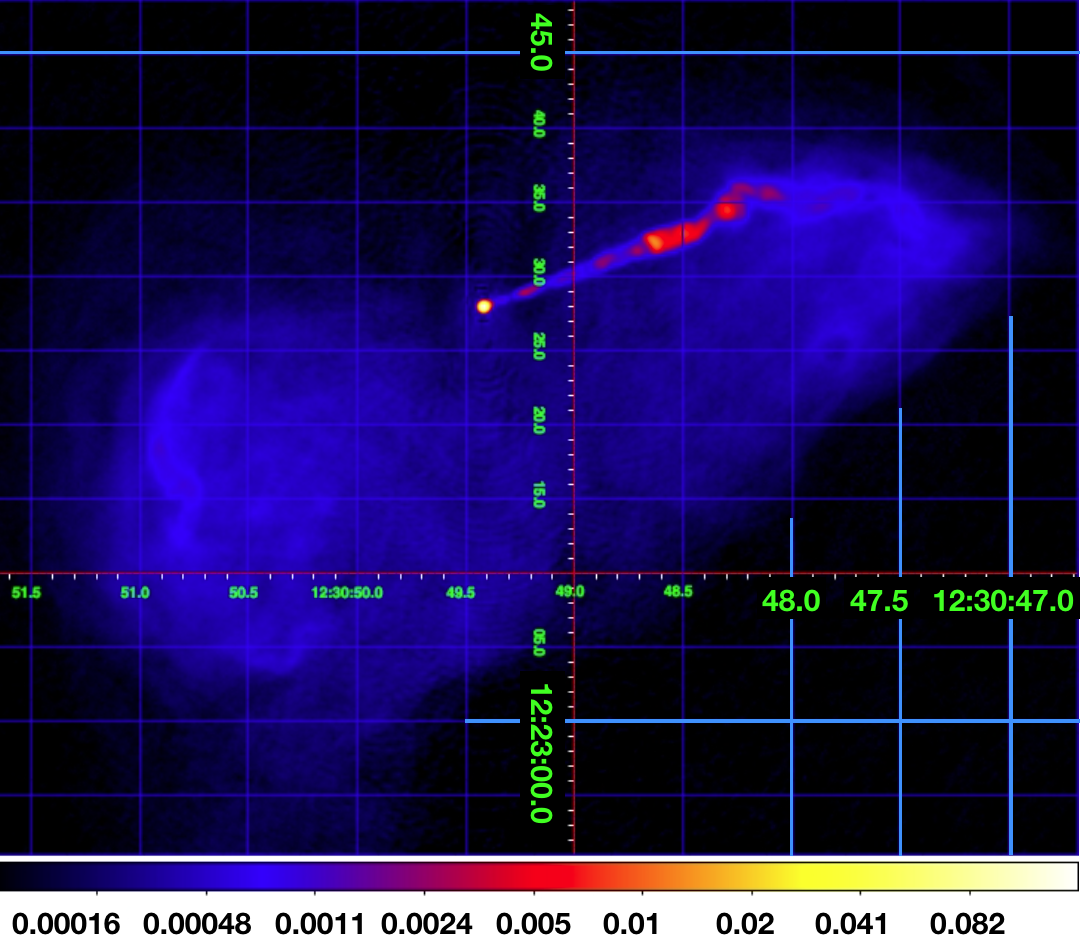}
	\caption{4.6 GHz image of M87; data from \citet{HOE89}. 
	The jet is at the edge of the radio lobe, an indication of precession.}
\end{figure}

\subsection{M87}
Low frequency radio imaging establishes the average direction of a previous outburst to be 
north-south \citep{Owea00}. The current jet direction is towards $PA \approx 300\deg$. 
The black hole of mass of $(6.6\pm0.4)\times10^9 M_\odot$ \citep{Gebhea11} 
is unlikely to have accreted a significant fraction of its mass in the last $5\times10^7$~yr since the last outburst. 
Hence, accretion has not changed the spin axis. 
The current jet is traced down to a few Schwarzschild radii and stays straight, 
hence there is no jet-cloud interaction. The fossil outburst has lasted too long so any 
clouds that would have affected the past jet would have been ablated too quickly to 
influence the long-term average symmetry. Hence, the difference between the symmetry axis 
of the older, large-scale lobes and the current jet direction is significant. 
A binary black hole could cause this in two ways: the current and previous outbursts 
could come from different black holes with different spin axes, or the spin axis has precessed to 
the different current location. In order to change direction between the outbursts, the precession period has to be less 
than about $10^8$~yr. The fact that the jet is straight over about 1~kpc from the core suggests a period longer
than about $10^5$ years. The asymmetry of the current source’s radio lobes (Fig.~\ref{fig:M87}) is 
reminiscent of simulations of precessing jets with source age similar to the precession period \citep{DonSmi16}. 
This suggests that the precession period is indeed comparable to the current source age, about 1~Myr. 
We conservatively adopt a range of $5\pm2$~Myr as precession period. M87 has a stellar core profile \citep{Cotea06}.

%
%


\bsp	
\label{lastpage}
\end{document}